\documentclass[useAMS, usenatbib]{mnras}
\usepackage{color}
\usepackage{booktabs}   % Fancier tables
\usepackage{tabularx}
\usepackage{float}
\usepackage{totcount}

\usepackage[dvipsnames, table]{xcolor} % adds colors to colorspace (https://en.wikibooks.org/wiki/LaTeX/Colors)
\usepackage{graphicx}
\usepackage{multirow}
\usepackage[section]{placeins}   % Force figures to stay in their sections (tex.stackexchange.com/q/279/22806)
\usepackage{subfig}
\usepackage{widetext}
\usepackage{upgreek}    % allow alternative greek letters like $\uptau$
\usepackage{enumitem}   % customize enumerate symbols
\setlist[enumerate]{leftmargin=*}   % Set left-margin of enumerate lists to match the edge

\usepackage{lipsum} % generate dummy text
\usepackage{amsmath}
\usepackage{amssymb}
\usepackage{subfiles}    % be able to compile the included subfiles

\usepackage{txfonts}     %more compact fonts closer to the ones used by MNRAS

\graphicspath{ {figures/} }

\newcommand{\tr}[1]{\textrm{#1}}
\newcommand{\trt}[1]{\textrm{\tiny{#1}}}

\newcommand{\msol}{\tr{M}_{\odot}}

\newcommand{\mbh}{M_\bullet}

\newcommand{\mstar}{M_\star}

\newcommand{\lr}[1]{\left(#1\right)}
\newcommand{\scale}[3][]{\lr{ \frac{#2}{#3} }^{#1}}

\newcommand{\E}[1]{\times\nobreak10^{#1}}
\newcommand{\ls}{\lesssim}
\newcommand{\gs}{\gtrsim}

\def\aj{Astron. J.}              		   		% Astronomical Journal
\def\apj{Astrophys. J.}       		        	 	% Astrophysical Journal
\def\apjl{Astrophys. J. Lett.}             		% Astrophysical Journal, Letters

\def\pasa{PASA}
\def\apjs{Astrophys. J., Suppl. Ser.}            % Astrophysical Journal, Supplement
\def\mnras{Mon. Not. R. Astron. Soc.}        % Monthly Notices of the RAS
\def\prd{Phys. Rev. D}      				% Physical Review D
\def\araa{Annu. Rev. Astron. Astrophys.}  % Annual Review of Astron and Astrophys
\def\nat{Nature}              				% Nature
\def\aap{Astron. Astrophys.}               		% Astronomy and Astrophysics

\newcommand{\secref}[1]{\textsection\ref{#1}}
\newcommand{\figref}[1]{Fig.~\ref{#1}}
\newcommand{\refeq}[1]{{Eq.~\ref{#1}}}
\newcommand{\tabref}[1]{{Table~\ref{#1}}}

\newcommand{\fnm}[1]{\footnotemark[#1]}
\newcommand{\fnt}[2]{\footnotemark[#1]{#2}}
\newcommand{\mc}[2]{\multicolumn{#1}{c}{#2}}
\newcommand{\mr}[2]{\multirow{#1}{*}{#2}}

\newcommand{\bmax}{b_\trt{max}}   % maximum impact parameter

\newcommand{\rlc}{\mathcal{R}_{lc}}
\newcommand{\rinfl}{\mathcal{R}_\trt{infl}}
\newcommand{\risco}{\mathcal{R}_\trt{isco}}
\newcommand{\rs}{R_\trt{s}}    % Stellar radius
\newcommand{\rb}{R_\trt{b}}    % Bound radius
\newcommand{\rh}{R_\trt{h}}    % Hard radius
\newcommand{\mpro}{m_\trt{p}}    % Proton Mass
\newcommand{\sigmat}{\sigma_\trt{T}}
\newcommand{\trel}{\tau_\trt{rel}}
\newcommand{\torb}{\tau_\trt{orb}}
\newcommand{\mce}{\mathcal{E}}
\newcommand{\tcross}{\tau_\trt{cross}}

\newcommand{\mdot}{\dot{M}}
\newcommand{\mdotedd}{\dot{M}_\trt{Edd}}
\newcommand{\mdotill}{\dot{M}_\trt{ill}}
\newcommand{\ledd}{L_\trt{Edd}}
\newcommand{\radeff}{\varepsilon_\trt{rad}}
\newcommand{\ms}{m_\star}
\newcommand{\rstarhalfmass}{R_{\star,\trt{1/2}}}
\newcommand{\omgw}{\Omega_\trt{GW}}   % ratio of GW energy density to critical density
\newcommand{\pyr}{\textrm{yr}^{-1}}
\newcommand{\ayr}{A_{\trt{yr}^{-1}}}        % GWB amplitude normalization at 1/yr
\newcommand{\atyr}{A_{{\scriptscriptstyle0.1} \trt{yr}^{-1}}}  % GWB amplitude normalization at 0.1/yr

\newcommand{\astropy}{\texttt{Astropy}}
\newcommand{\matplotlib}{\texttt{matplotlib}}
\newcommand{\numpy}{\texttt{NumPy}}
\newcommand{\scipy}{\texttt{SciPy}}
\newcommand{\ipython}{\texttt{ipython}}

\newcommand{\fcoal}{\mathcal{F}_\trt{coal}}
\newcommand{\fstall}{\mathcal{F}_\trt{stall}}
\newcommand{\frefill}{\mathcal{F}_\trt{refill}}
\newcommand{\flcsix}{$\frefill = 0.6$}
\newcommand{\flcten}{$\frefill = 1.0$}
\newcommand{\fedd}{f_\trt{Edd}}    % Mdot-eddington factor for limiting
\newcommand{\rsg}{\mathcal{R}_\trt{SG}}
\newcommand{\rsgmax}{\mathcal{R}_\trt{SG,Max}}
\newcommand{\dfatten}{f_\trt{DF,LC}}
\newcommand{\mchirp}{\mathcal{M}}     % Chirp-mass 
\newcommand{\mseed}{M_\trt{seed}}    % seed-mass BH
\newcommand{\diffco}{\mathcal{D}_{v^2}}   % Diffusion coefficient
\newcommand{\volcom}{V_\trt{c}}   % comoving volume

\newcommand{\bigt}{\scalebox{1.2}{\ensuremath{\uptau}}}
\newcommand{\thard}{\bigt_\tr{h}}
\newcommand{\thardgw}{\bigt_\tr{gw}}

\newcommand{\thardi}{\bigt_{\tr{r},i}}
\newcommand{\tdyn}{\bigt_\trt{dyn}}
\newcommand{\tgw}{\bigt_\tr{gw}}
\newcommand{\tgwi}{\bigt_{\trt{gw},i}}
\newcommand{\tvisc}{\bigt_v}
\newcommand{\tviscone}{\bigt_\tr{v,1}}
\newcommand{\tvisctwo}{\bigt_\tr{v,2}}
\newcommand{\rgap}{\lambda_\trt{gap}}
\newcommand{\rcrit}{\mathcal{R}_\trt{crit}}
\newcommand{\rselfgrav}{\lambda_\trt{sg}}

\newcommand{\reg}[1]{\textit{Region-#1}}
\newcommand{\egw}{\varepsilon_\trt{GW}}   % energy in GW
\newcommand{\fluxlc}{F_\trt{lc}}
\newcommand{\fluxflc}{F^\trt{full}_\trt{lc}}
\newcommand{\fluxsslc}{F^\trt{eq}_\trt{lc}}

\newcommand{\scell}[1]{\begin{tabular}{@{}c@{}}#1\end{tabular}}

\defcitealias{hkm09}{HKM09}
\defcitealias{bbr80}{BBR80}

\def\biblio{\bibliographystyle{mnras}\bibliography{\main/refs}}  

\newtotcounter{citnum} %From the package documentation
\def\oldbibitem{} \let\oldbibitem=\bibitem
\def\bibitem{\stepcounter{citnum}\oldbibitem}

\title[MBH Binaries in Dynamical Environments]{Massive Black Hole Binary Mergers in Dynamical Galactic Environments}
\author[L.Z.~Kelley et al.]{Luke Zoltan Kelley$^{1}$\thanks{E-mail:lkelley@cfa.harvard.edu},
	Laura Blecha$^{2}$,
	Lars Hernquist$^{1}$
\\
$^{1}$ Harvard University, Center for Astrophysics, Cambridge MA 02138 \\
$^{2}$ University of Maryland
}

\begin{document}

% Override 'biblio' command for this document (suppress subfiles bibliographies)
\def\biblio{}

%\date{Accepted 1988 December 15. Received 1988 December 14; in original form 1988 October 11}
\pagerange{\pageref{firstpage}--\pageref{lastpage}} \pubyear{2014}

\maketitle
\label{firstpage}

\begin{abstract} 
% This document contains \total{citnum}\ references.
Gravitational Waves (GW) have now been detected from stellar-mass black hole binaries, and the
first observations of GW from Massive Black Hole (MBH) Binaries are expected within the next
decade.  Pulsar Timing Arrays (PTA), which can measure the years long periods of GW from MBHB, have
excluded many standard predictions for the amplitude of a stochastic GW Background (GWB). We use
coevolved populations of MBH and galaxies from hydrodynamic, cosmological simulations ('Illustris')
to calculate a predicted GWB.  The most advanced predictions so far have included binary hardening
mechanisms from individual environmental processes.  We present the first calculation including all
of the environmental mechanisms expected to be involved: dynamical friction, stellar 'loss-cone'
scattering, and viscous drag from a circumbinary disk.  We find that MBH binary lifetimes are
generally multiple gigayears, and only a fraction coalesce by redshift zero.  For a variety of
parameters, we find all GWB amplitudes to be below the most stringent PTA upper limit of
$A_{\textrm{yr}^{-1}} \approx \nobreak 10^{-15}$.  Our fairly conservative fiducial model predicts
an amplitude of $A_{\textrm{yr}^{-1}} \approx \nobreak 0.4\times 10^{-15}$.  At lower frequencies,
we find $A_{0.1\,\textrm{yr}^{-1}} \approx \nobreak 1.5\times 10^{-15}$ with spectral indices
between $-0.4$ and $-0.6$---significantly flatter than the canonical value of $-2/3$ due to purely
GW-driven evolution.  Typical MBHB driving the GWB signal come from redshifts around $0.3$, with
total masses of a few times $10^9\,M_\odot$, and in host galaxies with very large stellar masses.
Even without GWB detections, our results can be connected to observations of dual AGN to constrain
binary evolution.

\end{abstract}

\begin{keywords}
quasars: supermassive black holes, galaxies: kinematics and dynamics
\end{keywords}

% ==============================================================================================
% = = = = = = = = = = = = = = = = = = = =   DOCUMENT   = = = = = = = = = = = = = = = = = = = = =
% ==============================================================================================

% Introduction
% ------------

% ==================================================================
% =======================   INTRODUCTION   =========================

\section{Introduction}
\label{sec:intro}

% ---------------------------------------------------------
% ------- Massive Black Hole Binaries and Hardening -------
% ---------------------------------------------------------

% MBH exist and galaxies merge
Massive Black Holes (MBH) occupy at least the majority of massive galaxies \citep[e.g.][]{soltan1982,
kormendy1995, Magorrian1998} which are also known to merge with each other as part of their typical lifecycles
\citep[e.g.][]{lacey1993, lotz2011, rodriguez-gomez2015}.  This presents two possibilities for the MBH of host
galaxies which merge: either they also merge, or they persist as multiples in the resulting remnant galaxies.
Naively, one might expect that BH \textit{must} undergo mergers for them to grow---as for halos and to some
degree galaxies in the fundamentally hierarchical Lambda-Cold Dark Matter model \citep[e.g.][]{white1991}. On
the contrary, the linear growth of black holes (i.e.~at most doublings of mass) is known to be woefully
inadequate to form the massive quasars observed at high redshifts \citep[e.g.][]{fan2006}, which require
exponential growth from Eddington\footnote{The Eddington accretion rate can be related to the Eddington
luminosity as $\mdotedd = \ledd/ \radeff c^2$, for a radiative efficiency $\radeff$, i.e.
$$\mdotedd = \frac{4\pi G M \mpro}{\radeff c \sigmat},$$ for a proton mass $\mpro$, and Thomson-Scattering
cross-section $\sigmat$.} (or super-Eddington) accretion \citep[e.g.][]{haiman2013}.  Assuming that the
energy fueling Active Galactic Nuclei (AGN) comes from accretion onto compact objects, the integrated luminosity
over redshifts requires a present day mass density comparable to that of observed MBH \citep{soltan1982}.
Coalescences of MBH are then neither sufficient nor necessary to match their observed properties \citep[e.g.][]{small1992}.

% Observations of dual/binary AGN
In the last decade the search for multi-MBH systems have yielded many with kiloparsec-scale separations
\citep[`Dual AGN', e.g.][]{comerford2012}---although they seem to represent only a fraction of the population
\citep{koss2012}---and even some triple systems \citep[e.g.][]{deane2014}.  At separations of kiloparsecs,
`Dual MBH' are far from the `hard' binary phase.  A `hard' binary is one in which the binding energy is
larger than the typical kinetic energy of nearby stars \citep{bt87}, and relatedly the binding energy tends to
increase (i.e.~the system `hardens') with stellar interactions \citep{hut1983}.  `Hard' is also used more
informally to highlight systems which behave dynamically as a bound system..  Only one true MBH Binary (MBHB,
i.e.~gravitationally bound) system has been confirmed: a resolved system in the radio galaxy 0402+379, at a
separation of $\sim 7 \textrm{ pc}$ by \citet{rodriguez2006}.  Even this system is well outside of the
`Gravitational Wave (GW) regime', in which the system could merge within a Hubble time due purely to GW emission.
There are, however, a growing number of candidate, unresolved systems with possible sub-parsec separations
\citep[e.g.][]{valtonen2008, dotti2009}.  Detecting---and even more so, excluding---the presence of MBHB is
extremely difficult because activity fractions of AGN (particularly at low masses) are uncertain, the spheres
of influence of MBH are almost never resolved, and the expected timescales at each separation are unknown.  It
is then hard to establish if the absence of MBHB observations is conspicuous.

% Theoretical considerations of mergers --- tease for later section
On the theoretical side, the picture is no more complete.
Pioneering work by \citet[][hereafter \citetalias{bbr80}]{bbr80} outlined the basic MBH merger process.
On large scales ($\sim\textrm{kpc}$), MBH are brought together predominantly by dynamical
friction---the deceleration of a body moving against a gravitating background.
Energy is transferred from the motion of the massive object to a kinetic thermalization of the background medium,
in this case the dark-matter, stellar and gaseous environment of MBHB host galaxies.  
As the binary tightens, stars become the primary scatterers.
Once the binary becomes hard ($\sim 10 \textrm{ pc}$), depletion of the `Loss Cone' (LC)---the region of
parameter space with sufficiently low angular momentum to interact with the MBHB---must be considered.
The rate at which the LC is refilled is likely the largest uncertainty in the merger process,
and determines the fraction of systems which are able to cross the so-called `final-parsec'
\citep[for the definitive review, see:][]{merritt2013}.  If binaries are able to reach smaller scales
($\lesssim 0.1 \textrm{ pc}$), gas-drag can contribute significantly to the hardening process in gas-rich
systems.  Eventually GW emission inevitably dominates at the smallest scales ($\lesssim 10^{-3} \textrm{ pc}$).

Since \citetalias{bbr80}, the details of the merger process have been studied extensively, largely focusing
on the LC \citep[for a review, see, e.g.][]{merritt2005}.  \citet{quinlan1996} and \citet{quinlan1997} made
significant developments in numerical, N-Body scattering experiments, allowing the `measurement' of binary
hardening parameters.  More advanced descriptions of the LC, applied to realistic galaxy density profiles,
by \citet{yu2002} highlighted the role of the galactic gravitational potential---suggesting that flattened
and strongly triaxial galaxies could be effective at refilling the LC and preventing binaries from stalling.
N-Body simulations have continued to develop and improve our understanding of rotating galaxies
\citep[e.g.][]{berczik2006}, and more realistic galaxy-merger environments and galaxy shapes
\citep[e.g.][]{khan2011, khan2013}.
The interpretation and usage of simulations have also developed
significantly, in step with numerical advancements, allowing for better understanding of the underlying
physics \citep[e.g.][]{sesana2015, vasiliev2015}.

Smoothed Particle Hydrodynamic (SPH) simulations of MBH binary dynamics in gaseous environments have also
been performed.  \citet{escala2005a, escala2005} showed that dense gaseous regions, corresponding to
ULIRG-like\footnote{Ulta-Luminous Infra-Red Galaxies (ULIRG) are bright, massive, and gas-rich --- all indicators
of favorable MBH merger environments.} galaxies, can be very effective at hardening binaries.  Similar SPH
studies have affirmed and extended these results to more general environments and MBHB configurations
\citep[e.g.][]{dotti2007, cuadra2009}.  While modern simulations continue to provide invaluable insights,
and exciting steps towards simulating MBHB over broad physical scales are underway \citep[e.g.][]{khan2016},
neither hydrodynamic nor purely N-Body simulations are close to simulating the entire merger process in its
full complexity\footnote{Resolving the interaction of individual stars with a MBHB over the course of the entire
merger process, for example, would require almost nine orders of magnitude contrast in each mass, distance, and time.}.

% ---------------------------------------------------
% ------- Gravitational Waves and Backgrounds -------
% ---------------------------------------------------

% GW-Era, LIGO and PTA
With our entrance into the era of Gravitational Wave (GW) astronomy \citep{ligo2016a}, we are presented with the
prospect of observing compact objects outside of both the electromagnetic spectrum and numerical simulations.
Direct detection experiments for gravitational waves are based on precisely measuring deviations in path length
(via light-travel times).  While ground-based detectors like the Laser Interferometer Gravitational-Wave Observatory
\citep[LIGO;][]{ligo2016b} use the interference of light between two orthogonal, kilometer scale laser arms, Pulsar
Timing Arrays \citep[PTA,][]{foster1990} use the kiloparsec scale separations between earth and galactic pulsars
\citep{detweiler1979}.  PTA are sensitive to GW at periods between the total observational baseline and
the cadence between observations.  These frequencies, roughly $0.1 - 10 \textrm{ yr}^{-1}$,
are much lower than LIGO---corresponding to steady orbits of MBHB with total masses between
$\approx 10^6 - 10^{10} \, \msol$, at separations of $\approx 10^{-3} - 10^{-1} \textrm{ pc}$
(i.e.~$1 - 10^6 \, \rs$\footnote{Schwarzschild radii, $\rs \equiv 2GM/c^2$.}).  The parameter space is shown in 
\figref{fig:gw_sep_freq}.

% GWBackground
Binaries produce GW which increase in amplitude and frequency as the orbit hardens,
up to the `chirp' when the binary coalesces.
MBHB chirps will be at frequencies below the LIGO band, but above that of PTA.  Future space-based
interferometers \citep[e.g.~eLISA;][]{elisa2013} will bridge the divide and observe not only the coalescence of
MBHB, but also years of their final inspiral.  The event rate of nearby, hard MBHB which could be
observed as individual `continuous wave' sources is expected to be quite low, and likely the first GW detections
from PTA will be of a stochastic GW Background (GWB) of unresolved sources \citep{rosado2015}.

% Derive GWB Spectrum
The shape of the GWB spectrum was calculated numerically more than two decades ago \citep{rajagopal1995}, but
\citet{phinney2001} showed that the characteristic GWB spectrum can be calculated analytically by considering the total 
energy emitted as gravitational waves, integrated over redshift.  For a complete and pedagogical derivation
of the GWB spectrum see, e.g., \citet{sesana2008}.  The `characteristic-strain', $h_c(f)$, can be calculated for a
finite number of sources, in some comoving volume $V_c$ (e.g.~a computational box), as,
	\begin{equation}
	\label{eq:gwb-naive}
	h_c^2(f) = \frac{4 \pi}{3 c^2} \left(2\pi f\right)^{-4/3}
				\sum_i \frac{1}{\left(1+z_i\right)^{1/3}} \frac{\left(G\mathcal{M}_i\right)^{5/3}}{V_c}.
	\end{equation}
Equation~\ref{eq:gwb-naive} is the simplest way to calculate a GW background spectrum,
requiring just a distribution of merger chirp-masses and redshifts. This type of relation is often written as,
	\begin{equation}
	h_c(f) = A_{0} \scale[-2/3]{f}{f_0},
	\end{equation}
which has become typical for GWB predictions, and usually normalized to $f_0 = 1 \textrm{ yr}^{-1}$
(with some $\ayr$).  The prediction of a GWB with spectral slope of $-2/3$ is quite general,
but does assume purely gravitational-wave driven hardening which produces a purely power-law evolution
in frequency.  The lack of high and low frequency cutoffs
is fortuitously accurate at the frequencies observable through PTA, which are well populated by
astrophysical MBHB systems.  Deviations from pure power-law behavior within this band, however,
are not only possible but expected---the degree of which, determined by how significant non-GW effects are,
is currently of great interest.

% ---------------------------------
% ------- GWB - Predictions -------
% ---------------------------------

Many predictions have been made for the normalization of the GWB based on extensions to the method
of \citet{phinney2001}.  The standard methodology is using Semi-Analytic Models\footnote{We use the term
`Semi-Analytic Model' loosely to refer to a \textit{realized population} constructed by an analytic prescription,
as apposed to derived from underlying physical models.} (SAM) of galaxy evolution, with prescribed MBHB mergers to
calculate a GWB amplitude.
Two of the earliest examples are \citet{wyithe2003}---who use analytic mass functions \citep{press1974} with observed
merger rates \citep{lacey1993}, and \citet{jaffe2003}---who use observationally derived galaxy mass functions,
pair fractions, and merger time-scales.  These studies find amplitudes of $\log \ayr = -14.3$ and $\log \ayr = -16$,
respectively which remain as upper and lower bounds to most predictions since then.  Monte Carlo realizations of
hierarchical cosmologies \citep{sesana2004} exploring varieties of MBHB formation channels
\citep[e.g.][]{sesana2008, sesana2013, roebber2016} have been extremely fruitful in populating and understanding the
parameter space, finding GWB amplitudes generally consistent with $\log \ayr \approx -15 \pm 1$.
\citet{sesana2016} find that accounting for bias in MBH-Host scaling relations moves SAM predictions towards the
lower end of this range at $\log \ayr = -15.4$.

    \begin{table*}
    \renewcommand{\arraystretch}{1.2}
    \setlength{\tabcolsep}{6pt}
    \begin{tabularx}{\textwidth}{@{}lccccc@{}}
    \toprule
							& GWB Amplitude\fnm{1}	& \mc{2}{Populations}			&					& Spectral Slope			\\
    Reference	 			& $\log \ayr$			& Galaxies		& Black holes	& MBHB Evolution	& [Deviations from $-2/3$]	\\
    \midrule
	\citet{jaffe2003}		& $-16$					& SAM			& SAM			& GW				& -							\\
	\citet{wyithe2003}		& $-14.3$				& SAM			& SAM			& GW				& -							\\
	\citet{kocsis2011}		& $-15.7 \pm 0.3$		& Cosmo-DM		& SAM 			& VD, GW 			& flattened, $f \ls 1 \pyr$ \\
	\citet{sesana2013}		& $-15.1 \pm 0.3$		& SAM			& SAM			& GW				& -							\\
	\citet{mcwilliams2014}	& $-14.4 \pm 0.3$		& SAM			& SAM			& DF, GW			& imposed cutoff, $f \ls 0.5\,\pyr$ \\
	\citet{ravi2014}   		& $-14.9 \pm 0.25$  	& Cosmo-DM		& SAM			& LC-Full\fnm{2}, GW& flattened $f\ls 10^{-1} \, \pyr$, cutoff $f \ls 10^{-2} \,\pyr$\\
    \citet{kulier2015}		& $-14.7 \pm 0.1$		& Cosmo-Hydro 	& SAM			& DF, GW			& -							\\
    \citet{roebber2016}		& $-15.2^{+0.4}_{-0.2}$	& Cosmo-DM		& SAM			& GW				& -							\\
    \citet{sesana2016}		& $-15.4 \pm 0.4$		& SAM			& SAM			& GW				& -							\\

    \bottomrule
    \end{tabularx}
    \begin{flushleft}
	\fnt{1}{Some values which were not given explicitly in the included references were estimated based
    on their figures, and thus should be taken as approximate.} \\
	\fnt{2}{In this case, the LC prescription is effectively always `Full'.} \\
	\vspace{-0.1in}
	\end{flushleft}
    \caption{Representative sample of previous predictions for the GW Background, with a basic summary of their
    implementation.  `Semi-Analytic Model (SAM)' is used loosely to refer to numerical models based on scaling
    relations and observed populations.  `Cosmo-DM' are cosmological Dark-Matter only (N-Body) simulations,
    while `Cosmo-Hydro' are hydrodynamic simulations including baryons.
	The physical evolution effects included are: Dynamical Friction (DF), Loss-Cone (LC) stellar scattering,
	Viscous Drag (VD) from a circumbinary disk, and Gravitational Wave (GW) radiation.  In this study we use
	populations of both galaxies and MBH which coevolved in the cosmological, hydrodynamic simulations `Illustris',
	and include all mechanisms of hardening (DF, LC, VD \& GW) in our models.}
    \label{tab:pta-predic}
    \end{table*}
    
% Environmental and higher order effects
More extensive models exploring deviations from the purely power-law GWB have also been explored.  For example,
at higher frequencies ($\gtrsim 1\,\pyr$) from a finite numbers of sources \citep{sesana2009}, or at lower
frequencies due to eccentric binary evolution \citep[e.g.][]{sesana2010}.  Recently, much work has focused on
the `environmental effects' outlined by \citetalias{bbr80}.  \citet{kocsis2011} incorporate viscous drag from a
circumbinary gaseous disk \citep[][hereafter \citetalias{hkm09}]{hkm09} on top of halos and mergers from the
dark-matter only Millennium simulations \citep{springel2005}, with MBH added in post-processing.   They find a
fairly low amplitude GWB, $\log \ayr \approx -16 \pm 0.5$, with a flattening spectrum below $\sim 1 \, \pyr$.
\citet{ravi2014} explore eccentric binary evolution in an always effectively refilled (i.e.~full) LC using
the Millennium simulation with the SAM of \citet{guo2011}.  They find $\log \ayr \approx -15 \pm 0.5$ with 
an turnover in the GWB below $\sim 10^{-2} \, \pyr$ and significant attenuation up to
$\sim 10^{-1} \, \pyr$.  Recently, both \citet{mcwilliams2014} and \citet{kulier2015} have implemented explicit
dynamical-friction formalisms along with recent MBH--Host scaling relations \citep{mcconnell2013} applied to
halo mass functions from Press-Schechter and the Millenium simulations respectively.  \citet{mcwilliams2014}
find $\log \ayr \approx 14.4 \pm 0.3$, and \citet{kulier2015} $\log \ayr \approx 14.7 \pm 0.1$, with both
highlighting the non-negligible fraction of binaries stalled at kiloparsec-scale separations.  Almost all
previous studies had assumed that all MBHB merge effectively.

% ----------------------------------
% ------- GWB - Upper Limits -------
% ----------------------------------

These predictions are summarized in \tabref{tab:pta-predic}.  While far from exhaustive,  we believe they are
a representative sample, with specific attention to recent work on environmental effects.  The amplitudes of the
predicted backgrounds are distributed fairly consistently around $\ayr \approx 10^{-15}$.
Assuming observational baselines of about $10$ yr, pulsar TOA accuracies of at least tens of microseconds are
required to constrain or observe a GWB with this amplitude
\citep[see, e.g.][]{blandford1984, rajagopal1995}.  Finding more millisecond pulsars
with very small intrinsic timing noise are key to improving GWB upper-limits, while increasing the total number
(and angular distribution) of pulsars will be instrumental for detections \citep{taylor2016}.

There are currently three ongoing PTA groups, the North-American Nanohertz Observatory for Gravitational-waves
\citep[NANOGrav,][]{arzoumanian2015a},
the European PTA \citep[EPTA,][]{desvignes2016}, and the Parkes PTA \citep[PPTA,][]{manchester2013a}.  Additionally, the
International PTA \citep[IPTA,][]{hobbs2010} aims to combine the data sets from each individual project, and has
recently produced their first public data release \citep{verbiest2016}.  Table~\ref{tab:pta-limits} summarizes the current
upper limits from each PTA.
These are the $2$--$\sigma$ upper bounds, based on both extrapolation to $\ayr$ along with that of the specific
frequency with the \textit{strongest constraint} assuming a $-2/3$ spectral index.
Overall, the lowest bound is from the PPTA, at $\ayr < 10^{-15}$, or in terms of the
fractional closure density, $\omgw(f = 0.2 \textrm{ yr}^{-1}) < \nobreak 2.3\E{-10}$ \citep{shannon2015}.

    \begin{table}\centering
    \renewcommand{\arraystretch}{1.2}
    \setlength{\tabcolsep}{4pt}
    \begin{tabular}{@{}lcccc@{}}
    \toprule
    			&				& \multicolumn{2}{c}{Strongest Constraint}	&							\\ \cmidrule(lr){3-4}
    PTA 		& $\ayr$		& $A_{f,0}$		& $f_0$ [$\textrm{yr}^{-1}$]& Reference					\\ \midrule
    European 	& $3.0\E{-15}$	& $1.1\E{-14}$	& $0.16$					& \citet{lentati2015}		\\
    NANOGrav	& $1.5\E{-15}$	& $4.1\E{-15}$	& $0.22$					& \citet{arzoumanian2015b}	\\
    Parkes		& $1.0\E{-15}$	& $2.9\E{-15}$	& $0.2$						& \citet{shannon2015}		\\
	IPTA		& $1.5\E{-15}$	& -				& -							& \citet{verbiest2016}		\\
    \bottomrule
    \end{tabular}
    \caption{Upper Limits on the GW Background from Pulsar Timing Arrays.  Values are given both at the standard
    normalization of $f = 1 \, \pyr$ in addition to the frequency and amplitude of the strongest constraint
    (when given).}
    \label{tab:pta-limits}
    \end{table}

Every existing prediction has been made with the use of SAM---mostly in the construction of the galaxy population,
but also in how black holes are added onto those galaxies.  SAM are extremely effective in efficiently creating large
populations based on observational relations.  Higher-order, less observationally constrained parameters can have
systemic biases however, for example galaxy merger rates \citep[see, e.g.][]{hopkins2010}---which are obviously critical
to understanding MBHB evolution.  Recently, \citet{rodriguez-gomez2015} have shown that merger rates from the cosmological,
hydrodynamic Illustris simulations \citep{vogelsberger2014a} show excellent agreement with observations, while differing
(at times substantially) from many canonical SAM.

% -----------------------------
% ------- Intro - Outro -------
% -----------------------------

\bigskip

% Paper/Project Outline
In this paper, we use results from the Illustris \citep{genel2014} simulations
(discussed in \secref{sec:ill}) to make predictions for the rates at which MBH form binaries, and evolve to coalescence.
Illustris provides the MBH population along with their self-consistently derived parent galaxies
and associated stellar, gaseous, and dark- matter components.  This is the first time that a hydrodynamic galaxy population,
with fully co-evolved MBH have been used to calculate a GWB spectrum.
We use these as the starting point for post-processed models of the unresolved merger dynamics themselves, including all of
the underlying hardening mechanisms: dynamical friction, loss-cone stellar scattering, viscous drag, and
GW evolution---again for the first time (\secref{sec:hard}) in a MBHB population calculation.
From these data, we make predictions for plausible GW backgrounds observable by PTA, focusing on the effects of
different particular mechanisms on the resulting spectrum such that future
detections and upper-limits can be used to constrain the physical merger process (\secref{sec:results}).

% Importance
In addition to the GWB, understanding the population of MBHB is also important for future space-based
GW observatories \citep[e.g.~eLISA][]{elisa2013}.  Solid predictions for binary timescales at different separations
will also be instrumental in interpreting observations of dual- and binary-AGN, in addition to offset and `kicked'
BHs \citep{blecha2016}.  Finally, MBHB could play a significant role in triggering stellar
Tidal Disruption Events \citep[TDE; e.g.~][]{ivanov2005, chen2009}, and explaining the distribution of observed TDE
host galaxies.

% Create a bibliography here, only if just this file is being compiled/built.
\biblio{}

%sec:intro

% ---------------------------------------------------------
% ------- Massive Black Hole Binaries and Hardening -------
% ---------------------------------------------------------

% Illustris
% ---------

% ===============================================================
% =======================   ILLUSTRIS   =========================

\section{The Illustris Simulations}
\label{sec:ill}

% Simulation overview
Illustris are a suite of cosmological, hydrodynamic simulations which have accurately reproduced both
large-scale statistics of thousands of galaxies at the same time as the detailed internal structures of
ellipticals and spirals \citep{vogelsberger2014a}.
Illustris---hereafter referring to Illustris-1, the highest resolution of three runs---is a cosmological box
of $106.5 \textrm{ Mpc}$ on a side, with $1820^3$ each gas cells and dark-matter particles.
The simulations use the moving, unstructured-mesh hydrodynamic code AREPO \citep{springel2010},
with superposed SPH particles \citep[e.g.][]{springel2005}
representing stars (roughly $1.3 \times 10^6 \, \msol$ mass resolution, $700 \textrm{ pc}$
gravitational softening length), dark-matter (DM; $6.3\times 10^6 \, \msol$, $\, 1.4 \textrm{ kpc}$),
and MBH (seeded at $M \approx 10^5 \, \msol$) and allowed to accrete and evolve dynamically.
Stars form and evolve, feeding back and enriching their local environments, over the course of the simulation
which is initialized at redshift $z=137$ and evolved until $z=0$ at which point there are over $3\times10^8$
star particles.

% Papers for Illustris Details
For a comprehensive presentation of the galaxy formation models
(e.g.~cooling, inter-stellar medium, stellar evolution, chemical enrichment) see the papers:
\citet{vogelsberger2013} and \citet{torrey2014}.  For detailed descriptions of the general results of the
Illustris simulations, and comparisons of their properties with the observed universe, see e.g.,
\citet{vogelsberger2014b}, \citet{genel2014}, and \citet{sijacki2015}.  Finally, the data for the Illustris
simulations, and auxiliary files containing the black hole data\footnote{The blackhole data files were made
public in late Sept.~2016.} used for this analysis, have been made publicly available online \citep[www.illustris-project.org;][]{nelson2015}.

% ---------------------------------------------------
% ------- Massive Black Holes and MBH Mergers -------
% ---------------------------------------------------

\subsection{The Black Hole Merger Population}
\label{sec:ill_bh}

% Blackhole Seeding and Accretion
Black holes are implemented as massive, collisionless `sink' particles seeded into sufficiency massive halos.
Specifically, halos with a total mass above $7.1 \times 10^{10} \, \msol$, identified using an on the fly
Friends-Of-Friends (FOF) algorithm, which don't already have a MBH are given one with a seed mass,
$\mseed = 1.42 \times 10^5 \, \msol$ \citep{sijacki2007}.
The highest density gas cell in the halo is converted into the BH particle.  The BH mass is tracked as an
internal quantity, while the particle overall retains a \textit{dynamical} mass initially equal to the total
mass of its predecessor gas cell \citep{vogelsberger2013}.  The internal BH mass grows by Eddington-limited,
Bondi-Hoyle accretion from its parent gas cell (i.e.~the total dynamical mass remains the same).  Once the
excess mass of the parent is depleted, mass is accreted from nearby gas particles---increasing both the
dynamical mass of the sink particle, and the internal BH-mass quantity.

% Blackhole repositioning and "mergers"
BH sink particles typically have masses comparable to (or within a few orders of magnitude of)
that of the nearby stellar and DM particles.  Freely evolving BH particles would then scatter
around their host halo, instead of dynamically settling to their center---as is the case physically.
To resolve this issue, BH particles in Illustris are repositioned to the potential minima of their host halos.
For this reason, their parametric velocities are not physically meaningful.
Black hole ``mergers'' occur in the simulation whenever two MBH
particles come within a particle smoothing-length of one another---typically on the order of a kiloparsec.
This project aims to fill in the merger process unresolved in Illustris.  In our model, an Illustris
``merger'' corresponds to the \textit{formation} of a MBH binary system, which we then evolve.
To avoid confusion, we try to use the term `coalescence' to refer to the point at which such a binary 
would actually collide, given arbitrary resolution.

Over the course of the Illustris run, 135 `snapshots' were produced, each of which include
internal parameters of all simulation particles.  Additional black-hole--specific output was also recorded
at every time-step, providing much higher time resolution for black hole accretion rates\footnote{These,
self-consistently derived mass accretion rates ($\mdot$) are used in our implementation of gas drag
(discussed in \secref{sec:visc}) as a way of measuring the local gas density.}, local gas densities
and most notably, merger events.  The entire set of mergers---a time and pair of BH masses---constitutes
our initial population of MBH binaries.   

The distribution of BH masses is peaked at the lowest masses.  Many of these black holes are
short-lived: their small, usually satellite, host of-matter halos often quickly merge with a nearby
neighbor---producing a BH `merger' event.  Additionally, in some cases, the identification of a particular
matter over-density as a halo by the FOF halo-finder, while transient, may be sufficiently massive to trigger
the creation of a new MBH seed particle.  This seed can then quickly merge with the MBH in a nearby massive
halo.  Due to the significant uncertainties in our understanding of MBH and MBH-seed formation,
it is unclear if and when these processes are physical.
For this reason we implement a mass cut on merger events, to ensure that each component BH has
$\mbh > \nobreak 10^6 \, \msol \approx \nobreak 10 \, \mseed$.  Whether or not these `fast-mergers'
are non-physical, the mass cut is effective at excluding them from our analysis.  The entire Illustris
simulation has $23708$ MBH merger events; applying the mass cut excludes $11291$ ($48\%$) of those,
leaving $12417$. We have run configurations without this mass cut, and the effects on the GWB
are always negligible.

There is very small population of MBH `merger' events which occur during close encounters (but not
true mergers) of two host-halos.  During the encounter, the halo-finder might associate the two constituent
halos as one, causing the MBHs to merge spuriously due to the repositioning algorithm.
These forced mergers are rare and, we believe, have no noticeable effect on the overall population of
thousands of mergers.  They are certainly negligible in the overall MBH-Halo statistical correlations
which are well reproduced in the Illustris simulations \citep{sijacki2015}.

% ----------------------------------
% ------- MBHB Host Galaxies -------
% ----------------------------------

\subsection{Merger Host Galaxies}
\label{sec:host-gals}

To identify the environments which produce the dynamical friction, stellar scattering, and viscous drag which
we are interested in, we identify the host galaxies of each MBH involved in the merger in the snapshot
preceding it, in addition to the single galaxy which contains the `binary' (at this point a single,
remnant MBH) in the snapshot immediately following the merger event.
$644$ mergers ($3\%$) are excluded because they don't have an associated galaxy before or
after the merger. To ensure that each host galaxy is sufficiently well resolved (especially important
for calculating density profiles), we require that a
sufficient number of each particle type constitute the galaxy.
Following \citet{blecha2016} we use a fiducial cut of 80 and 300 star and DM particles
respectively, and additionally require 80 gas cells.  This excludes $54$ of the remaining binaries.
We emphasize here that this remnant host galaxy, as it is in the snapshot following the Illustris
`merger' event, forms the environment in which we model the MBHB merger process.  In the future, we plan
to upgrade our implementation to take full advantage of the dynamically evolving merger environment:
including information from both galaxies as they merge, and the evolving remnant galaxy once it forms.

We construct spherically averaged, radial density profiles for each host galaxy and each particle/cell
type (star, DM, gas).  Because the particle smoothing lengths are larger than the MBHB separations of
interest, we extrapolate the galaxy density profiles based on fits to the inner regions.  Our fits use
the innermost eight radial bins that have at least four particles in them.  Out of the valid binaries,
$347$ ($1\%$) are excluded because fits could not be constructed---generally because the particles
aren't distributed over the required eight bins.  Successful fits typical use $\sim100$ particles,
with gas cell sizes $\sim 10^2 \textrm{ pc}$ and SPH smoothing lengths for stars and DM
$\sim 10^3 \textrm{ pc}$.

	% Fig 01 ------ Sample Subhalo Density Profiles and Fits
	\begin{figure}
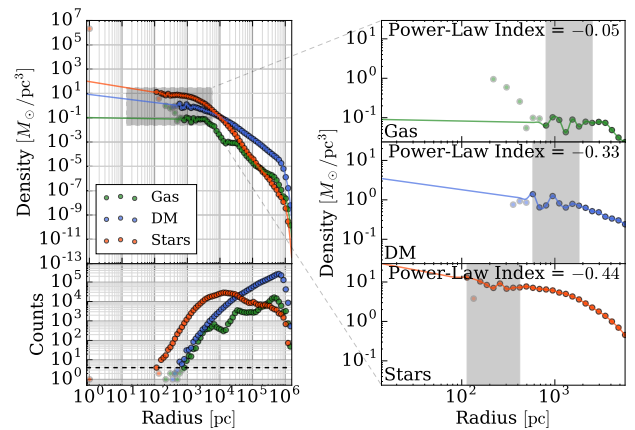

	\centering
	\includegraphics[width=1.0\columnwidth]{../figures/{{01_vd-pars-01_v0.9_fig11_subhalo-fits_20501_a1}}}
	\caption{Density profiles from a sample Illustris MBHB host galaxy.  Binned densities for each particle
	type are shown (upper-left) along with the number of particles/cells in each (bottom-left).
	Semi-transparent points are bins with less than four particles---the number required for consideration
	in calculating fits.  Zoom-ins are also shown separately for each particle type (right), with the eight
	inner-most bins with at least four particles shown in the shaded region.  Those bins were used to
	calculate fits, which are overplotted.  The resulting power-law indices used to extrapolate inwards
	are also shown.  Any galaxy without enough (eight) bins is excluded from our sample, in addition to
	the MBHB it contains.  This was the case for roughly $1\%$ of our initial population, almost entirely
	containing MBH very near our BH mass threshold ($10^6 \, \msol$).}
	\label{fig:subhalo-dens}
	\end{figure}

Density profiles for a sample Illustris MBHB host galaxy are shown in \figref{fig:subhalo-dens}.
The left panels show the binned density profiles for each particle type (top), and the number of particles
in each bin (bottom).  The semi-transparent points are those with less than the requisite four particles in
them.  The right panels show zoom-ins for each particle type, where the shaded regions
indicate the eight bins used for calculating fits.  The resulting interpolants are overplotted, with the
power-law index indicated.  While the four particles per bin, and eight inner bins generally provide for
robust fits, we impose a maximum power-law index of $-0.1$---i.e.~that densities are at least gently
increasing, and a minimum index of $-3$---to ensure that the mass enclosed is convergent.  Using these
densities we calculate all additional galaxy profiles required for the hardening prescriptions
(\secref{sec:hard}), e.g.~velocities, binding energies, etc.  When calculating profiles using our fiducial 
parameters, 2286 binaries ($10\%$) are excluded when calculating the distribution functions
(\secref{sec:lc}), usually due to significant nonmonotonicities in the radial density profile which are
incompatible with the model assumptions.  Overall, after all selection cuts, $9270/23708$ ($39\%$) of the
initial Illustris `merger' events are analyzed in our simulations.

	% Fig 02 ------ Histograms of MBH Binary Properties from Illustris
	\begin{figure}
	\centering
	\includegraphics[width=1.0\columnwidth]{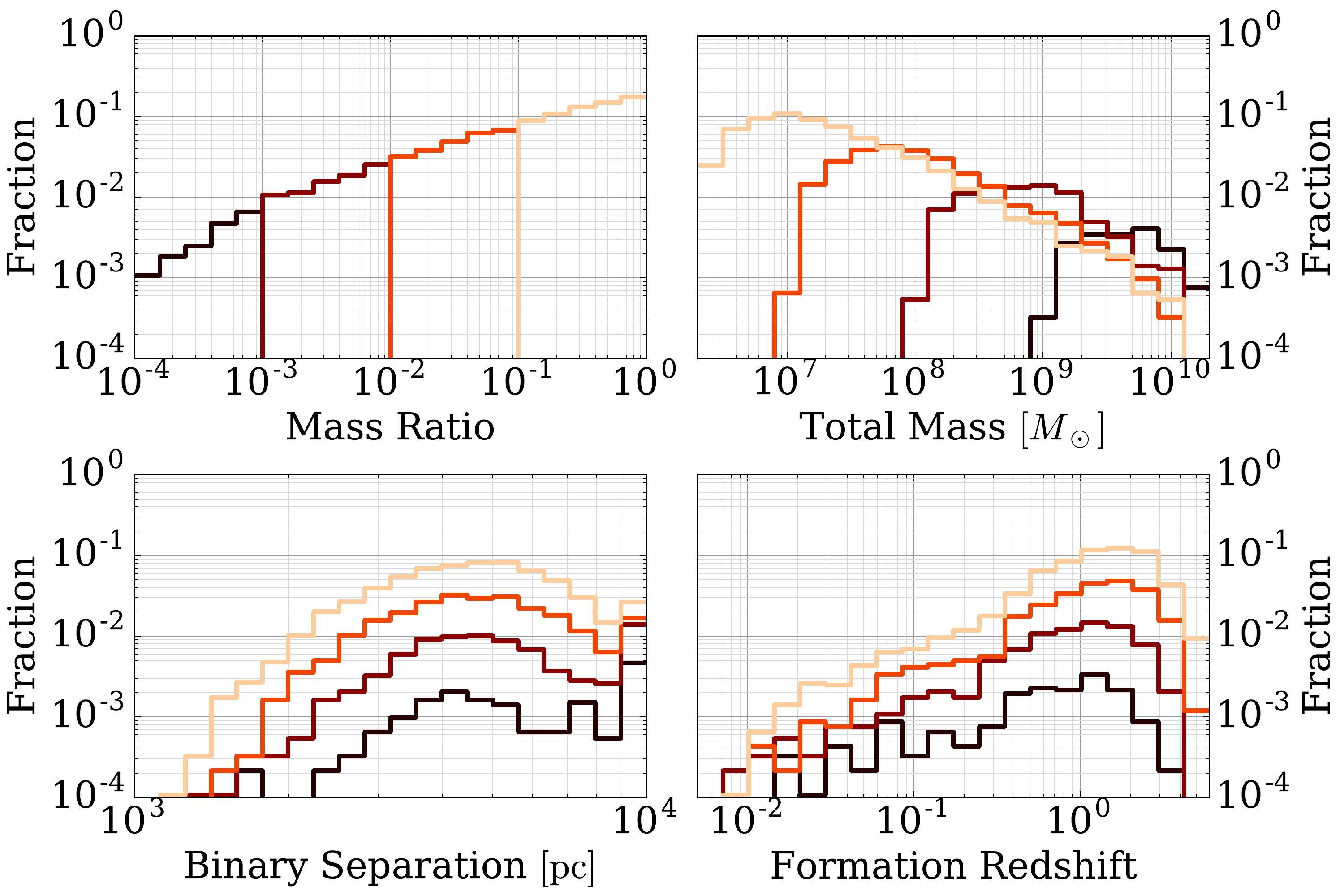}
	\caption{Properties of the Massive Black Hole Binaries from the Illustris population passing our
	selection cuts.  After selecting for MBH masses $M > 10^6 \, \msol$, and requiring the binary host
	galaxies to have sufficiently well resolved density profiles, 9270 of 23708 ($39\%$) systems remain.
	Distributions of mass ratio, total mass, initial binary separation (determined by MBH particle smoothing
	lengths), and formation redshifts (determined as the time at which particles come within a smoothing
	length of one-another) are shown.  The different lines (colors) correspond to different mass ratios
	which are strongly anti-correlated with total mass.}
	\label{fig:mbhb-dist}
	\end{figure}
	
Figure~\ref{fig:mbhb-dist} shows the properties of MBHB passing our selection cuts, grouped my mass ratio.
Mass ratio (upper-left panel) is strongly anti-correlated with total mass (upper-right) due to both 
selection effects (e.g.~at total masses just above the minimum mass, the mass ratio must be near unity),
and astrophysical ones (e.g.~the most massive MBH, in large, central galaxies tend to merge more often with
the lower mass MBH in small satellite galaxies).  Binary separations (lower-left) are set by the smoothing
length of MBH particles in Illustris.  Once two MBH particles come within a smoothing length of one another,
Illustris considers them a `merger' event---which corresponds to the `formation' (lower-right) of a binary
in the simulations of this study.
	
% Create a bibliography here, only if just this file is being compiled/built.
\biblio{}

%sec:ill
%sec:ill_bh
%sec:host-gals

% Hardening
% ---------

% Binary Hardening
% ================
\section{Binary Hardening Models}
\label{sec:hard}
Black-hole encounters from Illustris determine the initial conditions for the binary population
which are then evolved in our merger simulations.  Throughout the `hardening' process, where the binaries
slowly coalesce over millions to billions of years, we assume uniformly circular orbits.  In our models,
we use information from the MBHB host galaxies to implement four distinct hardening mechanisms
\citep[][hereafter \citetalias{bbr80}]{bbr80}: dynamical friction (DF), stellar scattering in the
`loss-cone' regime (LC)\footnote{Loss-cone scattering and dynamical friction are different regimes of the
same phenomenon, we separate them based on implementation.}, viscous drag (VD) from a circumbinary
gaseous-disk, and gravitational-wave radiation (GW).

% Dynamical Friction
% ------------------

% =====     Dynamical Friction     =====
% ======================================

\subsection{Dynamical Friction}
\label{sec:dyn_fric}

% Basic Idea
Dynamical friction is the integrated effect of many weak and long-range scattering events,
on a gravitating object moving with a relative velocity through a massive background.
The velocity differential causes an asymmetry which allows energy to be transferred
from the motion of the massive object to a kinetic thermalization of the background population.
In the case of galaxy mergers, dynamical friction is the primary mechanism of dissipating the
initial orbital energy to facilitate coalescence of the galaxies, generally on timescales
comparable to the local dynamical time ($\sim 10^8$ yr).  BH present in the
parent galaxies will tend to `sink' towards each other in the same manner \citepalias{bbr80}
due to the background of stars, gas and dark-matter (DM).
For a detailed review of dynamical friction in MBH systems, see \citet{am12}.

% Basic Equations
The change in velocity of a massive object due to a single encounter with a background particle
at a fixed relative velocity $v$ and impact parameter $b$ is derived \citep[e.g.][]{bt87}
following the treatment of \citet{chan42, chan43} by averaging the encounters over all possible angles to find,
	\begin{equation}
	\label{eq:dvpar}
	\Delta v = -2 v \frac{m}{M+m} \frac{1}{1 + (b/b_0)^2},
	\end{equation}
where the characteristic (or `minimum') impact parameter $b_0 \equiv G(M+m)/v^2$, for a primary object of
mass $M$, in a background of bodies with masses $m$.
The net deceleration on a primary mass is then found by integrating over distributions of stellar velocity
(assumed to be isotropic and Maxwellian) and impact parameters (out to some maximum effective distance $\bmax$)
which yields,
	\begin{equation}
	\label{eq:dyn-fric_b}
	\frac{d v}{dt} = - \frac{2 \pi G^2 (M+m) \rho}{v^2} \ln \left[1 + \left(\bmax/b_0\right)^2 \right],
	\end{equation}
for a background of mass density $\rho$.  The impact parameters are usually replaced with a constant---the
`Coulomb Logarithm',
$\ln \Lambda \equiv \ln \left(\frac{\bmax}{b_0}\right) \approx \frac{1}{2} \ln \left( 1 + \Lambda^2 \right)$,
such that,
	\begin{equation}
	\label{eq:dyn-fric_c}
	\frac{dv}{dt} = - \frac{2 \pi G^2 (M+m) \rho}{v^2} \ln \Lambda.
	\end{equation}

	% ---- Figure 03 : Dynamical Times (Radii vs. Masses)
	\begin{figure}
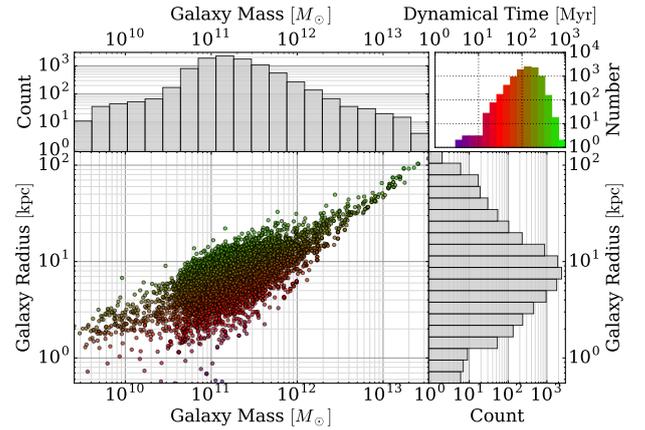

	\centering
	\includegraphics[width=\columnwidth]{../figures/{{03_vd-pars-01_v0.9_fig21_hard-df_f1}}}
	\caption{Stellar half-mass radii ($R_{\star,\trt{1/2}}$) and total mass within $2 \cdot R_{\star,\trt{1/2}}$
	for all	Illustris remnant host galaxies.  Values for each galaxy are colored by their dynamical times,
	calculated using \refeq{eq:dyn-time}, which are used for the `enhanced' DF masses.  The histogram in the
	upper right shows the distribution of dynamical times.  Galaxy masses and radii are peaked at about
	$10^{11} \, \msol$ and $10 \textrm{ kpc}$ respectively, with corresponding dynamical times around
	$100 \textrm{ Myr}$}
	\label{fig:dyn-times}
	\end{figure}

% DF Enhancement
In the implementation of \refeq{eq:dyn-fric_c}, we use spherically averaged density profiles from the
Illustris, remnant host galaxies.  Modeling a `bare', secondary MBH moving under DF through
these remnants would clearly drastically underestimate the effective mass---which, at early times is the
MBH secondary in addition to its host galaxy.  Over time, the secondary galaxy will be stripped by tidal
forces and drag, eventually leaving behind the secondary MBH with only a dense core of stars and gas
directly within its sphere of influence.  We model this mass `enhancement' by assuming that the effective
DF mass is initially the sum of the MBH mass ($M_2$), and that of its host galaxy ($M_{2,\trt{host}}$),
decreasing as a power-law over a dynamical time $\tdyn$, until only the MBH mass is left, i.e.,
	\begin{equation}
	\label{eq:df-enh}
	M_\trt{DF} = M_2 \left(\frac{M_2 + M_{2,\trt{host}}}{M_2} \right)^{1 - t/\tdyn}.
	\end{equation}
We calculate the dynamical time using a mass and radius from the remnant host galaxy.  Specifically, we use
twice the stellar half-mass radius $2 R_{\star,\trt{1/2}}$, and define $M_{2,\trt{host}}$ as the total mass
within that radius, i.e.,
	\begin{equation}
	\label{eq:dyn-time}
	\tdyn = \left[\frac{4 \pi \, (2 R_{\star,\trt{1/2}})^3}{3 G M_{2,\trt{host}}} \right]^{1/2}.
	\end{equation}

The galaxy properties and derived dynamical times for all MBHB host galaxies we consider are shown
in \figref{fig:dyn-times}.  Galaxy masses and radii are peaked at about $10^{11} \, \msol$ and
$10 \textrm{ kpc}$ respectively, with corresponding dynamical times around $100 \textrm{ Myr}$.
For comparison, we also perform simulations using a fixed dynamical time of $1 \textrm{ Gyr}$
for all galaxies, i.e.,~less-efficient stripping of the secondary galaxy.

% -----------------------------------------------------------
% ------- Impact Parameters and Explicit Calculations -------
% -----------------------------------------------------------

\subsubsection{Impact Parameters and Explicit Calculations}
\label{sec:df_explicit}

We have explored calculating the Coulomb logarithm explicitly, following \citetalias{bbr80} for the
maximum impact parameter such that,
	\begin{equation}
	\label{eq:bmax}
	\bmax(r) = 
		\begin{cases}
		\rs 							& \rs < r, \\
		\left(r/\rb\right)^{3/2} \rs 	& \rh < r < \rs, \\
		r \, \rh 						& r < \rh.
		\end{cases}
	\end{equation}
This, effective maximum impact parameter is a function of binary separation\footnote{Note that we use the term
`binary separation' loosely, in describing the separation of the two MBH even before they are gravitational bound.}
$r$---to account for the varying population of stars available for scattering and varying effectiveness of encounters.
\refeq{eq:bmax} also depends on the characteristic stellar radius $\rs$ \citepalias[`$r_c$' in][]{bbr80},
radius at which the binary becomes gravitationally bound, $\rb = \left[M/\left(N \ms\right)\right]^{1/3} \rs$,
and radius at which the binary becomes `hard', $\rh \equiv \left(\rb/\rs\right)^3 \rs$.
	
Not only is this formalism complex, but it often produces unphysical results.
For example, with this prescription the 
`maximum' impact parameter not infrequently becomes less than the `minimum', or larger than the distances
which interact in the characteristic timescales.  After imposing a minimum impact parameter ratio of
$\bmax / b_0 \geq 10$ (i.e.~$\ln \Lambda \geq 2.3$),
the results we obtained are generally consistent with using a constant coulomb-logarithm, with negligible effects on the
resulting merger rates and GW background.  We have also implemented an explicit integration over stellar distribution functions
(see: \secref{sec:lc}), and found the results to again be entirely consistent with \refeq{eq:dyn-fric_c}
which is both computationally faster and numerically smoother.
We believe the explicit impact parameter calculation is only valuable as a heuristic,
and instead we use $\ln \Lambda = 15$, consistent with detailed calculations \mbox{\citep[e.g.][]{am12}}.
Similarly, in the results we present, we take the local stellar density as that given by spherically symmetric
radial density profiles around the galactic center instead of first determining, then marginalizing over, the
stellar distribution functions.

% ----------------------------------
% ------- Applicable Regimes -------
% ----------------------------------

\subsubsection{Applicable Regimes}
\label{sec:df-regimes}

	% ---- Figure 04 : Dynamical Friction Hardening Rates (Gyr vs. Stellar vs. Bare; Gas vs. No)
	\begin{figure}
	\centering
	\includegraphics[width=\columnwidth]{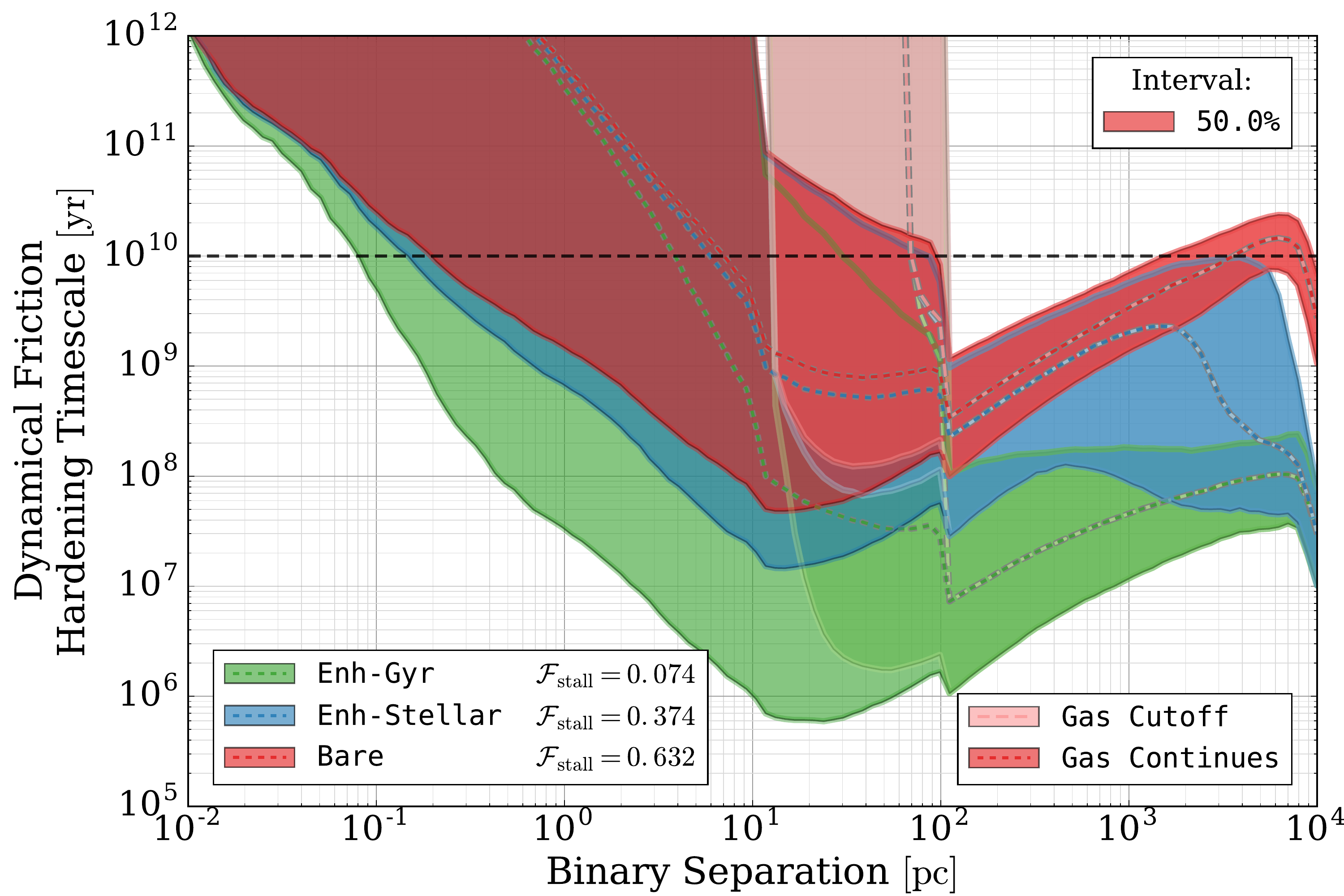}
	\caption{Dynamical Friction (DF) hardening timescales for 50\% of our MBHB around the median, under
	a variety of implementations.  Cases in which a `bare' secondary MBH migrates through the remnant
	host galaxy is compared to ones where the effective mass is \textit{enhanced} to the secondary's
	host galaxy, decreasing	as a power-law over the course of a dynamical time (`Enh'; see
	\refeq{eq:df-enh}).  The dynamical time is calculated using twice the `Stellar' half-mass radius
	(see \refeq{eq:dyn-time}) shown in blue, or using a fixed $1 \textrm{ `Gyr'}$ timescale shown in green.
	Allowing gaseous DF to continue below the attenuation radius $\rlc$ (`Gas Continues': darker regions,
	dotted lines) is compared to cutting off the gas along with stars and DM (`Gas Cutoff': lighter regions,
	dashed lines).  $\fstall$ is the fraction of mergers with mass ratio $\mu > 0.1$, remaining at
	separations $r > 10^2 \textrm{ pc}$, at redshift zero.}
	\label{fig:df-hard}
	\end{figure}
	
There is a critical separation at which the back-reaction of the decelerating MBH notably modifies
the stellar distribution, and the dynamical friction formalism is no longer appropriate.
Beyond this radius, the finite number of stars in the accessible region of parameter space
to interact with the MBH(B)---the `loss-cone', \citep[LC; see,][]{merritt2013}---must be considered
explicitly, discussed more thoroughly in \secref{sec:lc}.  The `loss-cone' radius
can be approximated as \citepalias{bbr80},
	\begin{equation}
	\rlc = \scale[1/4]{\ms}{M} \scale[9/4]{\rb}{\rs} \rs.
	\end{equation}
Stars and DM are effectively collisionless, so they can only refill the LC on a slow, diffusive
scattering timescale.  Gas, on the other hand, is viscous and supported thermally and by
turbulent motion which can equilibrate it on shorter timescales.  In our fiducial model, we assume that
for separations $r < \rlc$ the dynamical friction due to stars and DM is attenuated to low
values, but that of gas continues down to smaller separations.  We set the inner edge of gaseous DF
based on the formation of a (circumbinary) accretion disk on small scales (discussed in
\secref{sec:visc}).  The attenuation prescription given by \citetalias{bbr80} increases the dynamical
friction timescales by a factor,
	\begin{equation}
	\label{eq:df-atten}
	\dfatten = \scale[7/4]{m}{\mbh} N_\star \scale[27/4]{\rb}{\rs} \scale{\rlc}{r},
	\end{equation}
where $N_\star = \frac{1}{\mstar}\int_0^r 4 \pi r'^2 \rho_\star dr'$, is the number of stars available
to interact with the binary.  For all intents and purposes this negates the effectiveness of DF
for $r\lesssim \rlc$, such that without other hardening mechanisms (which become important at smaller 
scales), no MBHB would coalesce within a Hubble time.

% -------------------------------
% ------- Hardening Rates -------
% -------------------------------

\subsubsection{DF Hardening Rates}
\label{sec:df-gwb}

The resulting hardening timescales, $\thard = a / \left(da/dt\right)$, for our different DF
implementations are shown in \figref{fig:df-hard}.  We show evolution for `bare' MBH secondaries (red),
in addition to effective masses enhanced (`enh') by the secondary's host galaxy for a dynamical time calculated
using twice the `stellar' half-mass radii (blue), or with a fixed `Gyr' time scale (green).  In each case
we also compare between letting gas-DF continue below $\rlc$ (`Gas Continues': darker colors, dotted lines),
versus attenuating gas along with stars and DM (`Gas Cutoff': lighter colors, dashed lines).  As a metric of the
varying outcomes, we calculate the fraction of `stalled' major-mergers $\fstall$, defined as the number of
major mergers (mass-ratios $\mu \equiv M_2/M_1 > 0.1$; which are 6040 out of the full sample of 9270,
i.e.~$\sim 65\%$) remaining at separations larger than $100 \textrm{ pc}$ at redshift zero, divided
by the total number of major-mergers.  Attenuation of the DF begins below $100 \textrm{ pc}$, so $\fstall$
are unaffected by whether the gas DF is also cutoff.

For $r \gtrsim 100 \textrm{ pc}$, where the density of stars and especially dark matter dominate that of
gas, the hardening rates are the same with or without a separate treatment of gas.  Hardening differs
significantly however, between the `bare' and enhanced models---with the latter hardening more than an
order of magnitude faster at the largest separations\footnote{Recall that binary separations are initialized
to the MBH particle smoothing lengths, distributed between about $10^3$ and $10^4 \textrm{ pc}$, so the total
number of systems being plotted decreases over the same range.} ($\sim 10^4 \textrm{ pc}$).  After a
dynamical time, the `stellar' enhancement runs out and the hardening rate approaches that of a bare MBH
secondary by $\sim 10^3 \textrm{ pc}$.  Still, the enhanced mass over this time leads to a decrease in
the fraction of stalled binaries from $\sim63\%$ to $\sim 37\%$.  When the mass enhancement persists for
a gigayear---about a factor of ten longer than typical dynamical times---a large fraction of MBHB are able
to reach parsec-scale separations before tidal stripping becomes complete, leading to only $\sim7\%$
of major-mergers stalling at large separations.  The particular fraction of stalled systems is fairly
sensitive to the total mass and mass ratio cutoff, which we return to in \secref{sec:res_pops}.

Previous studies \citep[e.g.][]{ravi2014} have assumed that DF is very effective at bringing MBHB into
the dynamically `hard' regime (i.e.~instantly in their models), after which stellar interactions must be
calculated explicitly to model the remaining evolution.  For comparison, in our results we also include a
`Force-Hard' model in which we assume that all binaries reach the hard regime ($r = \rh$) over the
course of a dynamically time\footnote{calculated using the `Stellar' prescription: twice the stellar half
mass radius, and the mass within it.}.

It has long been suggested that MBHB could stall at kiloparsec-scale separations \citep[e.g.][]{yu2002},
but only recently has this effect been incorporated into population hardening models and studied specifically
\citep[e.g.][]{mcwilliams2014, kulier2015}.  By better understanding the timescales over which stalled systems
could be observable, we can use observed dual-AGN to constrain the hardening process and the event rates of
MBHB encounters.

% Create a bibliography here, only if just this file is being compiled/built.
\biblio{}

% Loss-Cone Scattering (Three-Body Interactions)
% ----------------------------------------------

% Loss-Cone Scattering (Three-Body Interactions)
% ==============================================

\subsection{Stellar Loss-Cone Scattering}
\label{sec:lc}

The population of stars that are able to interact (scatter) with the MBHB
are said to occupy the `loss-cone' \citep[LC,][]{merritt2013}, so-named because it describes
a conical region in parameter space. When stars are scattered out of the LC faster than they
can be replenished, the LC becomes depleted and the dynamical friction description,
which considers a relatively static background, becomes inconsistent. One approach to
compensate for this is to add an `attenuation' factor, as described in the previous section
(\secref{sec:dyn_fric}).  Physically, a steady state must be dynamically realized in which
stars are diffused into the outer edges of the LC via two-body relaxation at the same rate
at which stars are scattering out by the central MBH(B). The largest uncertainty in the MBHB
merger process is likely understanding the nature of this equilibrium state, and how it is
affected by realistic galaxy-merger environments.

The loss-cone has been extensively explored in both the context of MBH binary hardening,
as well as in tidal-disruption events (TDE).  The two cases are almost identical, differing
primarily in that for TDE calculations, only impact parameters small enough to cause
disruptions are of interest---while for binary hardening, weaker scattering events are
still able to extract energy from the MBH or MBHB.  For binary hardening, there is an additional
ambiguity in two subtly distinct regimes: first, where stars scatter with individual BHs,
decelerating them analogously to the case of dynamical friction (but requiring the LC
population to be considered explicitly).  Second, for a truly \textit{bound} binary, stars
can interact with the combined system---in a three-body scattering---and extract energy
from the binary pair together.  In our prescriptions we do not distinguish between these
cases, considering them to be a spectrum of the same phenomenon instead.

We use the model for LC scattering given by \citet{magorrian1999}, corresponding to a
single central object in a spherical (isotropic) background of stars.  We adapt this prescription
simply by modifying the radius of interaction to be appropriate for scattering with a binary
instead of being tidally disrupted by a single MBH.  This implementation is presented in
pedagogical detail in Appendix~\secref{sec:app-lc-calc}.  Scattering rates are calculated corresponding
to both a `full' LC (\refeq{eq:lc-flux-full}), one in which it is assumed that the parameter
space of stars is replenished as fast as it is scattered; in addition to a `steady-state' LC
(\refeq{eq:lc-flux-eq}), in which diffusive two-body scattering sets the rate at which stars
are available to interact with the binary.

\begin{figure}
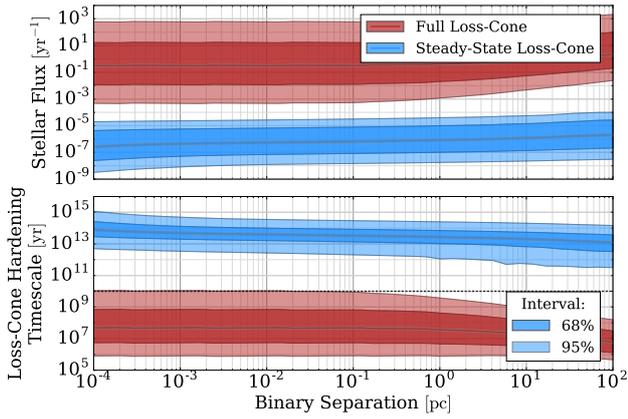

	\centering
	\includegraphics[width=\columnwidth]{../figures/{{05_vd-pars-01_v0.9_fig23_loss-cone_c1}}}
	\caption{Scattering rates and hardening timescales for full (red) and steady-state (blue)
	loss-cones.
	The bands represent $68\%$ and $95\%$ of the population around the median.  The difference
	between the two extremes of LC states is a stark six orders of magnitude, illustrating how
	strong of an effect the LC can have on MBHB mergers.  We use a simple,
	single parameter prescription to describe the state of the LC: the fraction, in log-space,
	between steady-state and full, $\frefill$ (see, \refeq{eq:frefill}).}
	\label{fig:lc-flux-hard}
	\end{figure}

% Resulting fluxes and hardening rates
The interaction rates (fluxes) of stars scattering against all MBHB in our sample are shown in
the upper panel of \figref{fig:lc-flux-hard} for both full (red, \refeq{eq:lc-flux-full}) and
equilibrium (blue, \refeq{eq:lc-flux-eq}) loss-cone configurations.  The interaction rates for
full LC tend to be about six orders of magnitude higher than equilibrium configurations.
The resulting binary hardening timescales are shown in the lower panel of
\figref{fig:lc-flux-hard}---reaching four orders of magnitude above and below a Hubble time.
Clearly, whether the LC is in the relatively low equilibrium state or is more effectively
refilled has huge consequences for the number of binaries which are able to coalesce within a
Hubble time.

	% ---- Figure 05 : Full vs. Steady Loss Cone Scattering and Hardening

% Things refilling the loss-cone faster
Many factors exist which may contribute to quickly refilling the loss cone.  In general, any
form of asymmetry in the potential will act as an additional perturber---increasing the
thermalization of stellar orbits.  The presence of a MBHB is premised on there having
been a recent galaxy merger---implying that significant asymmetries and aspherical
morphologies may exist.  Even ignoring galaxy mergers, galaxies themselves are triaxial
\citep[e.g][]{illingworth1977, leach1981}, many have bars \citep[e.g.][]{sellwood1993},
and in star forming galaxies there are likely large, dense molecular clouds
\citep[e.g.][]{young1991} which could act as perturbers.  Finally, because binary lifetimes
tend to be on the order of the Hubble time while galaxies typically undergo numerous merger
events \citep[e.g.][]{rodriguez-gomez2015}, subsequent merger events can lead to triple
MBH systems (see: \secref{sec:res_multi}) which could be very effective at stirring the
stellar distribution.  While there is some evidence that for galaxy-merger remnants the
hardening rate can be nearly that of a `full' LC \citep[e.g.][]{khan2011}, the
community seems to be far from a consensus \citep[e.g.][]{vasiliev2014}, and a purely
numerical solution to the LC problem is currently still unfeasible.

% Loss-Cone Refilling Parameter
In the future, we plan on incorporating the effects of triaxiality and tertiary MBH
to explore self-consistent LC refilling.  In our current models, we introduce an arbitrary
dimensionless parameter---the logarithmic `refilling fraction' $\frefill$ (in practice,
but not requisitely, between $[0.0, 1.0]$)---to logarithmically interpolate between the
fluxes of steady-state ($\fluxsslc$) and full LC ($\fluxflc$), i.e.,
	\begin{equation}
	\label{eq:frefill}
	\fluxlc = \fluxsslc \cdot \left(\frac{\fluxflc}{\fluxsslc}\right)^{\frefill}.
	\end{equation}

% Create a bibliography here, only if just this file is being compiled/built.
\biblio{}

% Viscous Hardening by Circumbinary-Disk
% --------------------------------------

% Viscous Hardening by Circumbinary-Disk
% ======================================

\subsection{Viscous Hardening by a Circumbinary Disk}
\label{sec:visc}

% Basic Idea of Circumbinary Gas-Drag
The density of gas accreting onto MBH can increase significantly at separations near the accretion
or Bondi radius, $\rb \equiv G \mbh / c_s^2$, where the sound-crossing time is comparable to the
dynamical time.  The nature of accretion flows onto MBH near and within the Bondi-radius are highly
uncertain, as observations of these regions are currently rarely resolved \citep[e.g.][]{wong2011}.
If a high density, circumbinary disk is able to form, the viscous drag (VD) can be a significant
contribution to hardening the binary at separations just beyond the GW-dominated regime
\citep[\citetalias{bbr80},][]{gould2000, escala2005}.  Galaxy mergers are effective at driving
significant masses of gas to the central regions of post-merger galaxies \citep{barnes1992},
enhancing this possibility.

% Introduce prescription --- Haiman, Kocsis and Menou 2009
We implement a prescription for VD due to a circumbinary accretion disk following
\citet[][hereafter \citetalias{hkm09}]{hkm09} based on the classic thin-disk solution of
\citet{shakura1973}, broken down into three, physically distinct regions \citep{shapiro1986}.
These regions are based on the dominant pressure (radiation vs.~thermal)
and opacity (Thomson vs.~free-free) contributions, such that the regions are defined as,

    \begin{enumerate}[label=\arabic*)]
    \item $r < r_{12}$, radiation pressure and Thomson-scattering opacity dominated;
    \item $r_{12} < r < r_{23}$, thermal pressure and Thomson-scattering opacity dominated;
    \item $r_{23} < r$, thermal pressure and free-free opacity dominated.
    \end{enumerate}

	% ---- Figure 06 : Distribution of Eddington Ratios for all MBHB
	\begin{figure}
	\centering
	\includegraphics[width=\columnwidth]{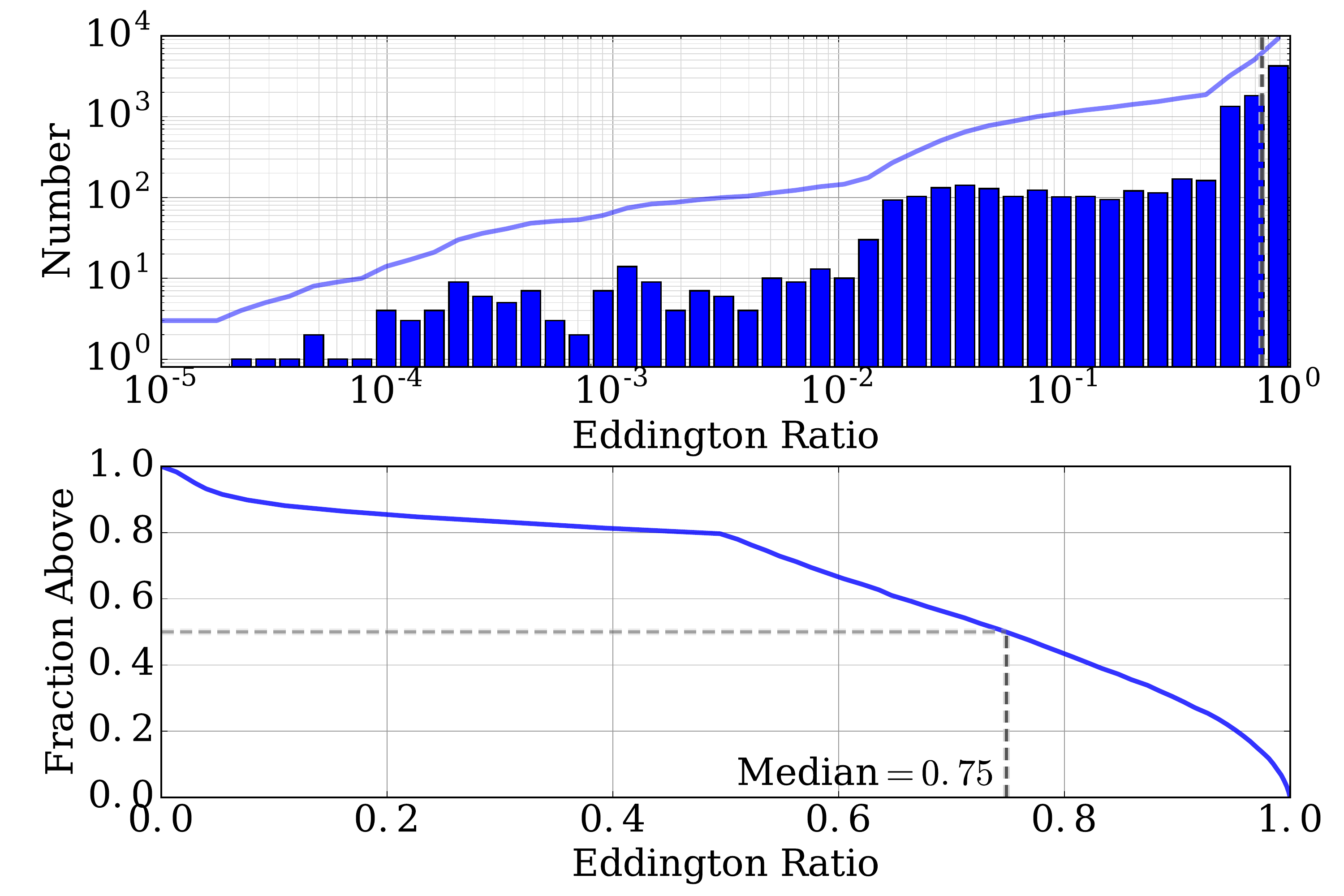}
	\caption{Accretion rates at the time of binary formation for all MBHB in our analysis.
	Values are measured as a fraction of the Eddington accretion rate,
	$\mdotedd \equiv \ledd/\radeff c^2$, where we use $\radeff = 0.1$.  Recall that in Illustris
	MBH merge when they come within a smoothing length of one another---corresponding to the
	formation of a binary in our models.  The accretion rates from Illustris are those of the
	resulting remnant MBH, and are limited to Eddington ratios of unity.  The upper panel shows
	the distribution of accretion rates (bars) and cumulative number (line), which are strongly
	biased towards near-Eddington values.  The lower panel shows the cumulative distribution of
	accretion rates above each value (note the different x-axis scaling).  The median Eddington
	ratio of $0.75$ is overplotted (grey, dashed line).}
	\label{fig:fedd-hist}
	\end{figure}
	
% Parameters
Recall that in Illustris, `mergers' occur when MBH particles come nearer than a particle smoothing
length, after which the MBH are combined into a single, remnant MBH.  We track these remnant
particles and use their accretion rates ($\mdot$) to calibrate the circumbinary disk's gas density.
The distribution of Eddington ratios ($\mdot/\mdotedd$)  for these remnants, at the time of their
formation, are presented in \figref{fig:fedd-hist}, showing a clear bias towards near-Eddington
accretion rates. MBH remnants tend to have enhanced accretion rates for a few gigayear after merger,
and have higher average accretion than general BHs \citep{blecha2016}.  In Illustris, MBH accretion
(and thus growth in mass) is always limited to the Eddington accretion rate.  We introduce a
dimensionless parameter $\fedd$ to modulate those accretion rates,
i.e., $\mdot = \nobreak \textrm{Min}\left[\mdotill, \fedd \mdotedd \right]$.

Otherwise, in the formalism of \citetalias{hkm09}, we use their fiducial parameter
values\footnote{Mean mass per electron, $\mu_e = 0.875$; viscosity parameter $\alpha = 0.3$;
radiative efficiency, $\radeff = 0.1$; temperature-opacity constant, $f_T = 0.75$; and
disk-gap size, $\rgap = 1.0$ (`$\lambda$' in \citetalias{hkm09}).} and assume an $\alpha$-disk
(i.e.~the viscosity depends on total pressure, not just thermal pressure as in a so-called
$\beta$-disk).  The outer disk boundary is determined by instability due to
self-gravity---measured as some factor times the radius, $r_Q$, at which the Toomre parameter
reaches unity, i.e.~$\rsg = \rselfgrav \, r_Q$.  In our fiducial model, $\rselfgrav = 1$,
and variations in this parameter have little effect on the overall population of binaries.
After marginalizing over all systems, changes to the different viscous-disk parameters tend
to be largely degenerate: shifting the distribution of hardening timescales and the GWB
amplitude in similar manners.

% Disk Regions, self-gravity cutoff (absolute upper limit)
VD hardening timescales tend to decrease with decreasing binary separations.  They thus tend to be
dominated by \reg{3}---at larger separations.  For near-Eddington accretion rates, however, \reg{2}
and especially \reg{3} tend to be self-gravity unstable, fragmenting the disk and eliminating VD
altogether.  For this reason, when high accretion rate systems have dynamically important disks,
they tend to be in \reg{1}.  In these cases, \reg{1} extends to large enough radii such that for
most masses of interest, GW emission will only become significant well within that region of the
disk.  Lower accretion rate systems are stable out to much larger radii, allowing many binaries
to stably evolve through \reg{2} and \reg{3}.  These regions also cutoff at smaller
separations, meaning that GW emission can become significant outside of \reg{1}.

Decreased disk densities mean less drag, but at the same time sufficiently high densities lead
to instability, making the connection between accretion rate and VD-effectiveness non-monotonic.
This is enhanced by gaseous DF, with an inner cutoff radius determined by the SG radius (see
\secref{sec:df-regimes}).  In other words, gaseous DF is allowed to continue down to smaller radii
when the outer disk regions become SG unstable.  We impose an additional, absolute upper-limit to
the SG instability radius of $\rsgmax = \nobreak 10 \textrm{ pc}$, i.e.
$\rsg = \textrm{Min}\left[\rselfgrav \, r_Q, \, \rsgmax\right]$,
to keep the outer-edge of disks physically reasonable.  

	% ---- Figure 07 : Fraction of Binaries in each Disk Region
	\begin{figure}
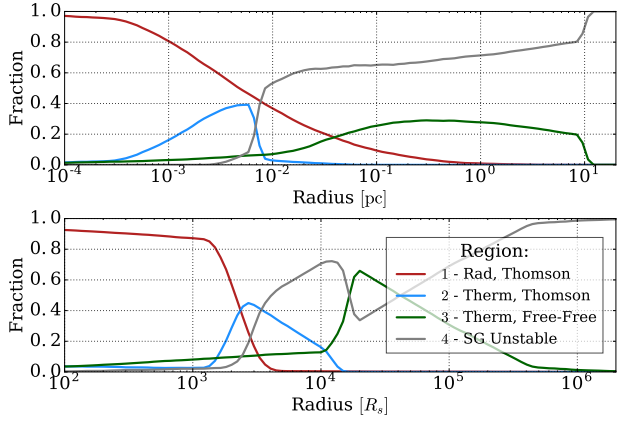

	\centering
	\includegraphics[width=\columnwidth]{../figures/{{07_vd-pars-01_v0.9_fig19_visc-disk_e2}}}
	\caption{Fraction of binaries in each circumbinary disk region as a function of radius.  Radii
	are given both physical units (upper panel) and Schwarzschild radii ($\rs$, lower panel),
	the latter highlighting the intrinsic scalings.  \reg{4} are locations in the disk which
	are unstable to self-gravity (`SG'), defined using the Toomre parameter for each of
	\reg{2} and \reg{3}.}
	\label{fig:disk-regs-fracs}
	\end{figure}
	
% Fraction of binaries in each region
\figref{fig:disk-regs-fracs} shows the fraction of Illustris binaries in the different regions of
the disk, for our fiducial model.  Only a fraction of MBHB spend time in \reg{2} and
\reg{3} disks, and even that is only for a small region of log-radius space.  While almost all
systems do enter \reg{1} by about $10^3 \, R_s$, GW hardening has, in general,
also become significant by these same scales.

In addition to the spatially-distinct disk regions, different types of migration occur
depending on whether the disk or the secondary-MBH is dynamically dominant
(analogous to the distinction between `planet-dominated' and `disk-dominated',
Type~II migration in planetary disks---see, \citetalias{hkm09}).  If disk-dominated,
the system hardens on the viscous timescale $\tvisc$,
whereas if the secondary is dominant---as is the typical case in our simulations,
the timescale is slowed by a factor related to the degree of secondary-dominance.  

	% ---- Figure 08 : VD Hardening Timescales by Heavy/Light, Major/Minor
	\begin{figure}
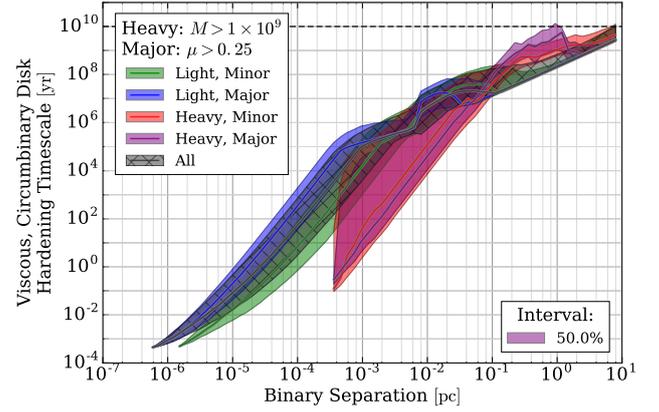

	\centering
	\includegraphics[width=\columnwidth]{../figures/{{08_vd-pars-01_v0.9_fig15_hard_f4}}}
	\caption{Hardening timescales due to circumbinary disk drag for binaries grouped by
	total mass and mass ratio.  Light and heavy MBHB are separated by total masses below and
	above $10^9 \, \msol$ respectively; and minor and major based on mass ratios ($\mu$) below
	and above $1/4$.  The median and surrounding $50\%$ intervals for all MBHB systems are shown in grey,
	showing that `light' systems dominate the bulk of the binary population.  Heavy, and
	especially heavy-major, systems tend to harden orders of magnitude faster than lighter ones.
	The light population (especially major) exhibits nonmonotonicities at intermediate
	separations ($\sim 10^{-3}$ -- $10^{-1} \textrm{ pc}$) indicative of changes between disk
	regions.  Heavy systems (especially minor), on the other hand, show much smoother hardening
	rates consistent with moving through primarily \reg{1}.}
	\label{fig:disk-time}
	\end{figure}

The resulting hardening timescales due to VD from a circumbinary disk are shown in
\figref{fig:disk-time}.  The more massive binaries (`heavy', $M > 10^9 \msol$) have orders of
magnitude shorter VD hardening times, but are quite rare.  The overall trend (grey,
cross-hatched) follows the less massive (`light') systems.  The changes in slope of the `light'
populations (especially `Major', $\mu \equiv M_2/M_1 > 1/4$) at separations larger than
$10^{-3} \textrm{ pc}$ are due to transitions in disk regions.  The `heavy' systems tend to have
SG unstable Regions~2 \& 3, and thus harden more smoothly, predominantly due to \reg{1}.

	% ---- Figure 09 : Median Hardening Timescales with and without VD, F_refill = 0.6 and 1.0
	\begin{figure}
	\centering
	\includegraphics[width=\columnwidth]{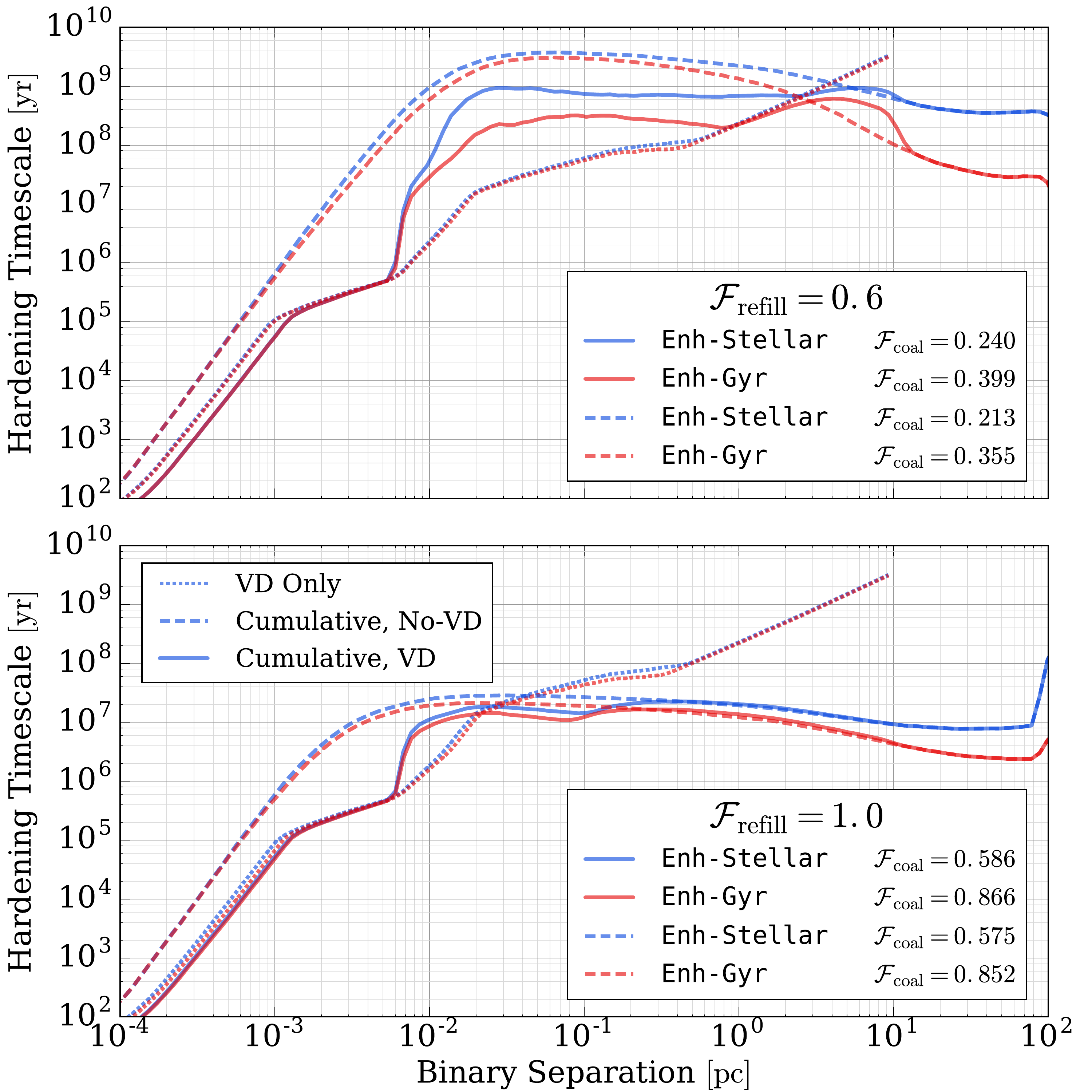}
	\caption{Median Hardening timescales with and without drag from a circumbinary Viscous Disk (VD)
	for different hardening models.
	Different LC refilling parameters ($\fcoal = 0.6$, upper; and $\fcoal = 1.0$, lower) are
	compared against DF models (`Enh-Stellar', blue; `Enh-Gyr', red).  Each line type shows
	a different hardening rate: viscous drag only (dotted), and cumulative hardening rates
	with VD (solid) and without VD (dashed).  With $\frefill = 0.6$, VD effects are apparent up to
	the disk cutoff, $\rsgmax = 10 \textrm{ pc}$, whereas for $\frefill = 1.0$, LC scattering
	dominates down to $\sim 10^{-2} \textrm{ pc}$.  Similarly, the presence of a circumbinary disk
	has a much more pronounced effect on the fraction of high mass-ratio ($\mu > 0.1$) systems which
	coalesce by redshift zero ($\fcoal$), which are indicated in the legends.  At very low
	separations the cumulative (with-VD) hardening rate is very nearly the purely VD rate, showing its
	importance down to very small scales.  At intermediate separations the `cumulative, VD' rate is
	intermediate between the `cumulative, No-VD' and the `VD only' model, showing that the disk is
	only present in some fraction of systems at those scales.}
	\label{fig:vd-comp-hard}
	\end{figure}

\figref{fig:vd-comp-hard} compares median hardening rates in simulations including VD (solid)
with those without a disk (dashed).  In the former case, the purely VD hardening rates are also
shown (dotted)---with the maximum disk cutoff $\rsgmax$, clearly apparent at $10 \textrm{ pc}$.
Different dynamical friction prescriptions are shown, with mass enhancements over dynamical times
calculated using the `stellar' method in blue, and fixed $1 \textrm{ Gyr}$ timescales in red (see
\secref{sec:dyn_fric}).  The upper panel shows a moderately refilled loss cone ($\frefill = 0.6$),
while  in the lower panel the LC is always full ($\frefill = 1.0$).  The effects of VD are clearly
apparent below a few $10^{-2} \textrm{ pc}$ in all models, and up to $\rsgmax = 10 \textrm{ pc}$
when $\frefill = 0.6$.  In the $\frefill = 1.0$ case, LC hardening dominates to much smaller
separations, making the VD effects minimal for the overall hardening rates.  This is echoed in the
changes in coalescing fractions\footnote{$\fcoal$, the fraction of systems with mass ratio
$\mu > 0.1$ which coalesce by $z=0$.} between the VD and No-VD cases: for $\frefill = 0.6$, VD
increases $\fcoal$ by $\sim 10\%$, while for $\frefill = 1.0$, it is only increased by $\sim 2\%$.

The circumbinary disk is SG unstable for many systems, and thus the median hardening rates including VD
are often intermediate between the purely VD timescales and simulations without VD at all
(e.g., seen between $\sim 5\E{-3}$ -- $1 \textrm{ pc}$ in the upper panel of \figref{fig:vd-comp-hard}).
At smaller separations ($\lesssim 10^{-3} \textrm{ pc}$), where LC is almost always subdominant to VD
and/or GW hardening, the VD decreases the median hardening timescales by between a factor of a few,
and an order of magnitude.  Notably, \figref{fig:vd-comp-hard} shows that the scaling of hardening rate with
separation below about $10^{-3} \textrm{ pc}$ is very similar between that of GW (which is dominant
in the No-VD case) and VD.  At these small scales, $\gtrsim 80\%$ of our binaries are in disk \reg{1}
(see \figref{fig:disk-regs-fracs}), which has a viscous hardening rate, $\tviscone \propto r^{7/2}$
\citepalias[][Eq.~21a]{hkm09}, compared to a very similar scaling for GW, $\tgw \propto r^4$ (see
\secref{sec:gw-hard}).  Thus, even when VD dominates hardening into the mpc-scale regime, we
don't expect the GWB spectrum from binaries in \reg{1} to deviate significantly from the canonical
$-2/3$ power-law.

	% ---- Figure 10 : VD Hardening Rates varying invidual Parameters
	\begin{figure}
	\centering
	\includegraphics[width=\columnwidth]{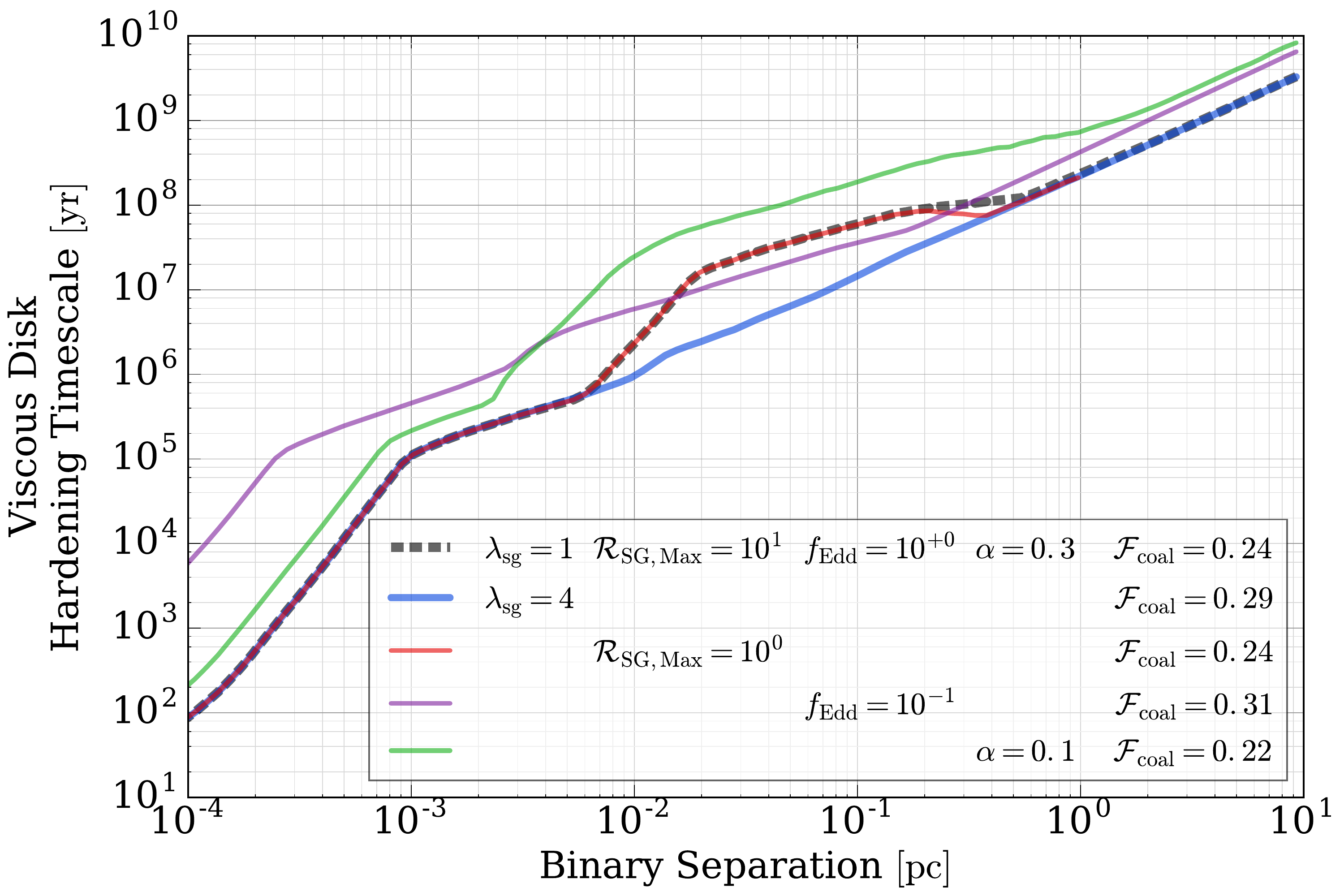}
	\caption{Median Hardening timescales comparing our fiducial Viscous Disk (VD) parameters (black,
	dashed) with other configurations.  The radius at which the disk becomes Self-Gravity (SG)
	unstable is $\rsg = \nobreak \textrm{Min}\left[\rselfgrav \, r_Q, \, \rsgmax\right]$, where
	$r_Q$ is the radius at which the Toomre parameter reaches unity.  $\rselfgrav$ scales the
	SG-unstable radius, while $\rsgmax$ is a maximum cutoff radius.  $\alpha$ is the standard disk
	viscosity parameter, and $\fedd$ limits the maximum accretion rate, i.e.
	$\mdot = \textrm{Min}\left[\mdotill, \fedd \mdotedd \right]$.
	While this variety of VD parameters produces hardening rates varying by two orders of magnitude,
	the resulting changes to the coalescing fraction $\fcoal$ is fairly moderate as VD is often
	subdominant to LC scattering at larger radii and to GW emission at smaller radii.  Effects on
	$\fcoal$ can be counterintuitive, for example decreasing the accretion rate (purple line)
	increases the median hardening timescale, but increases the coalescing fraction because the disk
	becomes SG-stable for a larger fraction of binaries.  Each model uses $\frefill = 0.6$, and the
	`Enh-Stellar' DF.}
	\label{fig:vd-comp-hard-pars}
	\end{figure}

Differences in median hardening timescales, solely due to viscous drag, are compared for a variety of
VD parameters in \figref{fig:vd-comp-hard-pars}.  A simulation with our fiducial disk parameters is
shown in dashed-black, and each color shows variations in a different parameter.  Decreasing
the viscosity of the disk ($\alpha$, green) amounts to a proportional increase in the hardening
timescale, and decrease in the coalescing fractions ($\fcoal$).  Decreasing the maximum disk
radius ($\rsgmax$, red) decreases the overall effectiveness of VD, but because gaseous DF continues
in its place, the coalescing fraction remains unchanged.  While $\rsgmax$ changes the maximum
disk radius, $\rselfgrav$ changes the radii at which \reg{2} \& \reg{3} become SG-unstable directly
(i.e.~even well within the maximum cutoff radius).  Increasing $\rselfgrav$ by a factor of four (blue)
significantly increases the number of MBHB with SG-stable \reg{2}\footnote{See the transition in
\figref{fig:disk-regs-fracs} between \reg{2} (light-blue) \& \reg{4} (grey) at
$\sim 10^{-2} \textrm{ pc}$.} between $\sim 10^{-2} \, \& \, 10^{-1} \textrm{ pc}$, increasing the
overall coalescing fraction.

Decreasing the accretion rates ($\fedd$, purple), and thus disk densities, increases the hardening
timescales similarly to changing the viscosity (green).  At the same time, significantly more systems
have stable outer disks.  This has the effect of increasing coalescing fractions noticeably, despite
the increased median hardening timescales.  In addition to increased outer-disk stability, the
transition between disk regions are also inwards.  A large number of MBHB at small separations
($\lesssim 10^{-3} \textrm{ pc}$) remain in disk \reg{2} instead of transitioning to \reg{1}. This
softens the scaling of hardening rate with separation to, $\tvisctwo \propto r^{7/5}$
\citepalias[][Eq.~21b]{hkm09}, which differs much more significantly from purely GW-driven evolution.

% Create a bibliography here, only if just this file is being compiled/built.
\biblio{}

% Gravitational-Wave Emission
% ---------------------------

% Gravitational-Wave Emission
% ===========================

\subsection{Gravitational-Wave Emission}
\label{sec:gw-hard}

Gravitational wave radiation will always be the dominant dissipation mechanism at the smallest binary
separations---within hundreds to thousands of Schwarzschild radii.  GW hardening depends only on the
constituent masses ($M_1$ \& $M_2$) of the MBHB, their separation, and the system's eccentricity.
The hardening rate can be expressed as \citep{peters1964},
	\begin{equation}
	\label{eq:dadt}
	\frac{da}{dt} = - \frac{64 \, G^3}{5 \, c^5} \frac{M_1 \, M_2 \left(M_1 + M_2\right)}{a^3}
					\frac{\left( 1 + \frac{73}{24} e^2 + \frac{37}{96} e^4 \right)}{\left(1 - e^2\right)^{7/2}},
	\end{equation}
where $a$ is the semi-major axis of the binary and $e$ is the eccentricity.  In our
treatment we assume that the eccentricities of all MBHB are uniformly zero, in which case, \refeq{eq:dadt}
can easily be integrated to find the time to merge,
	\begin{equation}
	\label{eq:tgw}
	\begin{split}
	% Don't use \tgw, that means a/(da/dt) for GW only...
	t_\trt{GW} & = \frac{5 c^5}{64 G^3} \frac{a_0^4 - \rcrit^4}{M_1 M_2 M}, \\
		 & \approx 10^{10} \textrm{ yr} \, \scale[4]{a_0}{0.01 \textrm{ pc}}
		 	\scale[-3]{M}{2\times 10^{7} \msol} \scale{2 + \mu + 1/\mu}{4},
	\end{split}
	\end{equation}
for a total mass $M = M_1 + M_2$, mass ratio $\mu \equiv M_2/M_1$, initial separation $a_0$ and critical
separation $\rcrit$.  In practice, we assume that the GW signal from binaries terminates at the
Inner-most Stable Circular Orbit (ISCO), at which point the binary `coalesces'; i.e.
$\rcrit = \risco(J=0.0) = 3 \rs$.  For an equal mass binary, with median Illustris MBH
masses\footnote{after typical selection cuts, described in \secref{sec:ill}.} of about $10^7 \, \msol$,
the binary needs to come to a separation of $\sim 0.01 \textrm{ pc}$ ($\sim 10^4 \, \rs$), to merge
within a Hubble time.  Characteristic timescales and separations for (purely) GW-driven inspirals
across total mass and mass ratio parameter space are plotted in \figref{fig:gw_char} of
Appendix~\ref{sec:app-figs}.  While the absolute most-massive MBHB can merge purely from
GW emission starting from a parsec, the bulk of physical systems, at $10^6 - 10^8 \, \msol$, must be
driven by environmental effects to separations on the order of $10^{-3} - 10^{-2} \textrm{ pc}$
($\sim 500 - 5000 \, \rs$) to coalesce by redshift zero.

% Create a bibliography here, only if just this file is being compiled/built.
\biblio{}

% Create a bibliography here, only if just this file is being compiled/built.
\biblio{}

%sec:hard
%sec:dyn_fric
%sec:df_explicit
%sec:df-regimes
%sec:df-gwb
%
%sec:lc
%sec:lc_relax
%sec:lc_distfunc
%sec:lc_lc
%
%sec:visc
%
%sec:gw-hard

% Results
% ----------

% =============================================================
% =======================   RESULTS   =========================

\section{Results}
\label{sec:results}

	% Fig 11 ------ Binary Hardening Landscape, By Mechanism
	\begin{figure}
	\centering
	\includegraphics[width=\columnwidth]{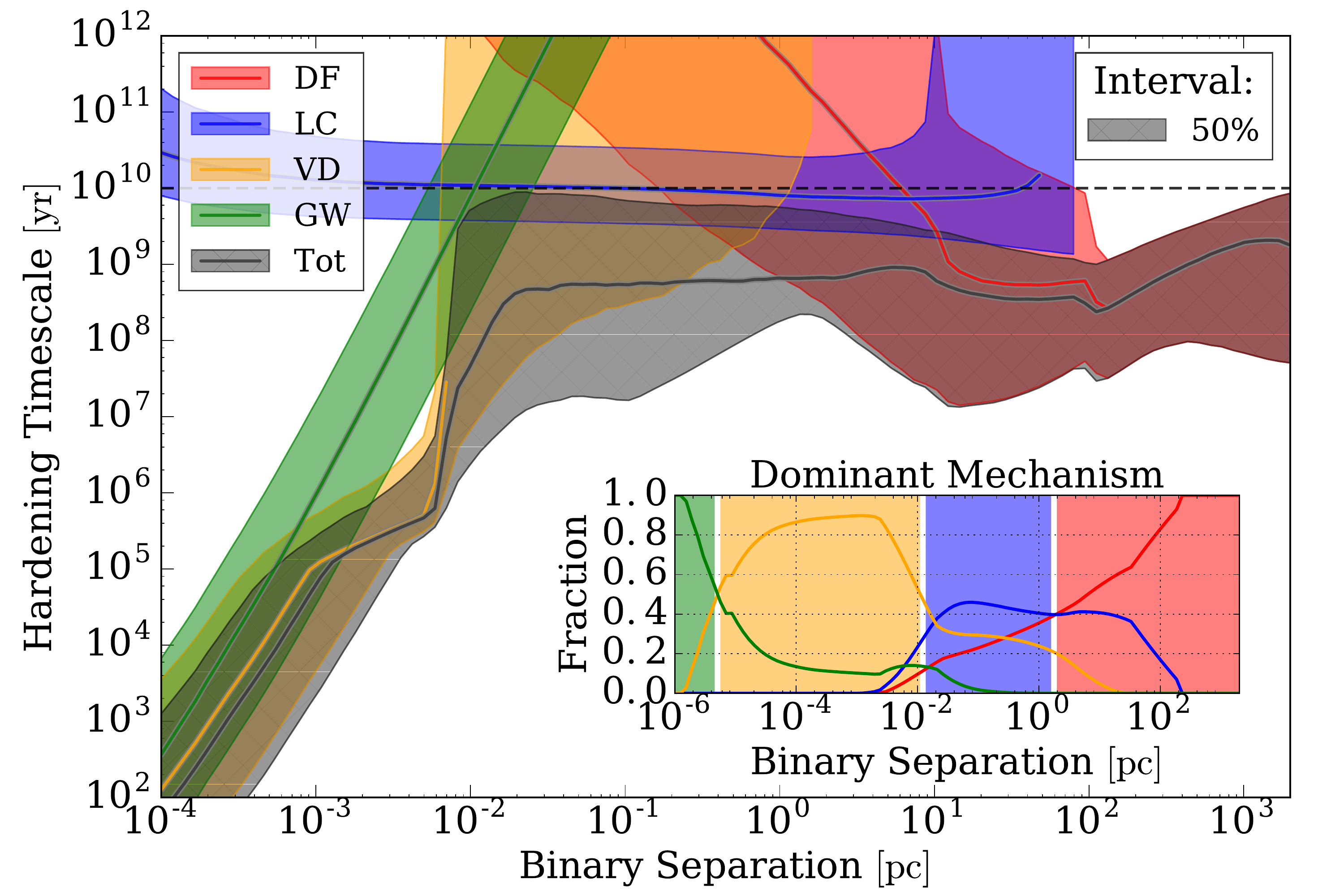}
	\caption{Binary hardening timescales versus binary separation by mechanism. Colored lines
	and bands show the median and 50\% intervals for Dynamical Friction (DF),
	Loss Cone (LC) scattering, Viscous Drag (VD), and Gravitational Wave (GW) emission
	with the total hardening rate shown by the grey, hatched region.
	The inset panel shows the fraction of binaries dominated by each mechanism.
	This simulation uses our fiducial parameters (e.g.~$\frefill = 0.6$), with `Stellar'
	DF-mass enhancement.  The binary hardening landscape is very similar to that outlined by
	\citetalias{bbr80}, but the details are far more nuanced.  For a comparison with alternate
	models, see \figref{fig:hard_set1}.}
	\label{fig:hard}
	\end{figure}

% Overall Hardening Rates, Dominant Hardening Mechanisms
The hardening timescales for all Massive Black Hole Binaries (MBHB) are plotted against binary
separation in \figref{fig:hard}, broken down by hardening mechanism.  This is a representative
model with a moderate loss-cone (LC) refilling fraction $\frefill = 0.6$ (see \secref{sec:lc}),
using the `Enh-Stellar' DF (see \secref{sec:dyn_fric}).  This is the fiducial model
for which we present most results, unless otherwise indicated.  The inset shows the fraction of
binaries with hardening rates dominated by each mechanism.  DF is most important at
large radii soon after binaries form, until LC scattering takes over at $\sim 1 \textrm{ pc}$.
The median hardening timescale remains fairly consistent at a few times $100 \textrm{ Myr}$,
down to $\sim 10^{-2} \textrm{ pc}$ at which point viscous drag (VD) drives the bulk of systems
until gravitational wave (GW) emission takes over at separations below $10^{-5} \textrm{ pc}$,
where the typical hardening timescale reaches years.  The landscape of hardening timescales
for alternative DF prescriptions and LC refilling fractions are shown in the appendix
(\figref{fig:hard_set1}).

% ----------------------------------------------
% ------ Binary Lifetimes and Coal Fracs -------
% ----------------------------------------------

\subsection{Binary Lifetimes}
\label{sec:res_life}

	% Fig 12 ------ 2D Hist of Lifetimes and Coalescing Fractions
    \begin{figure}
    \centering
        % Must have blank line between `\subfloat`
        \subfloat{\includegraphics[width=\columnwidth]{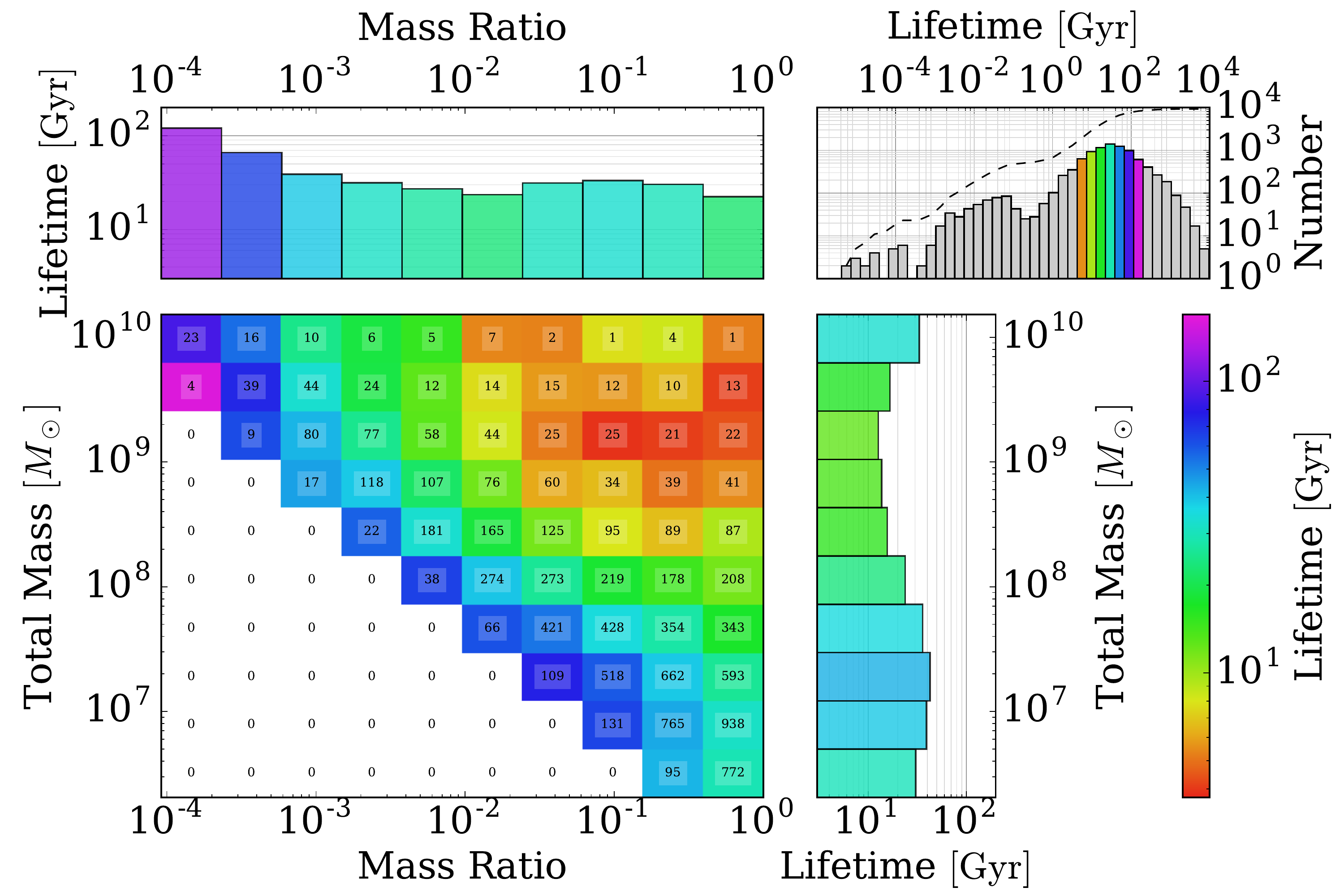}}
        
        \subfloat{\includegraphics[width=\columnwidth]{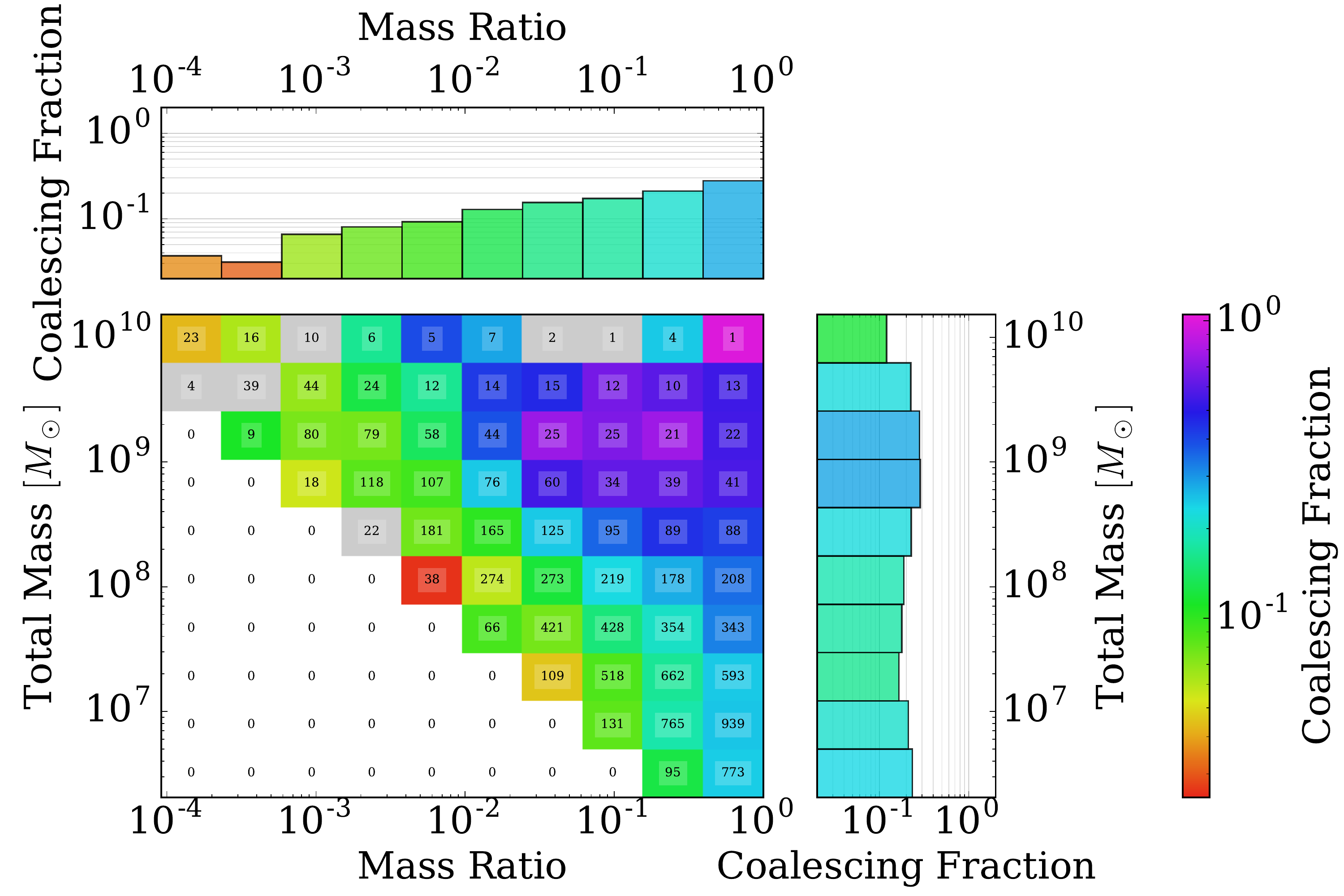}}
    \caption{Binary lifetimes (upper) and coalescing fractions (lower) for our fiducial model
    with a moderate DF and LC refilling (`Enh-Stellar' and $\frefill = 0.6$, respectively).
    The overall distribution of MBHB lifetimes are shown in the upper-right-most panel, with the
    cumulative distribution plotted as the dashed line.  The median lifetime is 
    $\sim 30 \textrm{ Gyr}$ overall, but is significantly shorter for MBHB with either high total
    masses, or nearly-equal mass ratios.  For this group, the coalescing fractions are near unity.
    Grey bins in the lower panel correspond to those with no binaries which coalesce by redshift
    zero.  While systems with the highest masses and mass ratios tend to have much shorter lifetimes,
    they also form at low redshifts with less time to coalesce.}
    \label{fig:life}
    \end{figure}

% Explicit figure description
Characteristic hardening timescales are often many $100\textrm{ Myr}$, and MBHB typically need to
cross eight or nine orders of magnitude of separation before coalescing.  The resulting lifetimes
of MBHB can thus easily reach a Hubble time.  \figref{fig:life} shows binary
lifetimes (upper panels) and the fraction of systems which coalesce by $z=0$ (lower panels) for our
fiducial model.
Systems are binned by total mass and mass ratio, with the number of systems in each bin indicated.
The plotted lifetimes are median values for each bin, with the overall distribution shown
in the upper-right-most panel.  Grey values are outside of the range of binned medians, and the cumulative distribution is given by the dashed line.

% Overall lifetime distribution and properties
The lifetime distribution peaks near the median value of $29 \textrm{ Gyr}$, with only $\sim7\%$
of lifetimes at less than $1 \, \textrm{ Gyr}$.  About $20\%$ of all MBHB in our sample coalesce
before redshift zero.  Systems involving the lowest mass black holes\footnote{Recall we require
MBH masses of at least $10^6 \, \msol$} (i.e.~down and left) tend towards much longer lifetimes.
Overall, lifetimes and coalescing fractions are only mildly correlated with total mass or mass
ratio, when marginalizing over the other.  For systems with total masses $M > 10^8 \, \msol$,
the coalescing fraction increases to $23\%$,
and for mass ratios $\mu > 0.2$, only slightly higher to $26\%$.  In general, examining slightly
different total-mass or mass-ratio cutoffs has only minor effects on lifetimes and coalescing
fractions.

There is a strong trend towards shorter lifetimes for simultaneously high total
masses and mass ratios (i.e.~up and right), where median lifetimes are \textit{only} a few
gigayear.  Considering both $\mu > 0.2$ and at the same time $M > 10^8 \, \msol$,
coalescing fractions reach $45\%$.  The handful of MBHB which coalesce after
$\lesssim 1 \textrm{ Myr}$ ($\sim0.3\%$) tend to involve MBH in over-massive galaxies
(i.e.~galaxy masses larger than expected from MBH-host scalings) with especially concentrated
stellar and/or gas distributions.  There are a handful of high mass ratio, and highest total mass
MBHB ($M \sim 10^{10} \, \msol$) systems showing a noticeable decrease in coalescing fraction.
These systems form at low redshifts and don't have time to coalesce despite relatively short
lifetimes.

% Difference with $\frefill = 1.0$ (in Appendix)
Lifetimes and coalescing fractions for an always full LC ($\frefill = 1.0$) are shown in
\figref{fig:life_lc10}, in Appendix~\ref{sec:app-figs}.  The median value of the lifetime
distribution shifts down to $\sim8\textrm{ Gyr}$, with $\sim24\%$ under $1 \, \textrm{ Gyr}$.  
The coalescing fractions increase similarly, and systems which are \textit{either} high mass ratio
($\mu \gs 0.2$) \textit{or} high total mass ($M \gs 10^8 \, \msol$) generally coalesce by
redshift zero.  Specifically, the coalescing fractions are $50\%$ and $61\%$ for all systems
and those with $\mu > 0.2$ respectively.  Considering only $M > 10^8 \, \msol$, fractions increase
to $54\%$ \& $99\%$.

	% Fig 13 ------ Lifetimes Cumulative 3x panels for DF; each LC
	\begin{figure}
	\centering
	\includegraphics[width=0.5\textwidth]{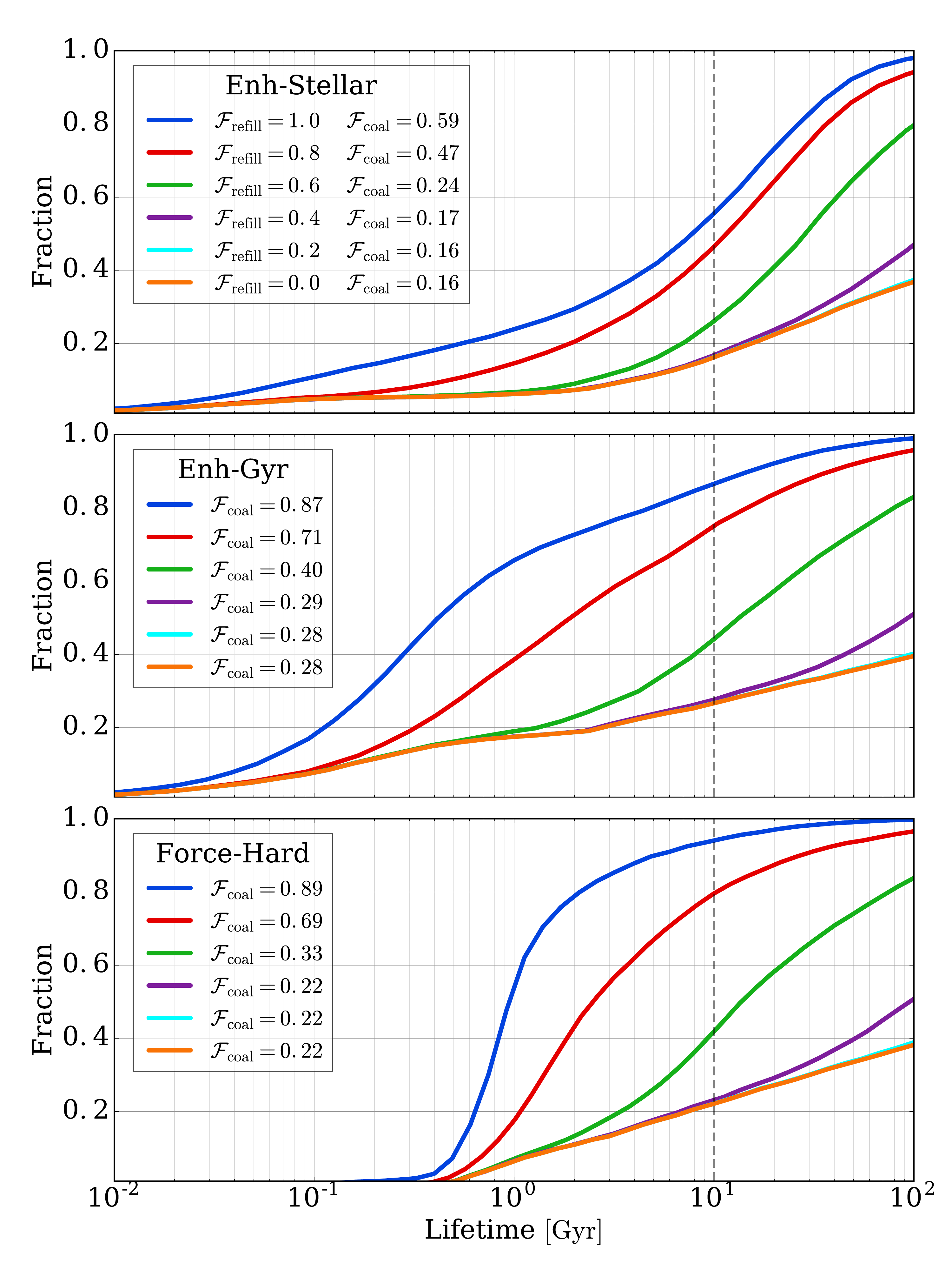}
	\caption{Cumulative distributions of binary lifetimes for a variety of DF and LC parameters.
	The `Enh-Stellar', `Enh-Gyr', and `Force-Hard' DF models are shown in each
	panel, and the colors of lines indicate the LC refilling fraction.  The fraction of high
	mass ratio ($\mu > 0.1$) systems coalescing by $z=0$ are given in the legend ($\fcoal$).
	$\frefill$ is the dominant factor determining the lifetime distribution, but the DF model
	significantly affects the earliest merging systems, and overall fraction of coalescing
	systems.}
	\label{fig:hard_df_lc}
	\end{figure}

Cumulative distributions of MBHB lifetimes are compared in \figref{fig:hard_df_lc} for a variety
of LC refilling factors (colors) and our three primary DF prescriptions (panels; see
\secref{sec:dyn_fric}).  The first two panels correspond to prescriptions where the effective
masses used in the DF calculation are the sum of the secondary MBH mass and the mass of its host
galaxy.  To model stripping of the secondary galaxy during the merger process, the effective mass
decreases as a power law to the `bare' MBH mass after a dynamical time.  The `Enh-Stellar' model
(upper) calculates the dynamical time at twice the stellar half-mass radius, and the mass there
enclosed.  The `Enh-Gyr' model (middle), on the other hand, uses a fixed $1 \mathrm{ Gyr}$
timescale---almost a factor of ten longer than the median `stellar'-calculated value.  Finally,
the `Force-Hard' model (lower), uses the `bare' secondary MBH mass, but the binary is forced to
the hard binary regime (generally $1$ -- $10 \textrm{ pc}$) over the course of a dynamical time
(calculated in the `stellar' manner).  Each color of line indicates a different LC refilling
fraction, from always full ($\frefill = 1.0$; blue) to the steady state ($\frefill = 0.0$;
orange).  The fraction of high mass-ratio ($\mu > 0.1$) systems which coalesce by redshift zero
($\fcoal$) are also indicated in the legends.

The high mass-ratio coalescing fractions tend to vary by almost a factor of four depending on the
LC state, while the varying DF prescriptions have less than factor of two effect.  Median
lifetimes change considerably, however, even between DF models, for example with $\frefill = 1.0$,
the median lifetime for the `Enh-Stellar' model is $7.7 \textrm{ Gyr}$, while that of
`Enh-Gyr' is only about $0.42 \textrm{ Gyr}$.  Apparently, with a full LC, DF at large scales tends
to be the limiting factor for most systems.  While the highest overall $\fcoal$ occurs for
`Force-Hard' \& $\frefill = 1.0$, it takes almost an order of magnitude longer for the first
$\sim 10\%$ of systems to coalesce than in either of the `Enh' models.  The effects of DF on the
lifetimes of the first systems to merge are fairly insensitive to the LC state.  There are thus cases
where DF can be effective at driving some systems to coalesce very rapidly.  At the same time,
for the bulk of systems, after hardening past kiloparsec scales the remaining lifetime can be quite
substantial.  For $\frefill \lesssim 0.6$, neither the precise LC refilling fraction nor the DF
model make much of a difference after the first $10$ -- $30\%$ of systems coalesce.  In these cases,
the most massive systems with high-mass ratios coalesce fairly rapidly regardless, but the smaller
more extreme mass-ratio systems take many Hubble times to merge.

	% Fig 14 ------ Binary Formation and Coalescence Over Redshift
	\begin{figure}
	\centering
	\includegraphics[width=0.5\textwidth]{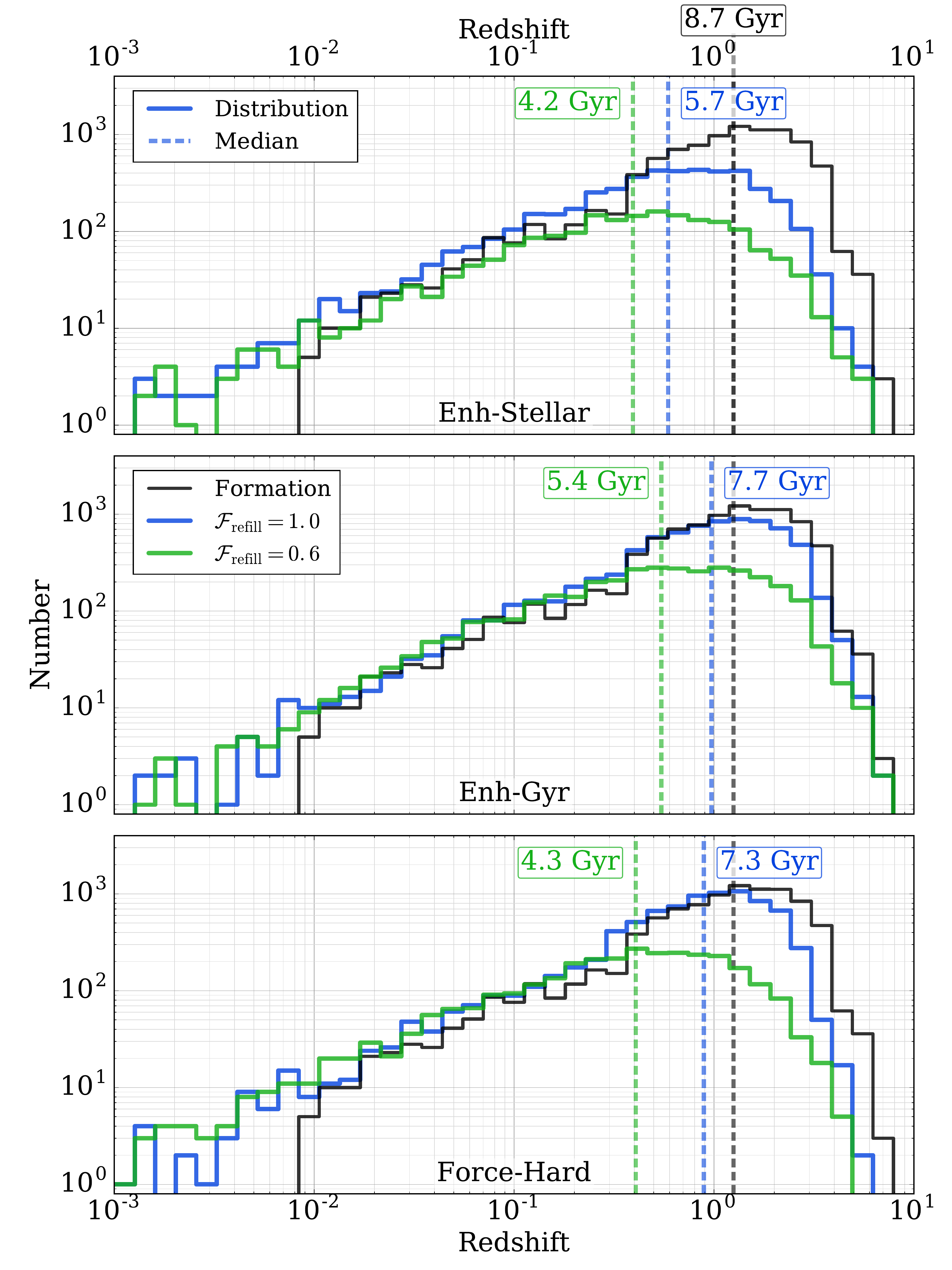}
	\caption{Distribution of MBH binary formation (black) and coalescence (colored) redshifts for
	different DF models (panels) and two LC refilling parameters: always full ($\frefill = 1.0$;
	blue) and our fiducial, moderately refilled ($\frefill = 0.6$; green) value.  The median
	redshift for each distribution is also plotted (dashed), with the corresponding look-back time
	indicated.  The minimum delay time between medians of formation and coalescence is
	$1 \textrm{ Gyr}$, but up to $4.5 \textrm{ Gyr}$ for our fiducial LC state and DF model 
	(`Enh-Stellar').}
	\label{fig:activity}
	\end{figure}

Figure~\ref{fig:activity} shows the distribution of formation (black) and coalescence (colored)
redshifts resulting from a variety of binary evolution models.  Each panel shows a different DF
prescription, and two LC refilling parameters are shown: always full, $\frefill = 1.0$ (blue), and
our fiducial, moderately refilled value of $\frefill = 0.6$ (green).  A handful of events have
been cutoff at low redshifts ($z < 10^{-3}$) where the finite volume of the Illustris simulations
and cosmic variance becomes important.  Median redshifts for each
distribution are overplotted (dashed), along with their corresponding look-back times.  The median
formation redshift for our MBHB is $z = 1.25$ (look-back time of $\sim 8.7 \textrm{ Gyr}$). For a
full LC and the stronger DF models, `Enh-Gyr' and `Force-Hard', the median coalescence redshifts
are delayed to $z \sim 1.0$ and $z \sim 0.9$ respectively---i.e.~by about a gigayear.  For our more
modest, fiducial DF prescription, `Enh-Stellar', even the full LC case still delays the median
coalescence redshift to $z \sim 0.6$, about $3 \textrm{ Gyr}$ after the peak of MBHB formations.
If the LC is only moderately refilled, the median redshifts are much lower: between $z\sim 0.4$
-- $0.6$.

% --------------------------------------------
% ------ Gravitational Wave Background -------
% --------------------------------------------

\subsection{The Gravitational Wave Background (GWB)}
\label{sec:res_gwb}

	% Fig 15 ------ Naive/Power-Law GWB with Predictions and Illustris
	\begin{figure}
	\centering
	\includegraphics[width=\columnwidth]{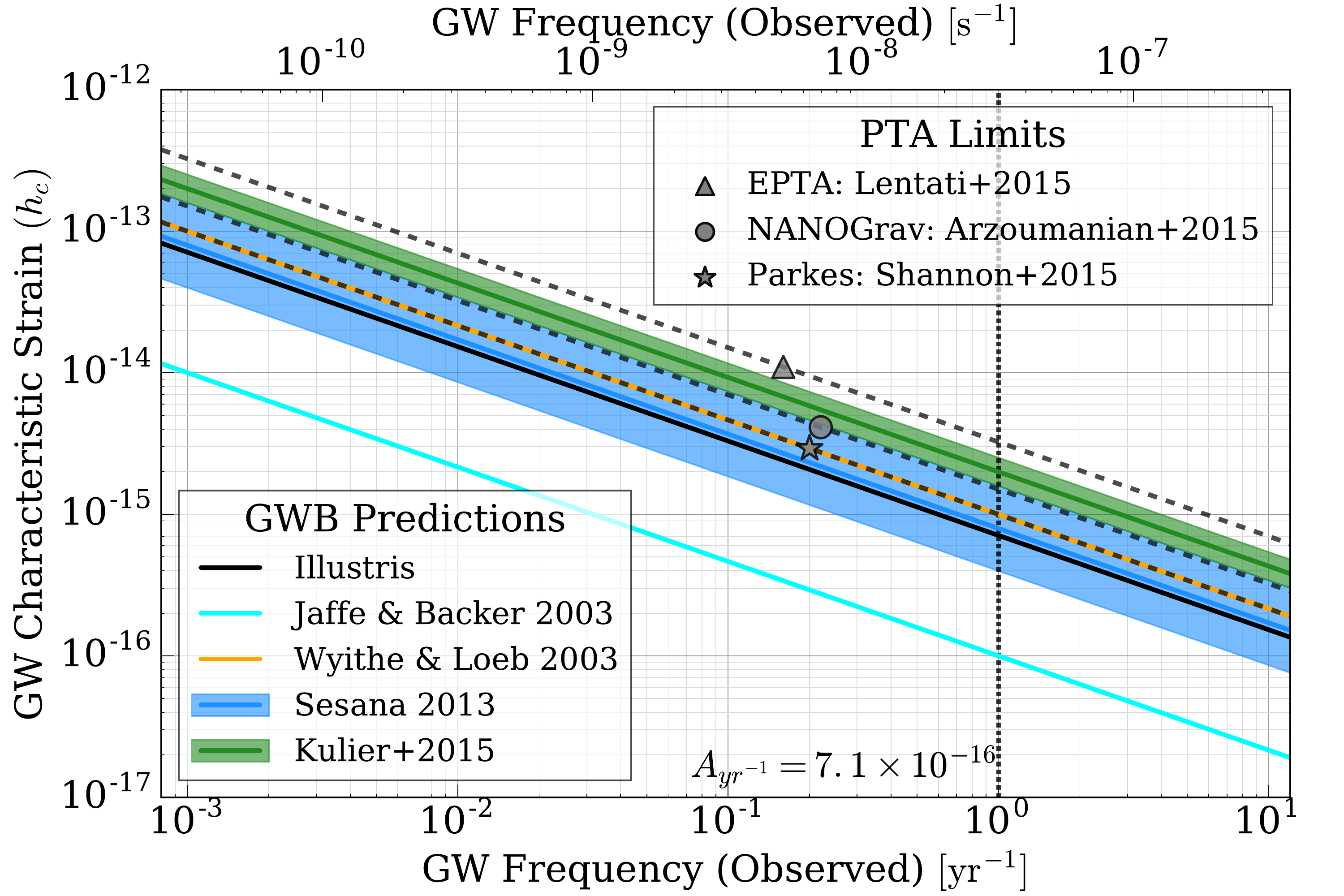}
	\caption{Stochastic Gravitational Wave Background spectrum produced by Illustris MBH binaries,
	assuming purely power-law evolution with all systems efficiently reaching the GW regime.
	Power-law predictions from the literature (described in \secref{sec:intro}) are presented for
	comparison, along with the most recent PTA upper limits.  The power-law spectrum resulting from
	the Illustris simulations is very consistent with previous results, and about $30\%$ below the
	most stringent observational upper limits.}
	\label{fig:gwb-naive}
	\end{figure}

% Naive/Power-Law GWB Predictions, Illustris version
In \secref{sec:intro} we have outlined the theoretical background for the existence of a
stochastic GWB, and introduced the formalism for calculating pure power-law spectra.
\figref{fig:gwb-naive} shows the purely power-law spectrum derived from Illustris MBH binaries,
assuming that all systems (passing our selection cuts outlined in \secref{sec:ill}) reach the GW
dominated regime and evolve purely due to GW emission.  Other representative power-law predictions
(see \secref{sec:intro}) and recent pulsar timing array (PTA) upper limits are included for comparison.
The Illustris prediction is completely consistent with the existing literature and about $30\%$
below the most recent PTA upper limits.  These consistencies validate the
Illustris MBHB population, and the prescriptions for the growth and evolution of individual MBH.

% GWB with the evolutionary details
Almost all of the details of binary evolution are obscured in purely power-law predictions
(i.e.~\refeq{eq:gw-energy-phinney}).  In particular, they imply that all MBHB instantly
reach the separations corresponding to the frequencies of interest, and evolve purely due to
GW-emission.  In reality, we've shown that the delay time distribution can be significant at
fractions of a Hubble time.  This has the important consequence that not all MBHB coalesce
(or even reach the PTA band) before redshift zero.  At the same time, the environmental effects
(e.g.~LC scattering) which are \textit{required} to bring MBH binaries to the relevant orbital
frequencies also decrease the time they emit in each band, attenuating the GW signal.

% ---- GWB Full Calculation - Formalism and Derivation ----

\subsubsection{Full GWB Calculation Formalism}

The GWB can be calculated more explicitly by decomposing the expression for
GW energy radiated per logarithmic frequency interval,
	\begin{equation}
	\label{eq:gw-energy-time-freq}
	\frac{d \egw}{d \ln f_r} = \frac{d \egw}{d t_r} \frac{dt_r}{d \ln f_r},
	\end{equation}
where the right-hand-side terms are the GW power radiated and the time spent in each frequency
band.  The latter term can be further rewritten using Kepler's law as,
	\begin{equation}
	\frac{d t_r}{d \ln f_r} = f_r \left(\frac{d f_r}{dt_r}\right)^{-1}
							= \frac{3}{2} \frac{a}{da/dt_r},
	\end{equation}
where `$a$' is the semi-major axis of the binary.  In this expression, we can identify the binary
`hardening time'\footnote{Sometimes called the `residence time' in the context of GW spectra.},
$\thard \equiv a / \left( da / dt_r\right)$.  For reference, the binary separations corresponding
to each GW frequency are shown in \ref{fig:gw_sep_freq}.  While the GW power radiated is
determined solely by the binary configuration (chirp mass and orbital frequency), the hardening time is
determined by both GW emission and the sum of all environmental hardening effects.
For more generalized binary evolution we can write,
	\begin{equation}
	\label{eq:gw-energy-emission_general}
	\frac{ d \egw}{d \ln f_r} = \frac{ d \egw}{d \ln f_r} \bigg|_\textrm{GW} \frac{\thard}{\tgw}.
	\end{equation}
This can be used to reformulate the GWB spectrum calculation\footnote{For a more complete derivation,
see \citet{kocsis2011}.} (\refeq{eq:gw-energy-phinney}) as,
	\begin{equation}
	h_c^2(f) = \frac{4 \pi}{3 c^2} \left(2 \pi f\right)^{-4/3}
				\int \frac{\left( G \mchirp \right)^{5/3}}{\left(1+z\right)^{1/3}}
				\frac{\thard}{\tgw}
				\frac{d^3 n}{dz \, d\mchirp \, d\mu} \, dz \, d \mchirp \, d\mu,
	\end{equation}
or for discrete sources,
	\begin{equation}
	\label{eq:gwb-full}
	h_c^2(f) = \frac{4 \pi}{3 c^2} \left(2 \pi f\right)^{-4/3}
				\sum_i \frac{\left( G \mchirp_i \right)^{5/3}}{\volcom \left(1+z_i\right)^{1/3}}
				\frac{\thardi}{\tgwi}.
	\end{equation}
Additional hardening mechanisms will decrease the hardening timescale, i.e.~$\thard/\tgw \leq 1$,
decreasing the GWB.  The purely power-law expression in \refeq{eq:gwb-naive} (and the Illustris
spectrum in \figref{fig:gwb-naive}) thus represents an
upper-limit to the GWB amplitude.  While non-GW mechanisms are required to bring MBH binaries close
enough to effectively emit gravitational waves, they also attenuate the amplitude of the GW
background.

% ---- GWB Full Calculation - Fiducial Results ----

\subsubsection{Fiducial Model Predictions}

	% Fig 16 ------ Fiducial, Full GWB Calculation 
	\begin{figure}
	\centering
	\includegraphics[width=\columnwidth]{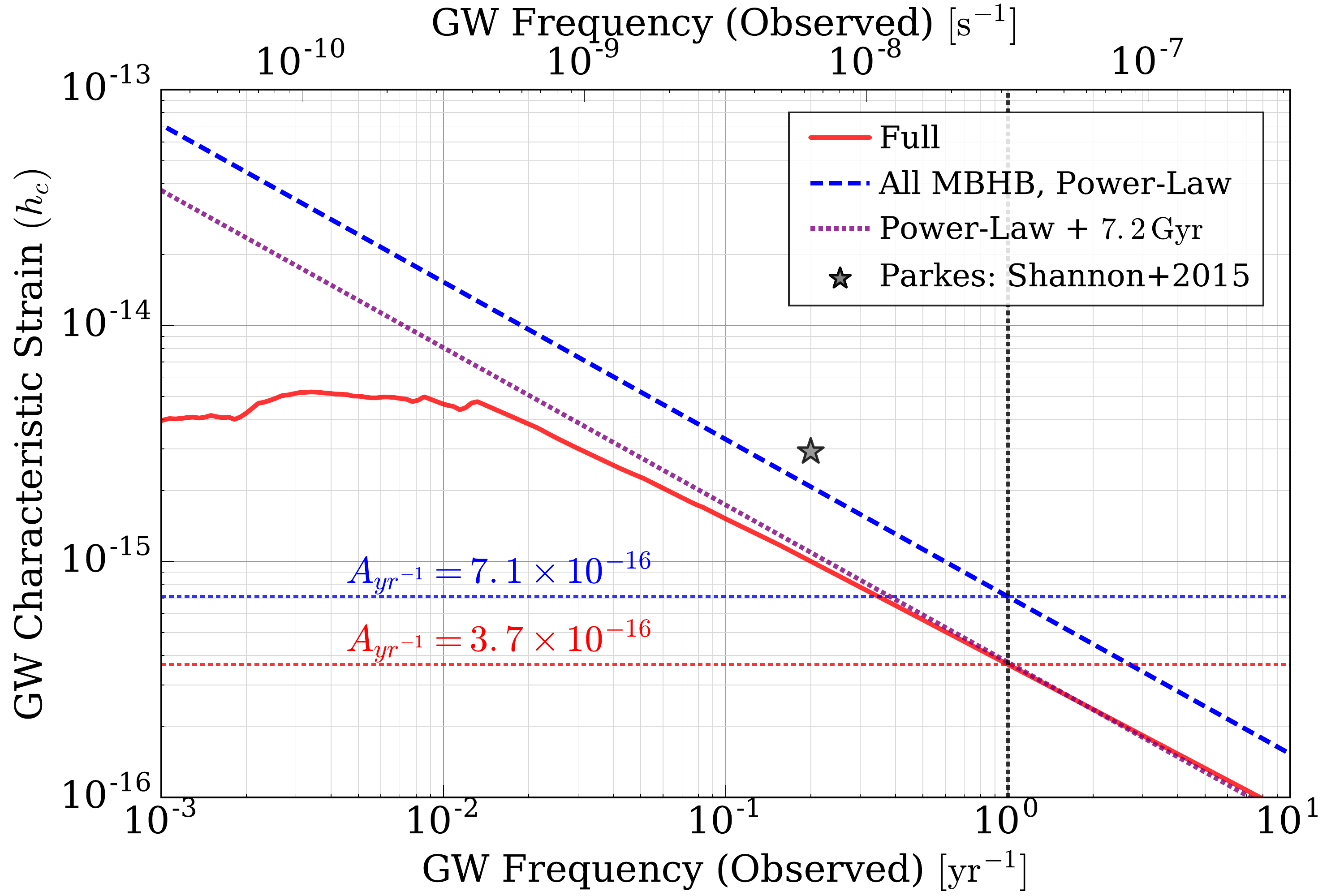}
	\caption{Stochastic Gravitational Wave Background calculated from Illustris MBH binaries.
	The `Full' calculation, shown in red, includes environmental effects from dynamical friction
	(`Enh-Stellar'), stellar scattering ($\frefill = 0.6$), and a viscous circumbinary disk.
	Purely power-law models are also shown, for all Illustris MBHB (blue, dashed) and only the
	MBHB which coalesce by redshift zero after being delayed for $7.2 \textrm{ Gyr}$ (purple,
	dotted).  The GWB strain amplitudes at the standard frequency of $1\,\pyr$ are given,
	showing that a complete model of MBHB evolution leads to a $\sim50\%$ decrease of the signal.
	The most stringent observational upper limits are also shown.}
	\label{fig:gwb-fiducial}
	\end{figure}

% Full GWB versus Pure Power Law
The stochastic Gravitational Wave Background (GWB) resulting from our fiducial model is
presented in \figref{fig:gwb-fiducial}.  The `Full' calculation (red, solid) uses \refeq{eq:gwb-full},
including the effects of DF, LC scattering, and VD in addition to GW emission.  This is compared
to a purely power-law model (blue, dashed), calculated with \refeq{eq:gwb-naive} and assuming
that all Illustris MBHB reach the PTA-band rapidly, and evolve solely due to GW-emission. The
amplitudes at $1 \, \pyr$ are indicated, showing that the full hardening calculation with an
amplitude of $\ayr \approx 3.7\E{-16}$ amounts to an almost $50\%$ decrease from the naive,
power-law estimate of $\ayr \approx 7.1\E{-16}$.

% Attenuation compared to power-law and delayed power-law
The amplitude of the full GWB calculation can be matched at $1 \, \pyr$ using the power-law
model by introducing a uniform delay time of $\sim 7.2 \, \textrm{ Gyr}$---such that the
systems which formed within a look-back
time of $7.2 \textrm{ Gyr}$ don't coalesce or reach the relevant frequency ranges.  This is
shown in \figref{fig:gwb-fiducial} (purple, dotted) as a heuristic comparison.  At frequencies
of the PTA band ($\sim 0.1 \, \pyr$) and higher, our full calculation very nearly matches the
$\ayr \propto \nobreak f^{-2/3}$ power-law.  A significant flattening of the spectrum is
apparent at and below a few $10^{-2} \, \pyr$, where environmental effects (e.g.~LC-scattering)
significantly increase the rate at which MBHB move through a given frequency band, decreasing
$\thard$ and thus attenuating the amplitude of the GWB.  The particular location and strength
of the spectral flattening (or turnover) depends on the details of the DF and especially LC
models.

% ---- GWB Comparison Of DF-LC Models ----

\subsubsection{GWB Variations with Dynamical Friction and Loss-Cone Model Parameters}

	% Fig 17 ------ Full GWB Spectra 3x Panels for DF; each LC
	\begin{figure}
	\centering
	\includegraphics[width=\columnwidth]{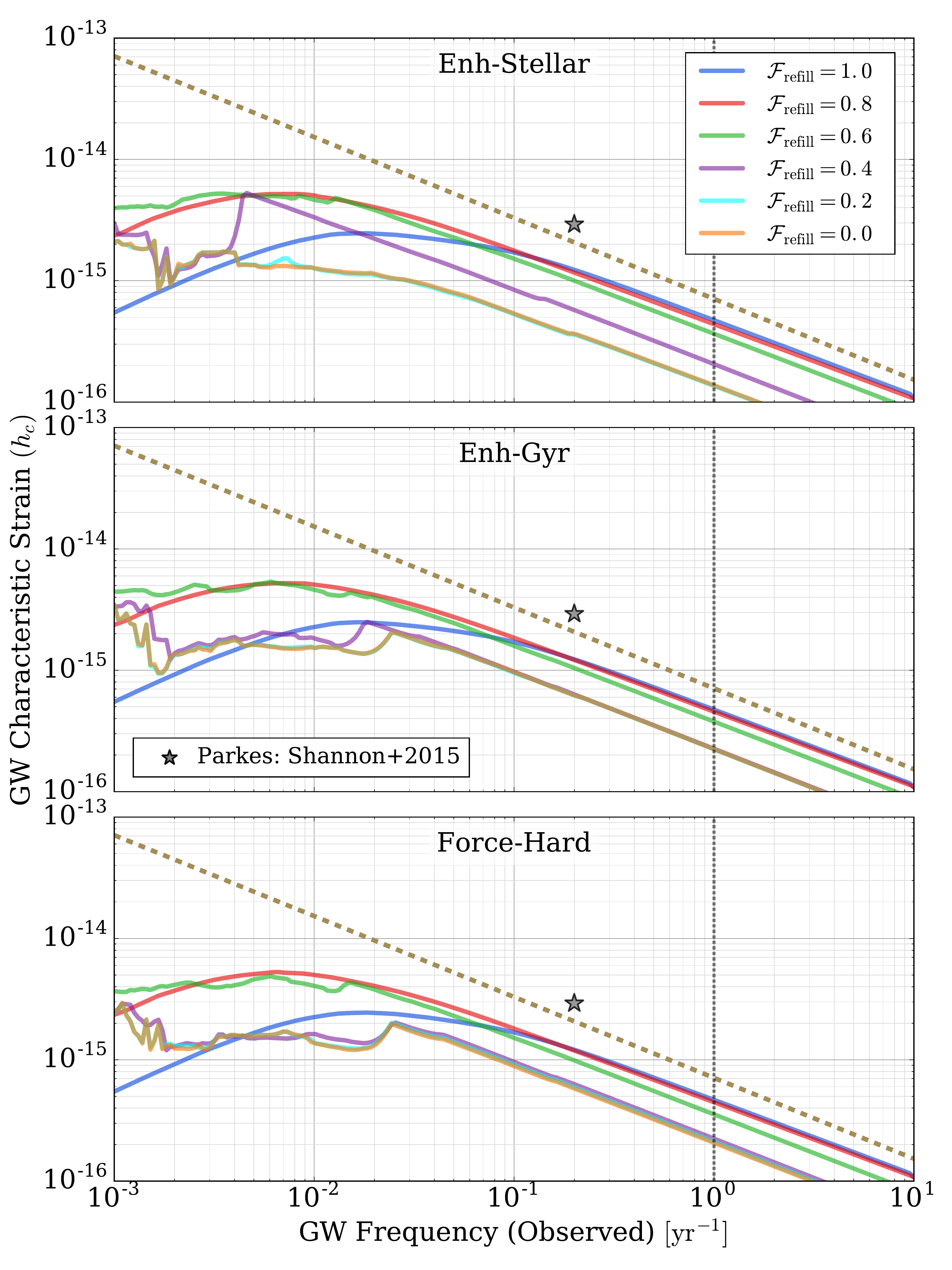}
	\caption{Comparison of GWB spectrum with variations in the LC refilling parameter and DF model.
	Each panel shows a different DF model, and each line-color a different $\frefill$.  Solid lines
	show the full GWB calculation while the dashed lines show the power-law model using all
	Illustris MBHB.  Variations in $\frefill$ have much stronger effects on
	the spectrum than the DF model, changing the location and strength of the spectral break at lower
	frequencies.  There tends to be a substantial jump in GWB amplitude between $\frefill = 0.4$
	and $0.6$, with more gradual variations on either side.  The full range of amplitudes
	at $1\,\pyr$ and $10^{-1} \, \pyr$ are \mbox{$\ayr = 0.14$ -- $0.47\E{-15}$} and
	\mbox{$\atyr = 5.4$ -- $17\E{-15}$}.}
	\label{fig:gwb-df-lc}
	\end{figure}

Figure~\ref{fig:gwb-df-lc} compares the GWB spectrum for different DF prescriptions (panels) and
LC refilling fractions (line colors).  The naive, power-law model is shown as the dashed line for
comparison, along with
the most stringent PTA upper limit.  Effects from variations in the DF prescription are strongly
subdominant to changes in the LC state.  The spectral shape is determined almost entirely,
and at times sensitively, to $\frefill$.  For $\frefill < 0.8$, the spectrum flattens at low frequencies,
whereas for higher values it becomes a turnover.  Even then, the location of the peak amplitude of the
spectrum changes by more than a factor of two between $\frefill = 0.8$ and $\frefill = 1.0$.

% Spectral Cutoff, Eccentric Evolution and Future Improvements
The cutoff seen in the full LC case ($\frefill = 1.0$) is very similar to that found by 
\citet{sesana2013b} (with ours $\sim 5$ times lower amplitude), who show that in the
scattering-dominated regime the GWB turns into a $h_c \propto f$ spectrum.  \citet{mcwilliams2014}
also find a spectral cutoff, but at an order of magnitude higher frequency and amplitude.
Unlike the
results of \citet{ravi2014}, the cutoffs in our predicted GWB spectra are always at lower
frequencies than will be reached by PTA in the next decade or so, likely because we assume
zero eccentricity in the binary evolution.  In the near future we hope to present results
expanded to include eccentric evolution, in addition to exploring `deterministic' or `continuous'
GW sources---i.e.~sources individually resolvable by future PTA observations.

For each DF case in \figref{fig:gwb-df-lc}, the GWB spectrum is almost identical between
$\frefill = 0.0, \, 0.2 \,\, \& \,\, 0.4$, with very little change in the coalescing fractions.
This is consistent with changes in the distribution of lifetimes from varying DF and LC parameters.
Looking at $f = 1 \, \pyr$, there is a sudden jump in amplitude with $\frefill = 0.6$, and a modest
increase in the coalescing fraction.  Between $\frefill = 0.6$ and $\frefill = 0.8$, on the other
hand, there tends to be a more modest increase in GWB amplitude, but a roughly factor of two
increase in $\fcoal$.  This contrast arrises from the changing population of MBHB which are brought
to coalescence from each marginal change in refilling fraction.  For an increase in $\frefill$, the
additional MBHB which are then able to coalesce tend to be the most massive of those which were
previously persisting.  Those, more massive systems, then have a larger effect on the GWB

	% Fig 18 ------ Strain vs. Coal-Frac for all models with double inset
	\begin{figure}
	\centering
	\includegraphics[width=\columnwidth]{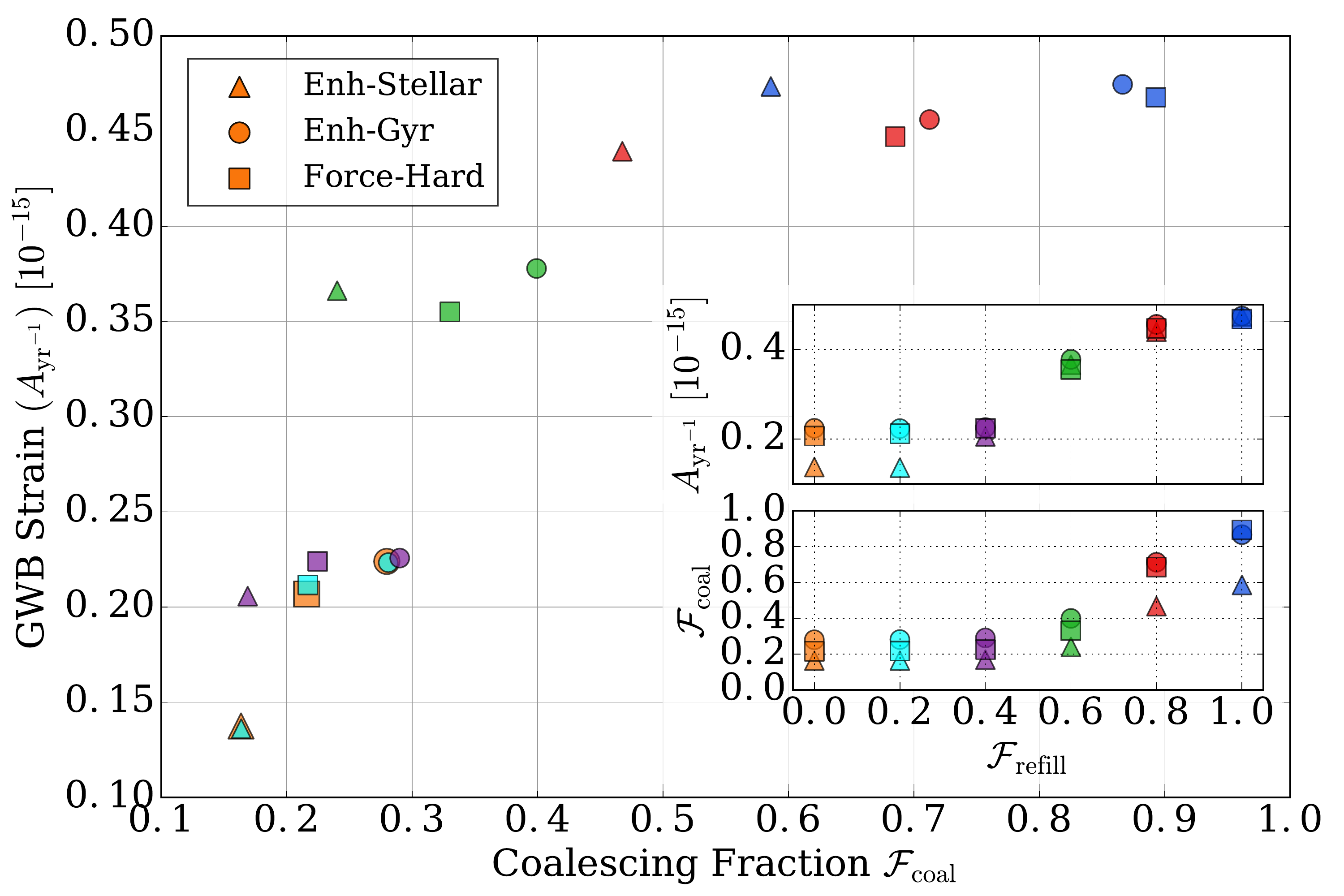}
	\caption{Dependence of GWB strain amplitude and binary coalescing fraction on LC refilling
	parameter and DF model.  The GW strain is measures at the canonical $f = 1 \, \pyr$, and
	the coalescing fraction is defined using the population of high mass-ratio $\mu > 0.1$
	systems.  Each symbol represents a different DF model, and each color a different LC refilling
	parameter.  $\ayr$ tends to increase monotonically with $\fcoal$, but plateaus above
	$\fcoal \gtrsim 0.5$.  The insets show how each $\ayr$ and $\fcoal$ change with $\frefill$.
	The strongest changes in $\frefill$ and $\ayr$ occur at slightly different values of the
	refilling fraction.  This is because for an increase in $\frefill$, the additional MBHB
	which are then able to coalesce tend to be the most massive of those which were previously
	persisting.  At $\frefill \approx 0.5$ there is a significant change $\ayr$, due to more
	massive and stronger GW emitting MBHB merger at that point.  At $\frefill \approx 0.7$,
	$\fcoal$ changes significantly due to less massive MBHB then being able to merger, and them
	constituting a larger portion of the binary population.}
	\label{fig:gwb-coal-frac}
	\end{figure}
	
Figure~\ref{fig:gwb-coal-frac} shows the strain at $1 \, \pyr$ ($\ayr$) versus coalescing fraction
for the same set of DF and LC models.  The colors again show different $\frefill$, and
now symbols are used for different DF prescriptions.  The GWB amplitude is strongly correlated with
coalescing fraction, but plateaus once roughly $50\%$ of high mass-ratio MBHB are coalescing.  Different
DF parameters have little effect on $\ayr$ but more noticeably affect $\fcoal$, in both cases
this is especially true at higher $\frefill$.  The inset panels show, independently, how $\ayr$ and
$\fcoal$ scale with $\frefill$ and DF model, reinforcing the previous points.  In general, as
$\frefill$ increases, lower total-mass MBHB systems are able to reach the PTA band, contribute to
the GWB and coalesce effectively.  At $\frefill \approx 0.5$, the large increase in $\ayr$ is driven
by massive MBH coming to coalescence, while at $\frefill \approx 0.7$, a large number of MBH
at lower masses are driven together, significantly increasing the coalescing fraction, but only
marginally increasing $\ayr$.

At the higher frequencies just discussed the GWB strain increases monotonically with $\frefill$
and coalescing fraction.
This is intuitive as increasing effectiveness of the LC means more MBHB are able to reach the
GW-regime and then coalesce.  \figref{fig:gwb-df-lc} shows that this trend is \textrm{not} the case
at lower frequencies (i.e.~$f \lesssim 10^{-1} \, \pyr$)---where the highest $\frefill$ show a 
decrease in the GWB amplitude.  This can be seen more clearly in \figref{fig:gwb-coal-frac_f-2},
which shows the GWB amplitude at $f=10^{-2} \, \pyr$ versus coalescing fraction.  The trend is
generally the same---strain increasing with $\frefill$---until $\frefill = 1.0$ at which point the GWB
amplitude drops significantly.  At these low frequencies, LC stellar scattering is effective enough
to significantly attenuate the GWB amplitude.  This reflects a fundamental tradeoff in the realization
of environmental effects: on one side bringing more MBHB into a given frequency bands, at the same
time as driving their evolution rapidly through it, and attenuating the GW signal.

% ---- GWB Comparison Of VD Models ----

\subsubsection{Effects of Circumbinary Viscous Drag on the GWB}

	% Fig 19 ------ Full GWB Spectra 2x Panels for LC, varying VD Parameters
	\begin{figure}
	\centering
	\includegraphics[width=\columnwidth]{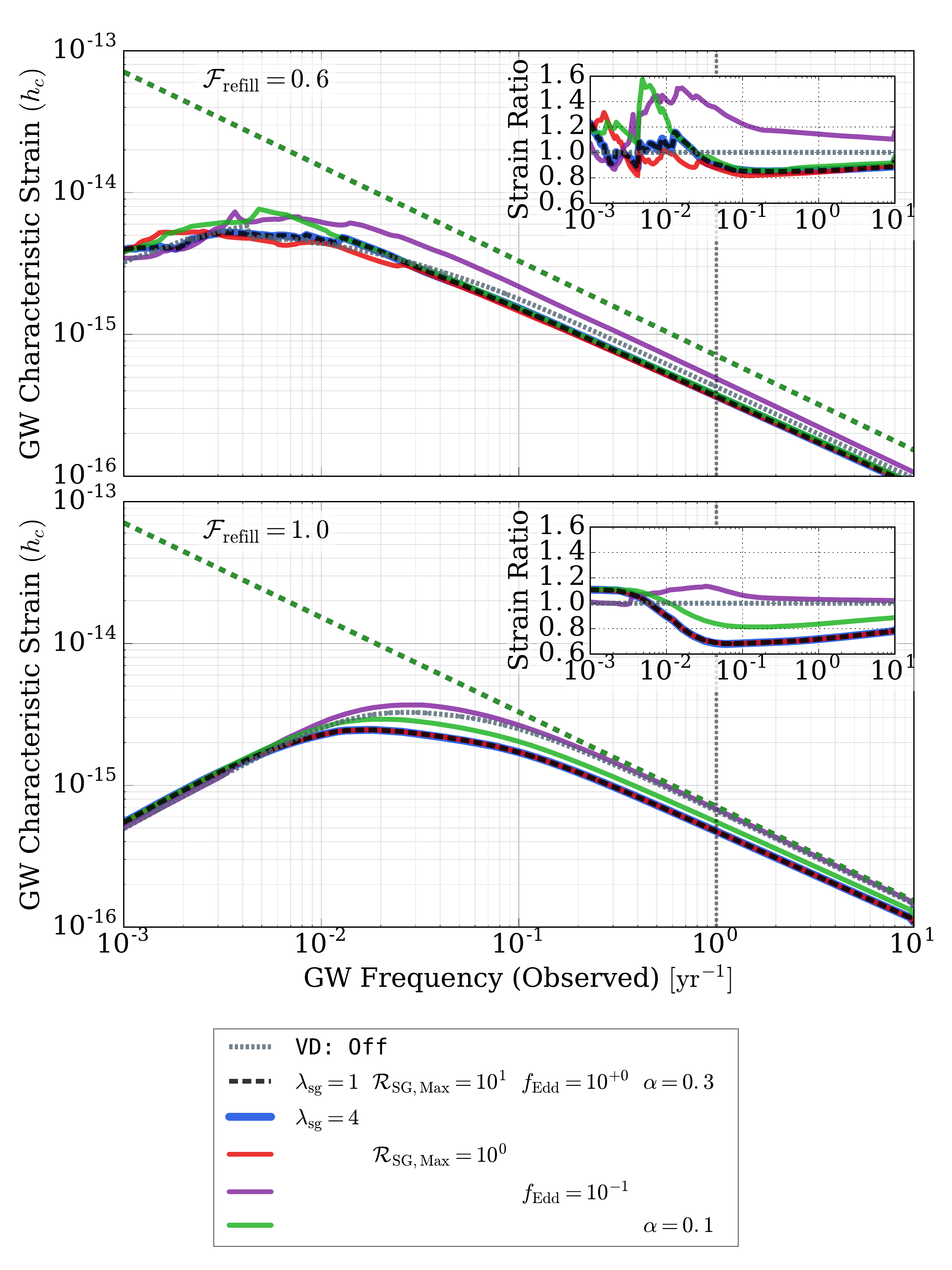}
	\caption{Gravitational wave background from varying viscous drag (VD) parameters.  Simulations
	with a variety of VD models are compared, with our fiducial model in dashed black, a model with
	no-VD in dotted grey, and each color of line showing changes to a different parameter.  For
	comparison, the power-law model using all Illustris MBHB is also shown.  The
	parameters modified are: the self-gravity (SG) instability radius ($\rselfgrav$), the maximum
	SG radius ($\rsgmax$), the maximum accretion rate ($\fedd$), and the alpha viscosity parameter
	($\alpha$).  The hardening rates for each of these models are shown in
	\figref{fig:vd-comp-hard-pars}.  The upper and lower panels show simulations for different LC
	refilling fractions. The inset panels show the ratio of GWB amplitude from each model to the
	`VD: Off' case, as a function of GW frequency.  Different VD parameters change $\ayr$
	by $10$ -- $40\%$, and the spectral slope at $1 \, \pyr$ by up to $\sim 10\%$.}
	\label{fig:gwb-vd}
	\end{figure}

The effects of viscous drag (VD) from a circumbinary disk are more subtle than those of DF and LC.
\figref{fig:gwb-vd} compares the GWB from our fiducial model (black, dashed) with a variety of VD
parameter modifications (colored lines) and to a simulation with VD turned off (grey, dotted).
All of these models use the `Enh-Stellar' DF, but because there is virtually no overlap in
between the VD and DF regimes, the results are very similar.  The
same VD parameters are explored as in the hardening rates shown in \figref{fig:vd-comp-hard}: modifying
the self-gravity (SG) instability radius ($\rselfgrav$), the maximum SG radius ($\rsgmax$), the
maximum accretion rate ($\fedd$), and the alpha viscosity parameter ($\alpha$).  The upper panel
shows a moderately refilled LC ($\frefill = 0.6$), while in the lower panel the LC is always
full ($\frefill = 1.0$).  The inset panels show the ratio of GWB strain from each model to that
of a `VD: Off' (i.e.~no disk) model.

The overall shape of the GWB spectrum and the location of the spectral turnover is again determined
almost entirely by the LC.  The circumbinary disk does affect an additional $10$ -- $40\%$ amplitude
modulation, tending to increase the amplitude at low frequencies ($\lesssim 10^{-2} \, \pyr$) and
decrease it at higher frequencies ($\gtrsim 10^{-1} \, \pyr$).  This reflects the same tradeoff
between bringing more MBHB into each frequency band, versus driving them more rapidly through them.
In the moderately (completely) refilled LC case, our fiducial VD model amounts to a $\sim 20\%$
($\sim 30\%$) decrease in $\ayr = h_c(f=1\, \pyr)$ and similarly at $h_c(f=10^{-1} \, \pyr)$.
 
At frequencies near the PTA band, the relationship between $\fcoal$ and the GWB amplitude can
be non-monotonic for VD variations, like with variations to the LC at low frequencies.  For
example, a comparison of Figures~\ref{fig:vd-comp-hard} \& \ref{fig:gwb-vd} shows that the
$\alpha = 0.1$ (green) model has the lowest fraction of high mass-ratio coalescences
(with $\fcoal = 0.22$, versus $\fcoal = 0.24$ for the fiducial model, and $\fcoal = 0.31$ for the
$\fedd = 0.1$ case) but an intermediate $\ayr$.

One striking feature of the GWB strain ratios is the clear variations in spectral index, even at
high frequencies.  This is especially true for the always full LC, where the slope of the GWB can
deviate by almost $10\%$ from the canonical $-2/3$ power-law.  The disk-less model (grey, dotted)
deviates by about $4\%$ ($3\%$) for $\frefill = 0.6$ ($\frefill = 1.0$) at $f = 1 \, \pyr$, due to
a combination of residual LC scattering effects and some binaries stopping emitting after coalesce
at varying critical frequencies.
In our fiducial model (black, dashed), the deviations are more significant at $6\%$ ($8\%$). As
different parameters make VD hardening more important at this frequency, the GWB amplitude
decreases, and the spectral index tends to flatten.  Our fiducial VD model tends to have among the
strongest spectral deviations.  Towards lower frequencies, where PTA are heading, the turnover in
the GWB spectrum becomes more significant, especially if the LC is effectively refilled.  At
$f = 10^{-1} \, \pyr$, for example, our fiducial model (`Enh-Stellar', $\frefill = 0.6$) gives a
spectral index of about $-0.6$, while for $\frefill = 1.0$ it becomes slightly flatter than $-0.4$.
A summary of GWB amplitudes and spectral indices are presented in \tabref{tab:results}, for a variety
of configurations.

	% Fig 20 ------ Hardening Landscape vs. Frequency, bans for each mechanism
	\begin{figure}
	\centering
	\includegraphics[width=\columnwidth]{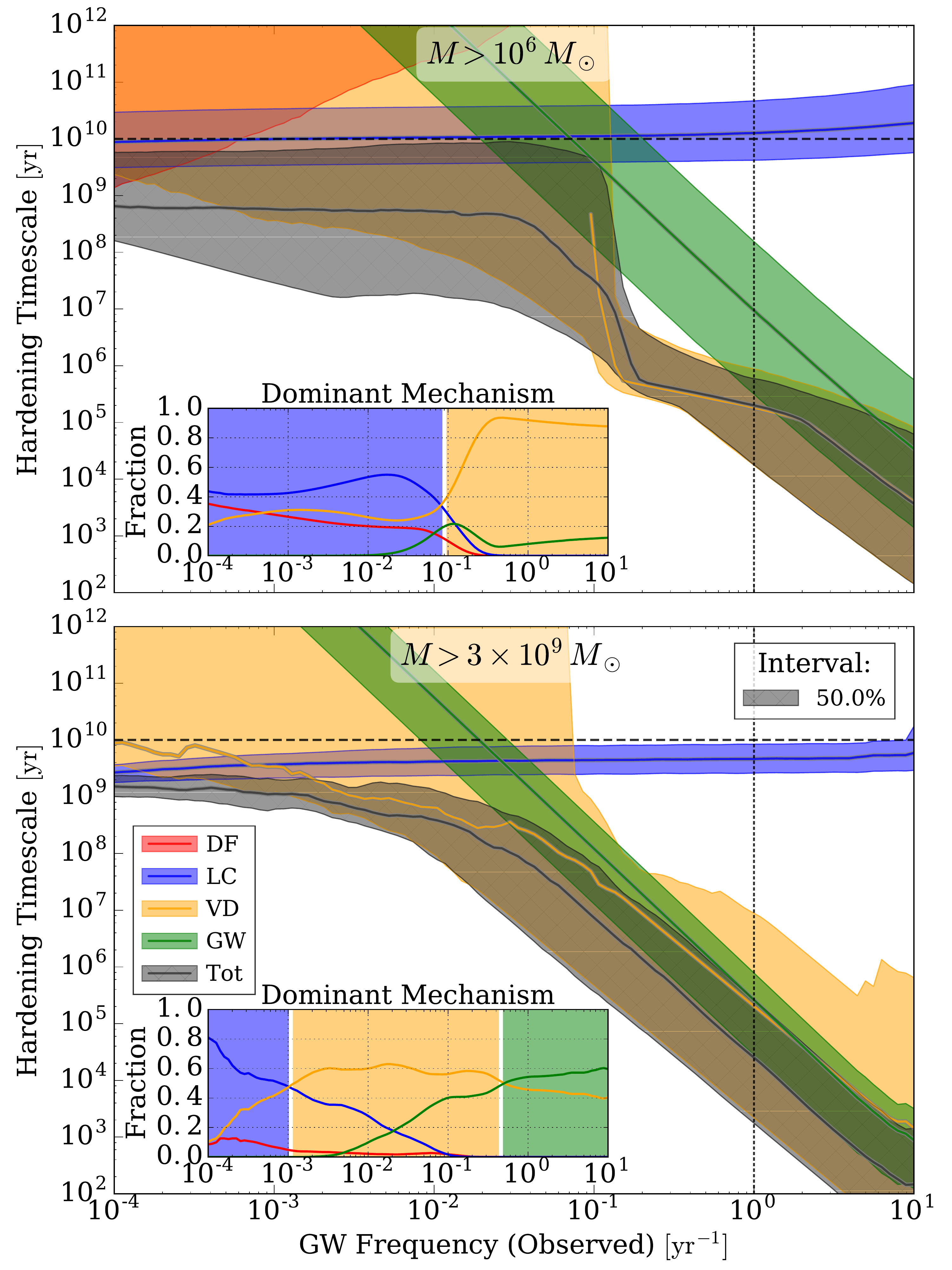}
	\caption{Binary hardening timescales versus GW frequency, by mechanism, for our fiducial model
	($\frefill = 0.6$, DF: `Enh-Stellar'). Lines and bands show the median and 50\% intervals for
	individual mechanisms (colors), along with the total hardening rate (grey, hatched).
	The inset panels show the fraction of binaries dominated by each mechanism, again versus
	frequency.  The upper panel shows all MBHB in our sample, while the lower panel includes
	only systems with total mass above $3\E{9} \msol$, roughly where the bulk of the GWB
	amplitude comes from.  Only for the high mass systems do the majority become dominated by
	GW emission at high frequencies, with VD still contributing substantially to the overall
	hardening timescales.}
	\label{fig:hard_freq}
	\end{figure}

For a given binary system, GW radiation will \textit{always} dominate at some sufficiently small
separation (high frequency) where the circumbinary disk dynamically decouples from the hardening
MBHB.  This does \textit{not necessarily} mean, however, that after considering a full ensemble of
MBHB systems, with a variety of masses, that there is any frequency band with a spectral slope
identical to the purely GW-driven case ($h_c \propto f^{-2/3}$).  Hardening rates as a function of
GW frequency are shown in \figref{fig:hard_freq}.  The upper panel includes all MBHB in our
sample\footnote{Recall that we select only MBH with masses $M > 10^6 \, \msol$.},
for which we see that VD remains dominant well above the PTA frequency band.  The high total-mass systems
($M > 3\E{9} \, \msol$)---which contribute the bulk of the GWB signal---are shown in the lower panel.
These binaries tend to be driven in roughly equal amounts by VD and GW hardening at the frequencies
where PTA detections should be forthcoming.

Figure~\ref{fig:hard_freq} (and \figref{fig:hard}) show that the typical hardening rates for VD are
very similar to that of GW radiation.  Indeed, as discussed in \secref{sec:visc}, the inner-most
disk region has hardening times $\tviscone \propto r^{7/2}$, while that of purely gravitational wave
emission is $\thardgw \propto r^4$.  Hardening rates for farther-out disk regions tend to deviate more
strongly from that of purely GW evolution, which could become more important for lower density disks.

The MBH accretion rates, which set the density of the
circumbinary disks in our models, are perhaps one of the more uncertain aspects of the Illustris
simulations, given that the accretion disk scale is well below the resolution limit and must
therefore rely on a sub-grid prescription.
Additionally, out of all possible configurations, the
fiducial disk parameters we adopt tend to produce fairly strong effects on the GWB.  If, for example,
a $\beta$-disk model is more accurate, or the $\alpha$-viscosity should be lower, the effects in the
PTA band will be more moderate \citep[see, e.g.,][]{kocsis2011}.
None the less, we consistently see GWB spectral indices
between $-0.6$ and $-0.65$ at $1 \, \pyr$, for a wide variety of model parameters.   While these
$\lesssim 10\%$ deviations may be entirely unobservable in PTA observations \citep[especially after
taking stochastic variations into account; e.g.][]{sesana2008}, it may need to be considered when
using priors or match-filtering for detecting a GWB.  More stringent observational constraints on
specifically post galaxy-merger AGN activity could be used to better calibrate the VD model.

% ---------------------------------------------------------
% ------ Coalescing and Persisting MBHB Populations -------
% ---------------------------------------------------------

\subsection{The Populations of MBHB}
\label{sec:res_pops}

% Environments from PTA upper limits
For the first time, we have used cosmological, hydrodynamic models which self-consistently evolve
dark matter, gas, stars and MBH, to more precisely probe the connection between MBHB mergers and
their environments.  Previous calculations (see \secref{sec:intro}) of the GWB using SAM prescribe
MBH onto their galaxies based on scaling relations.  The MBH population in Illustris, on the other
hand, co-evolves with, and shapes, its environment. These data are then much better suited to
analyze the details of MBHB and GW source populations, and their hosts.

	% Fig 21 ------ Source Properties of GWB MBH Binaries
	\begin{figure}
	\centering
	\includegraphics[width=\columnwidth]{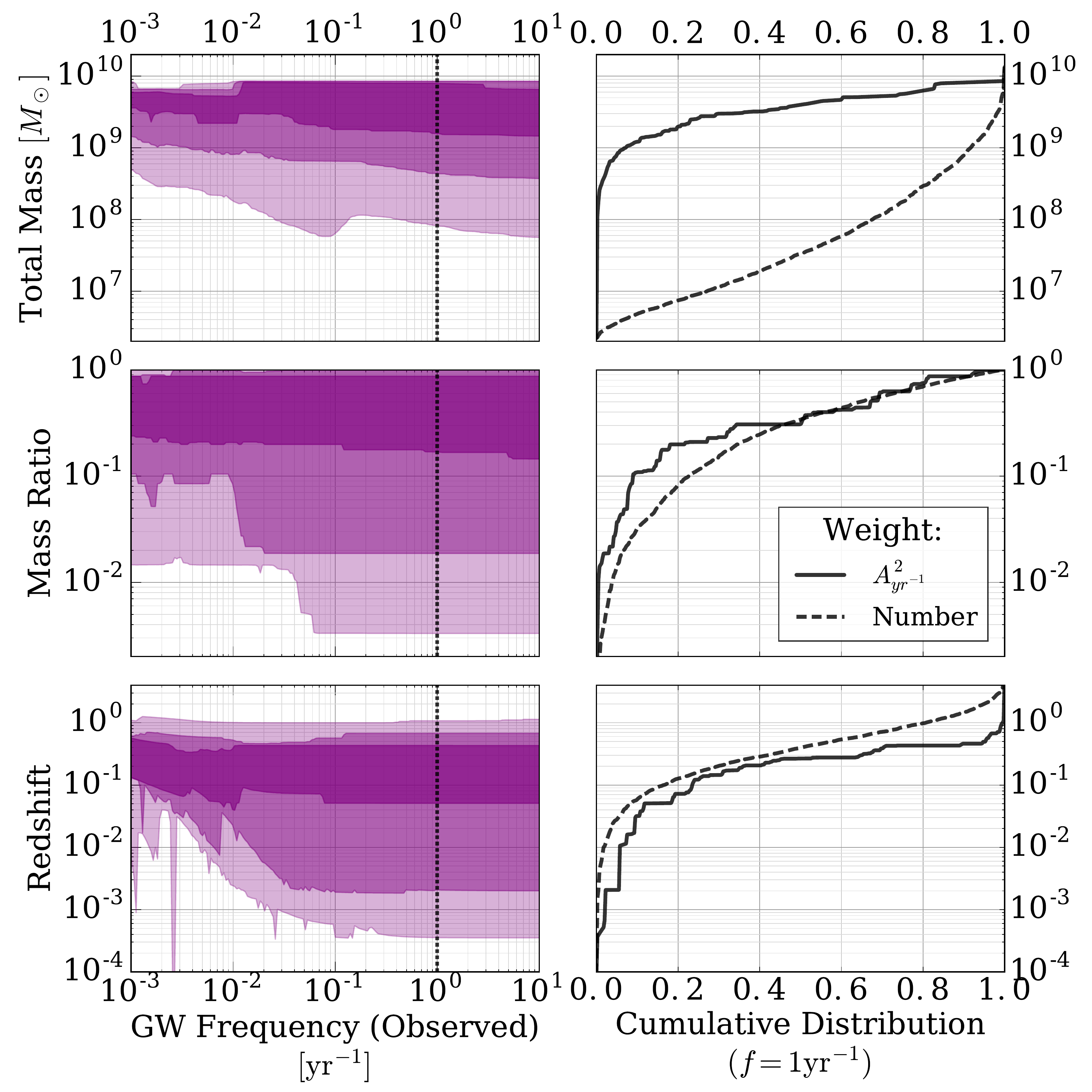}
	\caption{Population of MBHB contributing to the GWB.  The left column shows, from top to
	bottom, the MBHB total mass, mass ratio, and redshift, weighted by each system's contribution
	to the GWB amplitude.  Contours represent one-, two-, and three- sigma intervals.  The Right
	column shows cumulative distributions, at a frequency of $1 \, \pyr$, for the same parameters.
	The solid line weights by contribution to the GWB amplitude ($\ayr^2$) and the dashed line
	is the distribution of the number of sources \textit{contributing at $1\, \pyr$}.
	The median values by GWB-contribution are roughly
	constant over GW frequency, at $M \approx 4\times 10^9 \, \msol$, $\mu \approx 0.3$, and
	$z \approx 0.3$ for this, our fiducial model.  The overall distribution of sources moves
	noticeably to include lower masses, mass ratios, and redshifts at higher GW frequencies.
	The contribution from redshifts above $z \approx 0.4$ drops sharply, with $\lesssim 1\%$ of
	the GWB signal coming from $z > 1.0$, while still $\sim 20\%$ of all binaries emit there.}
	\label{fig:gwb-pop-mbhb}
	\end{figure}

% MBH Binary Source Properties
Figure~\ref{fig:gwb-pop-mbhb} shows the distribution of properties for sources contributing to
the GWB, from top to bottom: total mass ($M$), mass ratio ($\mu$), and redshift ($z$).  In the
left column, these properties are weighted by squared-strain\footnote{As seen in
\refeq{eq:gwb-full}, binaries contribute to the strain spectrum in quadrature.} for each source,
and the resulting \mbox{one-, two-, and three-sigma} contours are shown as a function of GW
frequency.  The right column shows the cumulative distribution over the same source properties,
weighted by $\ayr^2$ (solid), compared to the unweighted distribution of all sources
contributing at $f = 1 \, \pyr$. Strain-weighted sources tend to be at higher mass-ratios
and much higher masses.  While the fraction of all binaries rises fairly smoothly with total
masses between $10^7$ and $10^9 \, \msol$ (dashed, black line; top-right panel), $90\%$ of the GWB
is contributed (solid, black line) by binaries with total mass $\gtrsim 10^9 \, \msol$---simply
showing the strong dependence of the GW strain on the total system mass.  

The core contribution over all three parameters tends to remain fairly
constant over GW frequency, with median values around $M \approx 4\times 10^9 \, \msol$,
$\mu \approx 0.3$, and $z \approx 0.3$.  The tails of the distribution drop to noticeably lower
values when moving to higher frequencies.  This is especially pronounced in the redshift
distribution, where at frequencies of a few times $10^{-3} \, \pyr$ virtually all GWB-weighted
sources come from $z > 10^{-2}$, while at $f=1\, \pyr$, almost $10\%$ are below that redshift.
While $\sim20\%$ of binaries that reach $f = 1 \, \pyr$ come from redshift above $z=1$, they
only contribute $\sim0.5\%$ of the GWB amplitude.  Lower redshift and higher mass-ratio
systems do contribute somewhat disproportionately to the GWB amplitude, but their distributions are
altogether fairly consistent with the overall population.  The presence of a non-negligible
fraction of low redshift sources motivates the need to explore populations of MBHB in the local
universe which could be resolvable as individual `stochastic' sources, or contribute to angular
anisotropies in the GW sky.  An analysis of our results in this context is currently underway,
and the results will be presented in a future study.

	% Fig 22 ------ Source Properties of GWB MBH Host Galaxies
	\begin{figure}
	\centering
	\includegraphics[width=\columnwidth]{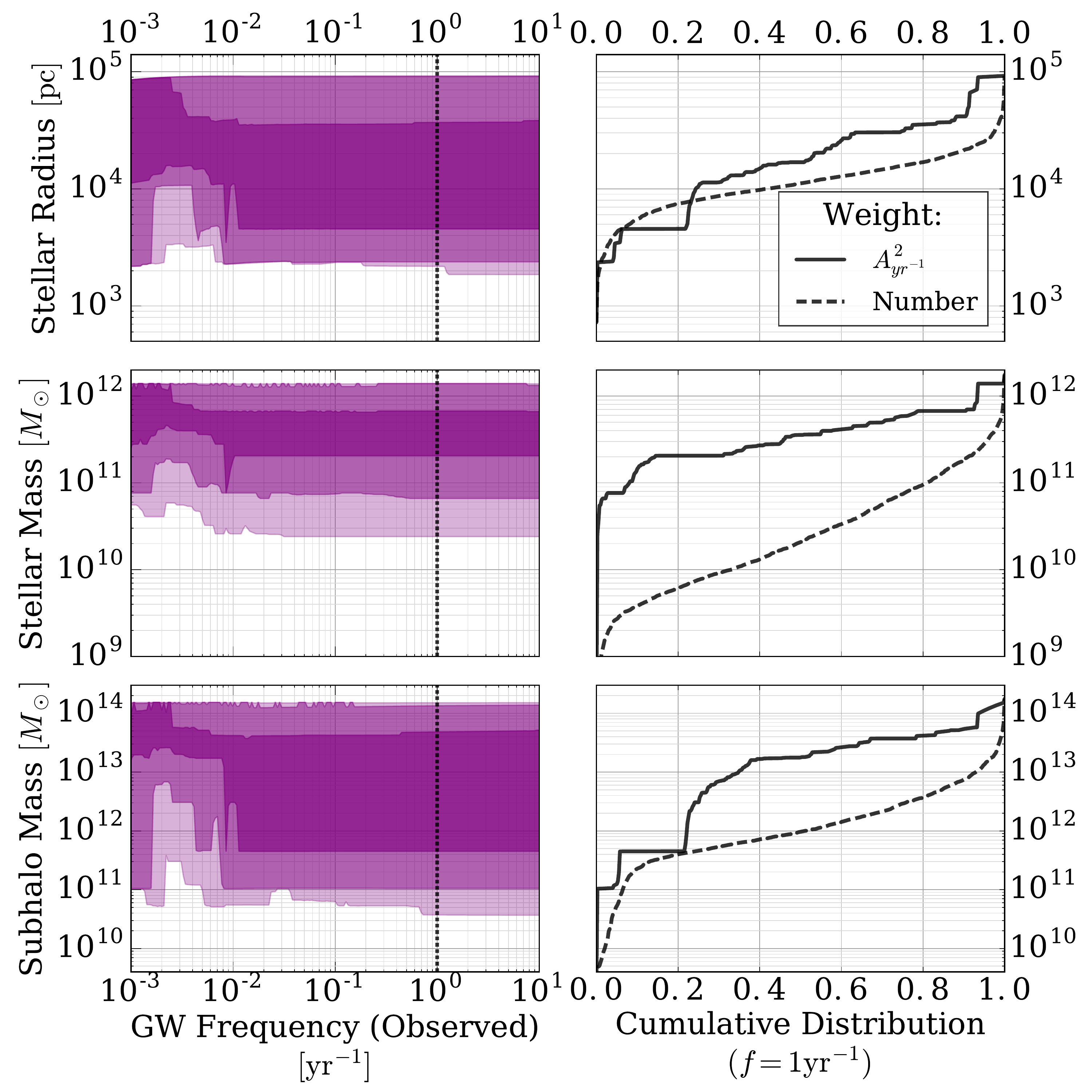}
	\caption{The properties of host galaxies for the population of MBHB contributing to the GWB.  From
	top to bottom, rows show the stellar (half-mass) radius, stellar mass (within twice the stellar
	half-mass radius), and subhalo mass (mass of all particles associated with the galaxy).
	The left column shows these parameters weighted by their resident MBHB's contribution to
	the GWB as a function of frequency.  The right column shows the cumulative distribution
	at $f = 1 \, \pyr$, both for contribution to GWB amplitude (solid) and by overall
	number (dashed).  The GWB comes from MBHB predominantly in galaxies which are over-sized and
	significantly over-massive---especially in stars.}
	\label{fig:gwb-pop-host}
	\end{figure}

% Host Galaxy Properties
As we move into the forthcoming era of PTA \textit{detections} it will be increasingly important to
use self-consistent hydrodynamic models to better understand the coupling of the MBH populations to
their host-galaxies and merger environments.  The Illustris host-galaxy properties of our MBHB, at
the time of binary formation, are presented in \figref{fig:gwb-pop-host}.  We show stellar
radius, stellar mass and `subhalo mass', and each of these
properties\footnote{The stellar radius is measured as the stellar half-mass radius ($\rstarhalfmass$);
the stellar mass is the mass of star particles within $R = 2 \, \rstarhalfmass$; and the subhalo mass
is the combined mass of all particles and cells associated with the host galaxy.  While these simulation
measurements are, of course, non-trivial to relate to their observational counterparts, they are
useful for relative comparisons.} is strongly biased towards higher values when weighting by GW strain.
In particular, the median, strain-weighted subhalo and stellar masses are each more than an order of
magnitude larger than the median of the host-galaxy population by number.  The bias is exceedingly
strong for stellar mass, where $\sim90\%$ of the GWB amplitude is contributed by only $\sim 20\%$ of
MBHB host galaxies.

Following the galaxies that host MBHB to observe their parameters at the times they
contribute significantly to the GW spectrum will be important for any future multi-messenger
observations using PTA or predicting and deciphering anisotropies in the GWB
\citep{taylor2013, mingarelli2013, taylor2015}.  Better understanding host galaxy properties as they evolve in time
could also be useful in understanding whether `offset' AGN (those distinctly separated from the
morphological or mass-weighted center of their galaxies) are due to binarity (i.e.~a recent, or
perhaps not so recent, merger) or possible due to post-coalesce GW `kicks'
\citep[e.g.][]{blecha2016}.

% Challenges of Dual/Binary AGN
Of great observational interest is the presence \citep[e.g.][]{comerford2015}, or perhaps
conspicuous absence \citep[e.g][]{burke-spolaor2011}, of dual and binary AGN.  The observational
biases towards finding or systematically excluding MBH binaries with electromagnetic observations
are extremely complex.  None the less, understanding the characteristic residence times of binaries
at different physical separations, the types of host galaxies they occupy, and the probability they
will be observable (e.g.~via the amount of gas available to power AGN activity) is crucial to
backing out the underlying population, and placing empirical constraints on models of MBHB inspiral.
A systematic study of this topic using these data is currently underway (Kelley et al. in prep.).

	% Fig 23 ------ Cumulative Persisting Fractions 3x2 
	\begin{figure}
	\centering
	\includegraphics[width=\columnwidth]{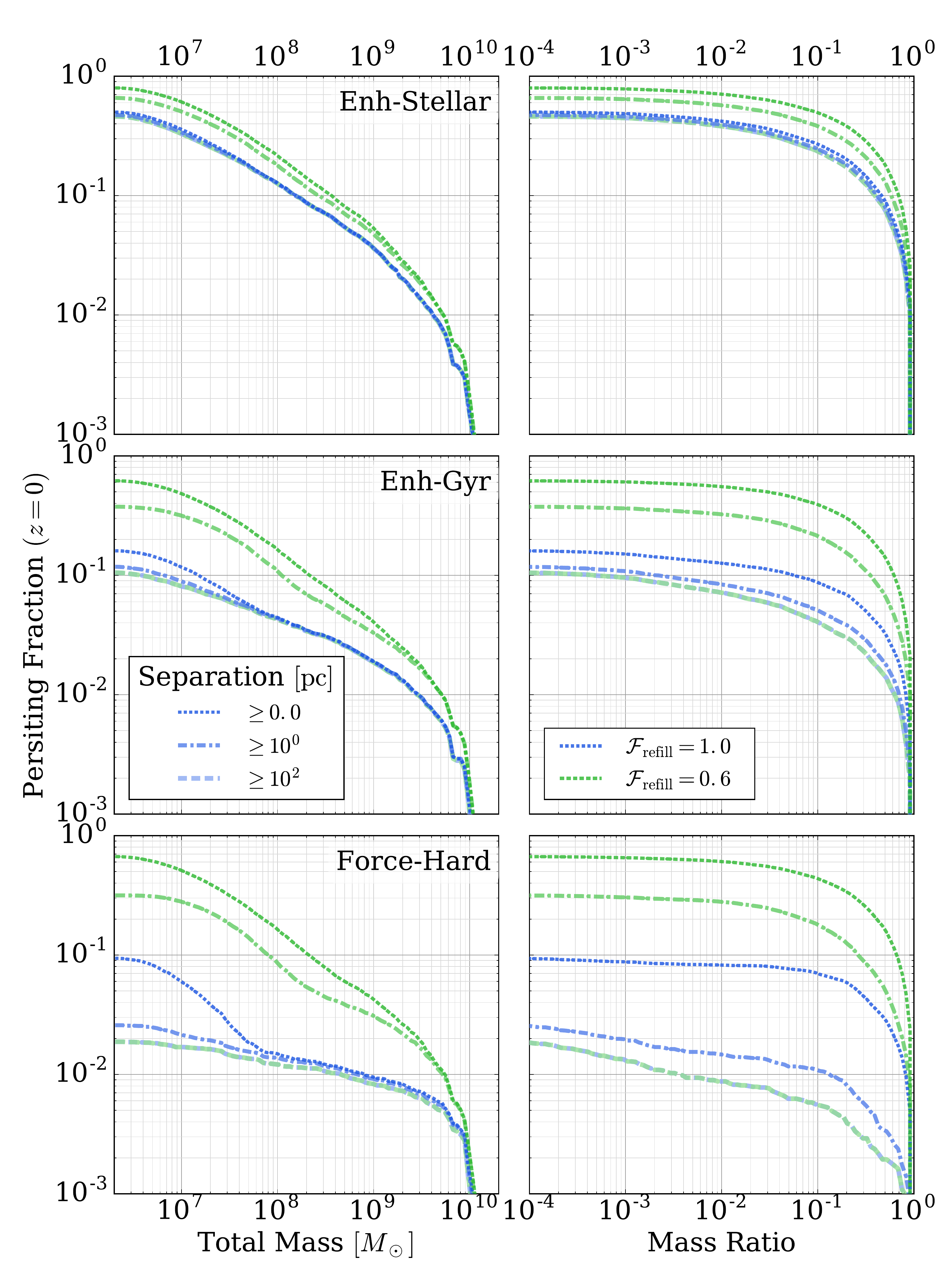}
	\caption{Fraction of binaries which persist at redshift zero as a function of total mass (left
	column) and mass ratio (right column).  Three different DF models are compared, from top to
	bottom: `Enh-Stellar', `Enh-Gyr', and `Force-Hard'; and in each case, LC refilling parameters
	of $\frefill = 0.6$ (green) and $\frefill = 1.0$ (blue) are compared.  Different line
	patterns show binaries with different separations: all
	separations ($r \geq 0.0$, dotted), $r \geq 1 \textrm{ pc}$ (dot-dashed), and
	$r \geq 10^2 \textrm{ pc}$ (dashed). Note that in each panel the $r \geq 10^2 \textrm{ pc}$
	distributions are indistinguishable between LC parameters as the LC only takes effect at and
	below about $10^2 \textrm{ pc}$.
	The fraction of persisting systems is very strongly dependent on both
	DF and LC model.  For `Enh-Stellar', the most conservative DF case, a large fraction of
	systems remain in the DF regime ($r \sim 10^2 \textrm{ pc}$), before LC scattering can have
	a significant effect. `Force-Hard', on the other hand, represents an approximately optimal
	DF at large scales, and shows a corresponding dearth of wide separation systems.
	Observations of the true fraction of systems at these separations could
	strongly constrain the efficiency of these hardening mechanisms.}
	\label{fig:mbhb-pers}
	\end{figure}

The fraction of MBHB which persist (i.e.~remain uncoalesced) at redshift zero are shown in
\figref{fig:mbhb-pers} as a function of total mass (left column) and mass ratio (right column).
Three different separation criteria are shown in each panel: $r > 0.0$ (i.e.~any persisting MBHB;
dark, dotted), $r > 1 \textrm{ pc}$ (medium, dot-dashed), and $r > 10^2 \textrm{ pc}$ (light,
dashed).  Each row corresponds to
a different DF model, and line colors vary by LC refilling fractions.  In general, persisting
fractions fall rapidly with increasing total mass and moderately with increasing mass ratio, until
nearly equal-mass systems where the persisting fractions plummet.

The specific persisting fraction depends quite sensitively on both $\frefill$ and DF model.  The
`Enh-Stellar' model has by far the most persisting systems, and relatively slight variance with
either separation criteria or $\frefill$.  For our fiducial model with $\frefill = 0.6$, $80\%$
of all binaries persist, with only weak trends with either total mass or mass ratio: $77\%$
with $M > 10^8 \,\msol$, and $74\%$ with $\mu > 0.2$.  For systems fulfilling both requirements,
the persisting fraction drops more noticeably to $55\%$.

The $r > 10^2 \textrm{ pc}$ population in particular, is almost solely determined by DF, as the
other hardening mechanisms take effect only at smaller scales.  At these large separations,
persisting fractions for our fiducial model are  $46\%$ (all), $45\%$ ($M > 10^8 \, \msol$), and
$33\%$ ($\mu > 0.2$), but for both high total-mass and mass-ratio, the widely-separated persisting
fraction drops dramatically to only $1\%$.  If the DF is more effective, as in the `Enh-Gyr' model,
these fractions decrease significantly to $11\%$ (all), $15\%$ ($M > 10^8 \, \msol$), $6\%$
($\mu > 0.2$), and $1\%$ ($M > 10^8 \, \msol$ \& $\mu > 0.2$).  A summary of persisting fractions
at both $r > 10^2 \textrm{ pc}$ and $r > 1 \textrm{ pc}$, for mass combinations
and DF \& LC models are presented in \tabref{tab:results}.

% ----------------------------
% ------ MBH Multiples -------
% ----------------------------

\subsection{MBH Triples}
\label{sec:res_multi}

% Long lifetimes motivate consideration of triples
The long characteristic lifetimes we see in our MBHB populations, and the (at times) substantial
number of systems which remain at large separations, immediately begs the question of how often
a third MBH (i.e.~second galaxy merger) could become dynamically relevant.  For $\frefill = 1.0$,
the median lifetimes of our MBHB tend to be comparable to the median time between binary formation
events, and for $\frefill = 0.6$, they are almost an order of magnitude longer.
After selection cuts (see \secref{sec:ill}), $37\%$ of our MBH binaries have subsequent `merger'
events (i.e.~a second `merger' is recorded by Illustris involving an MBH meeting our
selection criteria).  In our implementation, each of those binary systems are evolved completely
independently, even if parts of their evolution are occurring simultaneously\footnote{Recall that
in Illustris, after the initial `merger' event, only a single remnant MBH particle remains.}.
With this caveat, we can still consider, very simplistically, in how many systems the \textit{second}
binary overtakes the first as they harden.  Out of the binaries with subsequent events, $83\%$
($31\%$ of all binaries) are overtaken, $76\%$ ($28\%$ overall) of those before redshift zero,
and $42\%$ ($16\%$) with $z > 0.0$ and mass ratio $\mu > 0.2$.

The tendency for subsequent binaries to cross in our simulations likely reflects systems' ability
to increase noticeably in total mass over the course of the merger process.  We emphasize
that this is a very simplistic and preliminary investigation.  If, for example, MBH remnants tend
to receive significant `kicks' after merger, the resulting fractions could change significantly.
None the less, the apparent commonality of candidate multiples suggests that the role of triples
should be investigated more thoroughly.

% Hierarchical treatment of triples
It is unclear how such triple systems should be treated, even in a simple semi-analytic manner
\citep[see, however,][]{bonetti2016}.
The conventional wisdom of triple system dynamics is that the lowest mass object will be ejected,
while the more massive pair become bound in a binary \citep[e.g.][]{hills1975}.  Such `exchange'
interactions are motivated primarily from stochastic scattering events, like those which may occur
between stars in dense stellar environments.  In these cases, the system can be viewed as nearly
dissipationless, and their initial encounter is effectively stochastic.  It is our premise,
however, that the environments and dynamics of MBH multiples
are heavily dissipational.  For example, consider an initial pair of MBH which encounter at kpc
scales, on a hyperbolic orbit.  If the system quickly circularizes, and hardens to scales of
$1-100$ pc, then a third MBH which encounters the system---again at kpc scales---may similarly
settle into an outer, roughly-circular orbit forming a hierarchical system.  In such a situation,
secular instead of scattering dynamics, such as the Kozai-Lidov mechanism \citep{lidov1962,
kozai1962} or resonant migration may be more appropriate than traditional three-body scatterings.
In this case, the outer MBH in the triple system may accelerate the hardening of the inner binary,
driving it to coalescence \citep[e.g.][]{blaes2002}.  This may be less likely in gas-rich
environments which could effectively damp eccentric evolution, but here gas-driven inspiral will
likely cause rapid coalescence in any case.

% Scattering treatment of triples
MBH triples forming hierarchically with low to moderate eccentricities may evolve in a resonant
fashion.  If, on the other hand, environmental effects sufficiently enhance (or preserve initially
high) eccentricities of MBHB, then the resulting highly radial orbits may strongly intersect.  In
that case, a more stochastic-scattering-like regime may indeed still be appropriate. Numerous studies
have suggested that environmental effects can indeed enhance MBHB eccentricity
\citep[e.g.][]{quinlan1996, sesana2010, ravi2014}.  If the interaction between MBH triples (or
even higher-order multiples) is indeed most similar to scattering, then the simplest prescription
of removing the lowest-mass BH, with or without some additional hardening of the more massive pair,
may still be appropriate \citep[e.g.][]{hoffman2007}.  An ejected MBH which may later fall back to
the galactic center, while of great observational interest in and of itself, is likely less important
for GW emission \textit{per se}.  

% Triple observations
The observation of a
triple-AGN system could provide insight into the type of system they form (i.e.~hierarchical
vs.~scattering), and their lifetimes.  Additionally, MBH ejected by three-body
interactions could be observable as offset AGN, and possible confused with binary MBH, or ones
`recoiling' from previous coalescences \citep[e.g.][]{blecha2011}.  We have assumed that recoiling
systems do not significantly affect our populations, effectively assuming that kicks are
small---which is expected for spin-aligned MBHB.  This is motivated by studies which have shown that
gravitational torques from circumbinary disks, such as those we consider, can be effective at
aligning spins on timescales significantly shorter than a viscous time
\citep[e.g.][]{bogdanovic2007, dotti2010, miller2013}.

The MBH populations from the Illustris simulations are well suited for this problem, as they
accurately follow the histories and large-scale environments of MBHB systems and host galaxies.
As we are currently working on implementing eccentric evolution into our simulations, we plan to
explore multi-MBH systems in more detail.  This framework will also allow for the treatment of
kicked MBH resulting from random spin orientations, if for example the spins of a substantial
fraction of MBHB occur in gas-poor environments in which they may not be aligned.

% ------------------------------
% ------ Summary / Outro -------
% ------------------------------

\section{Conclusions and Summary}
\label{sec:res_summary}

% Summary of what we've done
For the first time, we have used the results of self-consistent, hydrodynamic cosmological
simulations, with a co-evolved population of Massive Black Holes (MBH) to calculate the plausible
stochastic Gravitational-Wave Background (GWB) soon to be detectable by Pulsar Timing Arrays (PTA).
We have also presented the first simultaneous, numerical treatment of all classes of MBH Binary
(MBHB) hardening mechanisms, discussing the effects of each: dynamical friction, stellar
(loss-cone) scattering, gas drag from a viscous circumbinary disk, and gravitational wave emission.

% versus with previous results
The most advanced previous studies have included only individual environmental effects, for example,
calculating dynamical friction (DF) timescales to determine which systems will contribute to the GWB
\citep{mcwilliams2014}, or attenuating the GWB spectrum due to loss-cone (LC) stellar-scattering
\citep{ravi2014}.  We explicitly integrate each of almost ten thousand MBH binaries, from galactic
scales to coalescence, using self-consistently derived, realistic galaxy environments and MBH
accretion rates.  We thoroughly explore a broad parameter space for each hardening mechanism to
determine the effects on the MBHB merger process, the lifetimes of systems, and the resulting GWB
spectrum they produce.

% Lifetime and Coalescing Fractions
The resulting lifetimes of MBHB that coalesce by redshift zero are usually gigayears, while that of
low total-mass and extreme mass-ratio systems typically extend well above a Hubble time.  In our
fiducial model, with a modest DF prescription (`Enh-Stellar') and moderately refilled LC
($\frefill = 0.6$), the median lifetime of MBHB with total masses $M > 10^8 \, \msol$ is
$17 \textrm{ Gyr}$, with $23\%$ coalescing before redshift zero.  Massive systems that also have
high mass ratios, $\mu > 0.2$, merge much more effectively, with a median lifetime of
$6.9 \textrm{ Gyr}$ and $45\%$ coalescing at $z>0$.  Increasing the effectiveness of the LC
drastically decreases system merger times.  For an always full LC ($\frefill = 1.0$), the lifetime
of massive systems decreases to $4.9 \textrm{ Gyr}$ and $0.35 \textrm{ Gyr}$ for systems with
$M > 10^8 \, \msol$ and all mass ratios, and those with $\mu > 0.2$ respectively.  The coalescing
fractions in these cases doubles to $54\%$ and $99\%$.  A summary of lifetimes and coalescing
fractions for different models is presented in \tabref{tab:results}.

% Dual-AGN, Persisting Fractions
The growing number of dual-MBH candidates \citep[e.g.][]{deane2014, comerford2015} presents the
opportunity to constrain binary lifetimes and coalescing fractions observationally.  For most of
our models, only about $1\%$ of MBHB with total masses $M > 10^8 \, \msol$ and mass $\mu > 0.2$
remain at separations $r > 10^2 \textrm{ pc}$ at redshift zero.  At smaller separations,
$r > 1 \textrm{ pc}$, the fractions are dependent on model parameters, but in general between
$1$ -- $40\%$.  Tabulated persisting fractions are included in \tabref{tab:results} for a variety
of models and situations.  Observational constraints on these fractions can narrow down the
relevant parameter space of hardening physics.  Accurate predictions for dual-MBH observations
must fold in AGN activity fractions and duty cycles, and their correlations with binary merger
lifetimes.  A comprehensive study of dual-AGN observability predicted by our models, over
redshift and different observational parameters, is currently underway. 

% Redshift distribution
In addition to measuring the fraction of MBH in associations (e.g.~dual AGN) as a function of
separation, the redshift distribution of dual-MBH can be useful in understanding their evolution.
The Illustris simulations, for example, give a median MBHB formation redshift\footnote{Recall that
MBHB `formation', in this context, corresponds to two MBH coming within a few kpc of eachother.} of
$z \approx \nobreak 1.25$.  Depending on the parameters of the hardening models, the median coalescing
redshift can be anywhere between $z\approx 0.4$ -- $1.0$, with $z\approx0.6$ suggested by our
fiducial model.

% Lowest predicted GWB still close to limits
Without electromagnetic observations, GWB detections and upper limits can also be
used to inform our understanding of MBH evolution \citep[e.g.][]{sampson2015}.
Even if the fraction of systems which coalesce
is quite low, the most massive and high mass-ratio systems, which produce the strongest GW, are
difficult to \textit{keep} from merging.  In a simulation with the weakest hardening rates
(`Enh-Stellar'; $\frefill = 0.0$) only $\sim12\%$ of all binaries coalesce by redshift zero,
but the GWB amplitude at $f = 1\, \pyr$ is still $0.2\E{-15}$---only about a factor of five
below the most recent upper limits\footnote{Previous studies have shown that including eccentric
evolution can significantly decrease the GWB amplitude \citep{ravi2014}, so we caution that the
weakest GWB observed in our simulations, which \textit{do not} include eccentric evolution,
may not be a robust lower limit.  We are currently exploring the effects of eccentricity, and
altered MBH-Host scaling relations on the minimum plausible GWB---to be presented in a future study.}.

% Fiducial predictions at 1/yr
In our fiducial model, we use a moderate LC refilling rate ($\frefill=0.6$) which increases 
the number of MBHB contributing to the GWB at $1\,\pyr$, producing an amplitude of
$\ayr \approx \nobreak 0.4\E{-15}$.  Increasing the effectiveness of DF and/or LC scattering tends
to increase the amplitude further.  Our fiducial model also includes fairly strong viscous drag (VD)
from circumbinary disks, which decreases the time MBHB spend emitting in each frequency band,
and thus attenuating the GWB.  This effect tends to be more subtle, producing GWB attenuation of
about $15\%$.  In general, for a fairly broad
range of parameters, our simulations yield GWB amplitudes between \mbox{$\sim 0.3 - 0.6\E{-15}$}.
A GWB amplitude of $\ayr \approx \nobreak 0.4\E{-15}$ is less than a factor of three below current
detection limits---a parameter space which will likely be probed by PTA within the next decade.

% Most optimistic predictions still below upper-limits
The most stringent PTA upper limits of $\ayr \lesssim 10^{-15}$ \citep{shannon2015}
have already excluded a broad swath of previous predictions.  Many of those models assume that
binary hardening is very effective, with all MBHB quickly reaching the PTA band and emitting
an unattenuated signal---i.e.~evolving purely due to GW-emission, without additional environmental
hardening effects.  Following the same procedure, to calculate an upper-limit to the GWB based on
our population of MBHB, we find a GWB amplitude of $\ayr \approx 0.7\E{-15}$---slightly below
the PTA limit.  The Illustris simulation volume is very large for a hydrodynamic simulation,
but it lacks the very-rare, most massive MBH in the universe ($\gs 10^{10} \, \msol$) which could
slightly increase our predicted GWB amplitude---although, likely a correction on the order of
$\sim 10\%$\footnote{The effects of simulation volume on the predicted GWB amplitude should be
studied more carefully to confirm this estimate.} (Sesana, private communication).  None the less,
our upper limit suggests that the current lack of PTA detections shouldn't be interpreted as a
missing signal.

% Comparison and issues with previous models
Our upper-limit value of $\ayr \approx 0.7\E{-15}$ falls just within the lower end of some recent
studies \citep[e.g.][]{ravi2014, roebber2016}, but is generally lower than much of the previous literature
\citep[see e.g.~\tabref{tab:pta-predic}, and Fig.~2 of][]{shannon2015}.  Likely, this is at-least partly
because the MBH merger rates derived from Illustris are based directly
on simulated galaxy-galaxy merger rates.  The bulk of existing calculations have either used
inferences from (dark matter only) halo-halo mergers which may have systematic issues
\citep[see, e.g.,][]{rodriguez-gomez2015}, or observations of galaxy merger rates which have
uncertain timescales.  This upper-limit is based on optimistic, GW-only evolution.  In our fiducial
model, the signal is lower by $\sim 50\%$ due primarily to the moderately refilled LC, and mildly
due to VD attenuation.

% Spectral shape, low frequencies
Variations in the rate at which the stellar LC is refilled has the strongest effect on the shape
and amplitude of the GWB spectrum in our simulations, especially at low frequencies.  PTA
observations are moving towards these frequencies, as the duration of their timing baselines increase.
Unlike at higher frequencies where scattering increases the number of MBHB contributing to the GWB,
at $10^{-1} \, \pyr$, for example, effective LC refilling leads to attenuation of the GWB from
accelerated binary hardening.  Here, our spectra tend to lie at amplitudes between
$1.5$ -- $2.5\E{-15}$, with spectral indices between about $-0.4$ and $-0.6$---a significant
deviation from the canonical $-2/3$ power-law.  At frequencies lower still ($f \ls 10^{-2} \, \pyr$),
the effective LC scattering produces a strong turnover in the GWB spectrum.
A summary of GWB amplitudes and spectral indices is presented in \tabref{tab:results}
for both $f = 1 \, \pyr$ and $10^{-1} \, \pyr$, and numerous hardening models.

% Nature of sources contributing to GWB
In our fiducial simulation, we find that the median contribution to the GWB comes from binaries
at a redshift of $z \approx \nobreak 0.3$, with total masses $M \approx \nobreak 10^9 \, \msol$,
and mass ratios $\mu \approx 0.3$.  The co-evolved population of MBH and galaxies in Illustris
allows us to also examine typical host-galaxy properties for the first time.  Galaxies
containing MBHB contributing strongly to the GWB are noticeably larger and more massive galaxies.
The median stellar mass of galaxies, weighted by GWB contribution, is about $3\E{11} \msol$---more
than an order of magnitude larger than the median stellar mass for all galaxies.

% Importance of Triples
Based on the merger trees and binary lifetimes produced from our simulations, we have also shown
that the presence of higher-order MBH multiples could be an non-negligible aspect of MBH evolution.
The simplest examination suggests that triples could be important in about $30\%$ of MBHB in our
simulations.  In future work, we will explore these triple systems in more detail, as well as the
effects of nonzero eccentricity and post-merger recoils.  We also hope to implement more
self-consistent LC refilling, and more comprehensive tracking of the changing galactic environment.

% Summary
In summary,
\begin{itemize}
%    Lifetimes
\item \textbf{MBH binary lifetimes tend to be multiple $\textrm{Gyr}$, even for massive systems}.
While massive and high mass ratio systems are likely rare at very large separations, observations
of dual MBH at $r > 1 \textrm{ pc}$ can be used to constrain the merger physics.
%    GWB Amplitude
\item \textbf{The GWB amplitude predicted by our models is $\mathbf{A_{1\,\pyr} \approx \nobreak
0.4\E{-15}}$, with a range of about $\mathbf{0.3}$ -- $\mathbf{0.6\E{-15}}$ for different hardening
parameters.}  At lower frequencies, we find $A_{0.1\,\pyr} \approx 1.5$ -- $2.5\E{-15}$, with
spectral indices between $-0.4$ \& $-0.6$---a noticeable deviation from the canonical $-2/3$
power-law.
%    versus PTA
\item \textbf{We find that the lack of PTA detections so far is entirely consistent with our MBH
population}, and does not require environmental effects.  At the same time, our most conservative
models yield a GWB amplitude of $\ayr = 0.2\E{-15}$.  While incorporating non-zero
eccentricities may further suppress our GWB predictions, our simulations suggest that if PTA limits
improve by a factor of 3--4 and no detection is made, our understanding of galaxy and MBH evolution
may require revision.
%    source properties
\item \textbf{The median redshift and total mass of MBHB sources contributing to the GWB are
$\mathbf{z\approx 0.3}$ and a few $\mathbf{10^9\,\msol}$}, while the median coalescence time of all
systems tends towards $z \approx 0.6$.  Observations of the redshift distribution and host galaxy
properties of dual-MBH can be informative for our understanding of binary evolution.
%    Triples
\item \textbf{Our simulations suggest that up to $\mathbf{30\%}$ of binaries could involve the
presence of a third MBH.}  The role of MBH triples is currently unclear, but should be explored and
included in future simulations.
\end{itemize}

% Outro
The environments around MBHB form a complex and interwoven parameter space with additive and often
degenerate effects on the GWB. Better constraints on MBH--host correlations, combined with
increasingly strict upper limits on the GWB amplitude, will soon tightly constrain the efficiency
with which MBHB are able to coalesce.  That efficiency then determines the fraction of galaxies
which should have observable dual or binary AGN, providing an additional test of our most
fundamental assumptions of MBH/galaxy growth and co-evolution.  We believe that our results, and
similar analyses, can be used to leverage GWB observations along with dual and offset AGN to
comprehensively understand the MBH population and their evolution.  These exotic binaries involve
a plethora of dynamical processes which are still poorly understood, but affect our most
fundamental assumptions of black holes in the universe and thus the evolution of galaxies over
cosmological time. Right now, we are entering the era of GW astronomy, and with it, a direct view
of BH physics and evolution on all scales.

% Create a bibliography here, only if just this file is being compiled/built.
\biblio{}

%sec:results
%sec:res_gwb

% Acknowledgments
% ---------------
\section*{Acknowledgments}
We are grateful to Julie Comerford, Jenny Greene, Daniel Holz, Maura McLaughlin, and Vikram Ravi for fruitful
suggestions and discussions.  Advice from Alberto Sesana has been extremely helpful throughout, and especially
in beginning this project.  We are also thankful to the facilitators, organizers, and attendees of the NANOGrav,
spring 2016 meeting at CalTech, especially Justin Ellis, Chiara Mingarelli, and Stephen Taylor.  We also thank
the anonymous referee for numerous, very constructive comments about the manuscript.

This research made use of \astropy, a community-developed core Python package for Astronomy \citep{astropy2013},
in addition to \scipy~\citep{scipy}, \ipython~\citep{ipython}, \numpy~\citep{numpy2011}.  All figures were
generated using \matplotlib~\citep{matplotlib2007}.

% Bibliography
% ------------

\let\oldUrl\url
\renewcommand{\url}[1]{\href{#1}{Link}}

\bibliographystyle{mnras}
\bibliography{refs}

% Appendices
% ----------

\appendix

\section{Additional Figures of Gravitational Wave Scalings and Alternative Model Results}
\label{sec:app-figs}

	% Fig A1 ------ GW Freq vs. Separation
	\begin{figure}
	\centering
	\includegraphics[width=1.0\columnwidth]{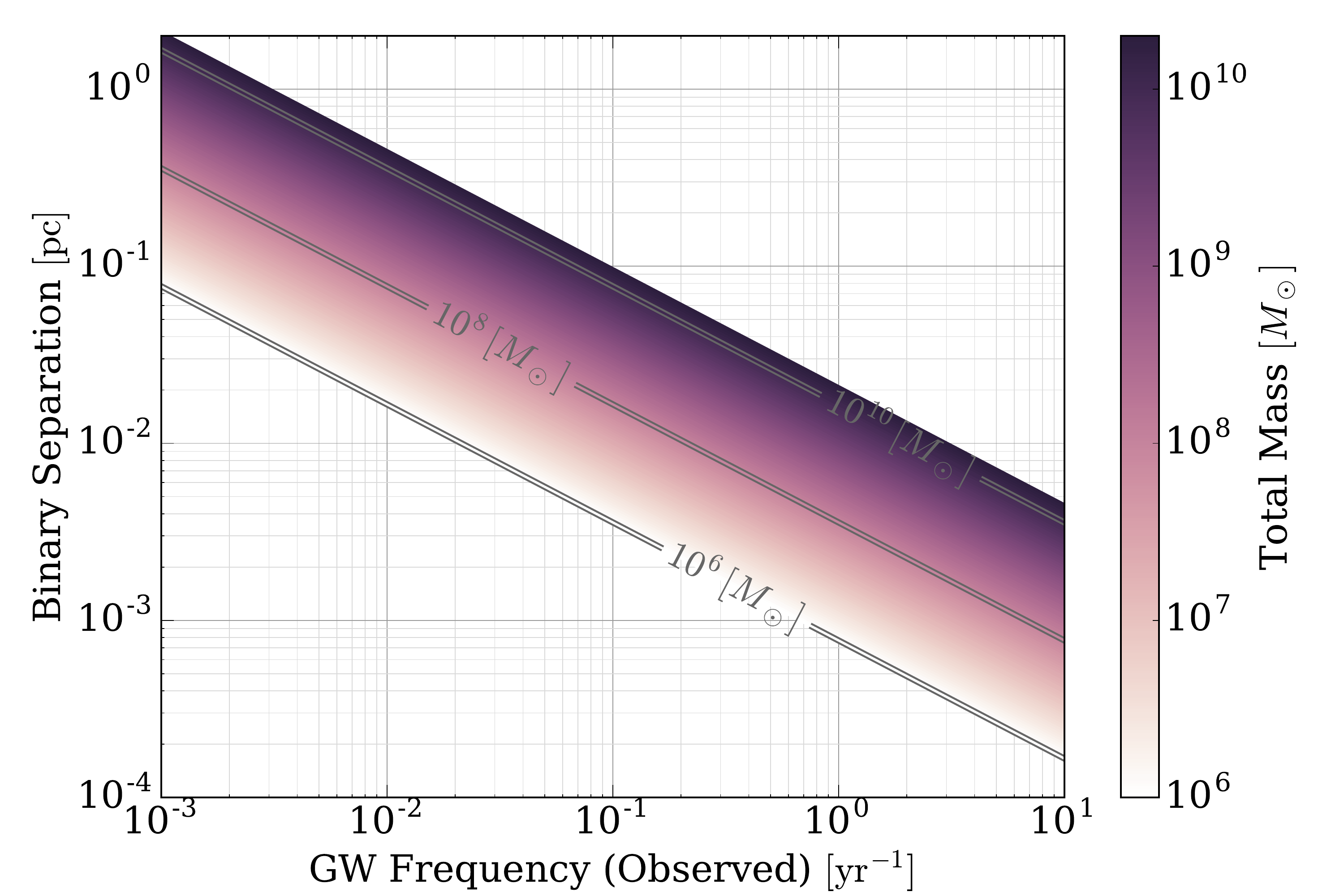}
	\caption{Binary separation versus gravitational wave frequency for the total MBHB 
	masses used in our simulations.  GW frequency is twice the orbital frequency, and
	is redshifted traveling from the source to observer; the values plotted are at $z=0$.
	For the PTA band, roughly $0.1$--$1 \, \pyr$,
	binaries contribute anywhere between about $10^{-3}$ and $10^{-1} \, \textrm{pc}$.
	Most of the GWB amplitude come from higher mass systems, towards larger separations.}
	\label{fig:gw_sep_freq}
	\end{figure}

	% Fig A2 ------ Analytic GW Scales
	\begin{figure*}
	\centering
	\includegraphics[width=1.4\columnwidth]{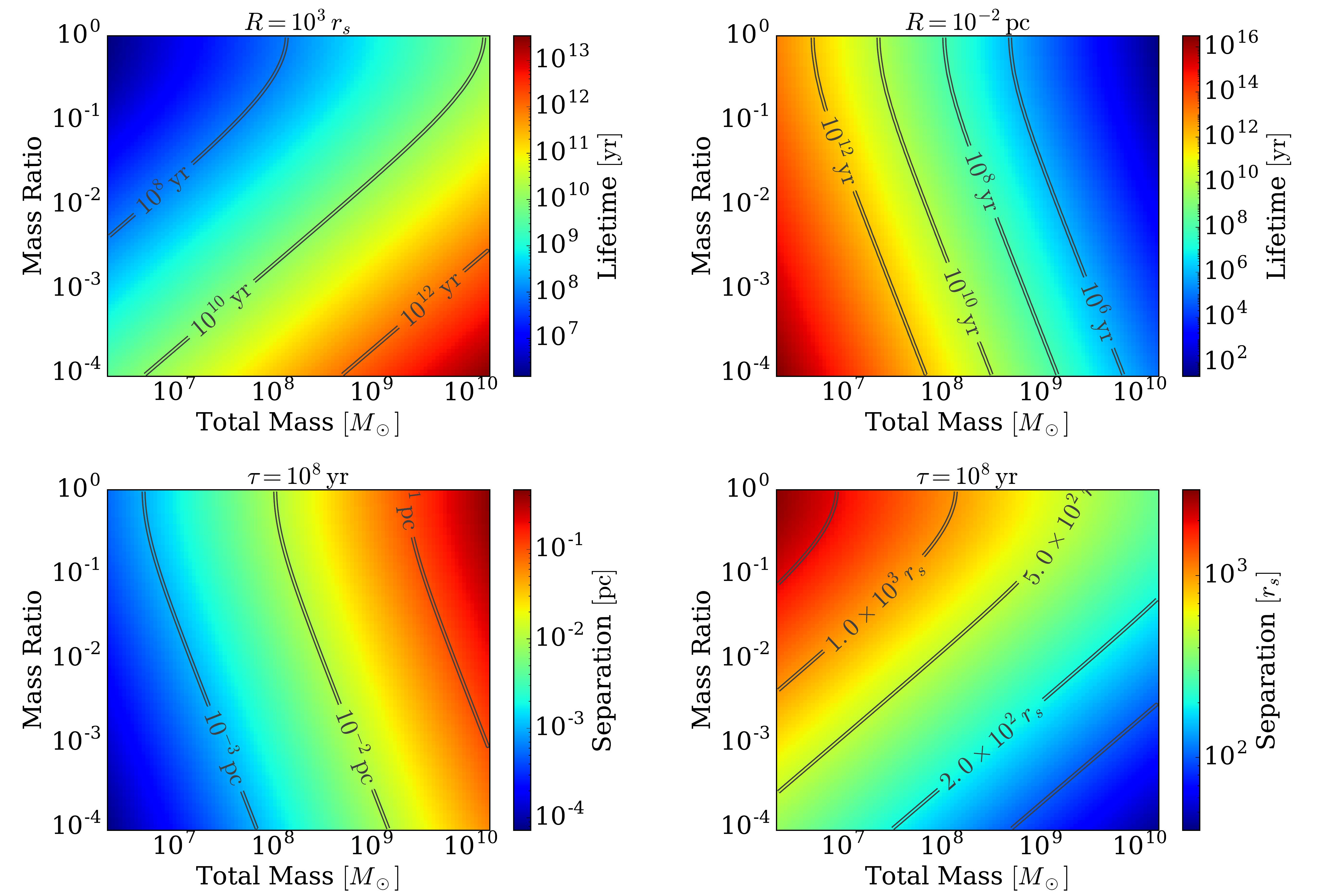}
	\caption{Characteristic lifetimes and separations for coalescence due to purely GW emission
	over the relevant parameter space of mass ratio and total mass.  The upper panels are colored by time
	to coalescence for fixed separations: $R = 10^3 \, \rs$ (left), and $R = 10^{-2} \textrm{ pc}$
	(right).  The lower panels are colored by separation to coalesce within $10^{8} \textrm{ yr}$,
	in units of parsecs (left) and Schwarzschild radii ($\rs$, right).}
	\label{fig:gw_char}
	\end{figure*}

	% Fig A3 ------ Hardening Bands vs separation by DF and LC
	\begin{figure*}
	\centering
	\includegraphics[width=\textwidth]{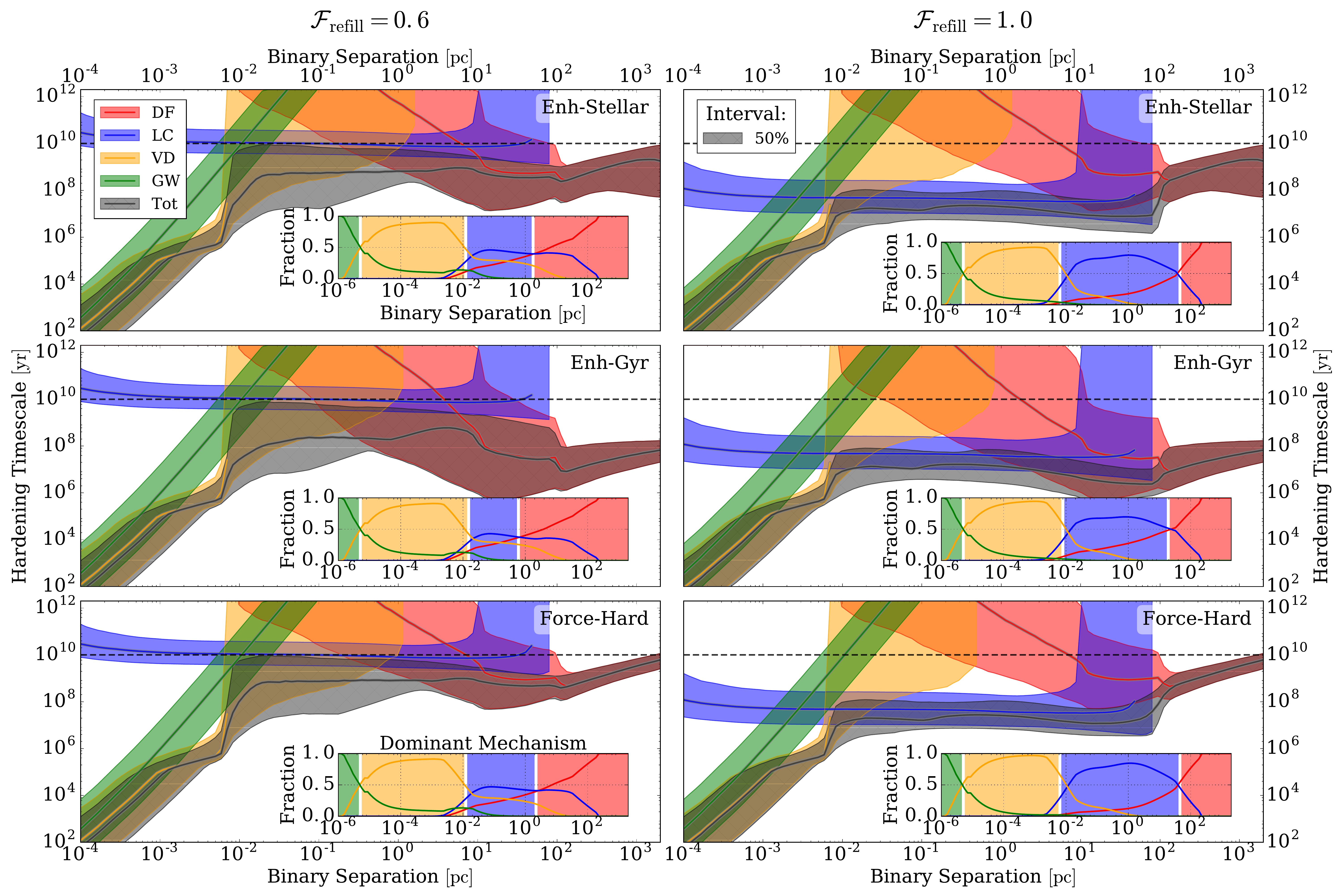}
	\caption{Binary hardening timescales versus separation for different
	DF models (rows) and LC refilling fractions (columns).  Colored lines and bands show the
	median and 50\% intervals for each hardening mechanism: Dynamical Friction (DF),
	Loss-Cone (LC) scattering, Viscous Drag (VD), and Gravitational Wave (GW) emission
	with the total hardening rate shown by the grey, hatched region.  The inset panels shows
	the fraction of binaries dominated by each mechanism, also as a function of separation.
	\textit{\textbf{Note:} in the `Force-Hard' model, the binary separation is artificially altered
	faster than the timescale shown (between about $\gs 10^2 \textrm{ pc}$).  While the
	DF hardening timescale is still relatively long, the binaries are forcibly hardened at a
	faster rate.}}
	\label{fig:hard_set1}
	\end{figure*}

	% Fig A4 ------ Lifetimes and Coalescing Fractions for LC=1.0
    \begin{figure*}
    \centering
        \subfloat{\includegraphics[width=\columnwidth]{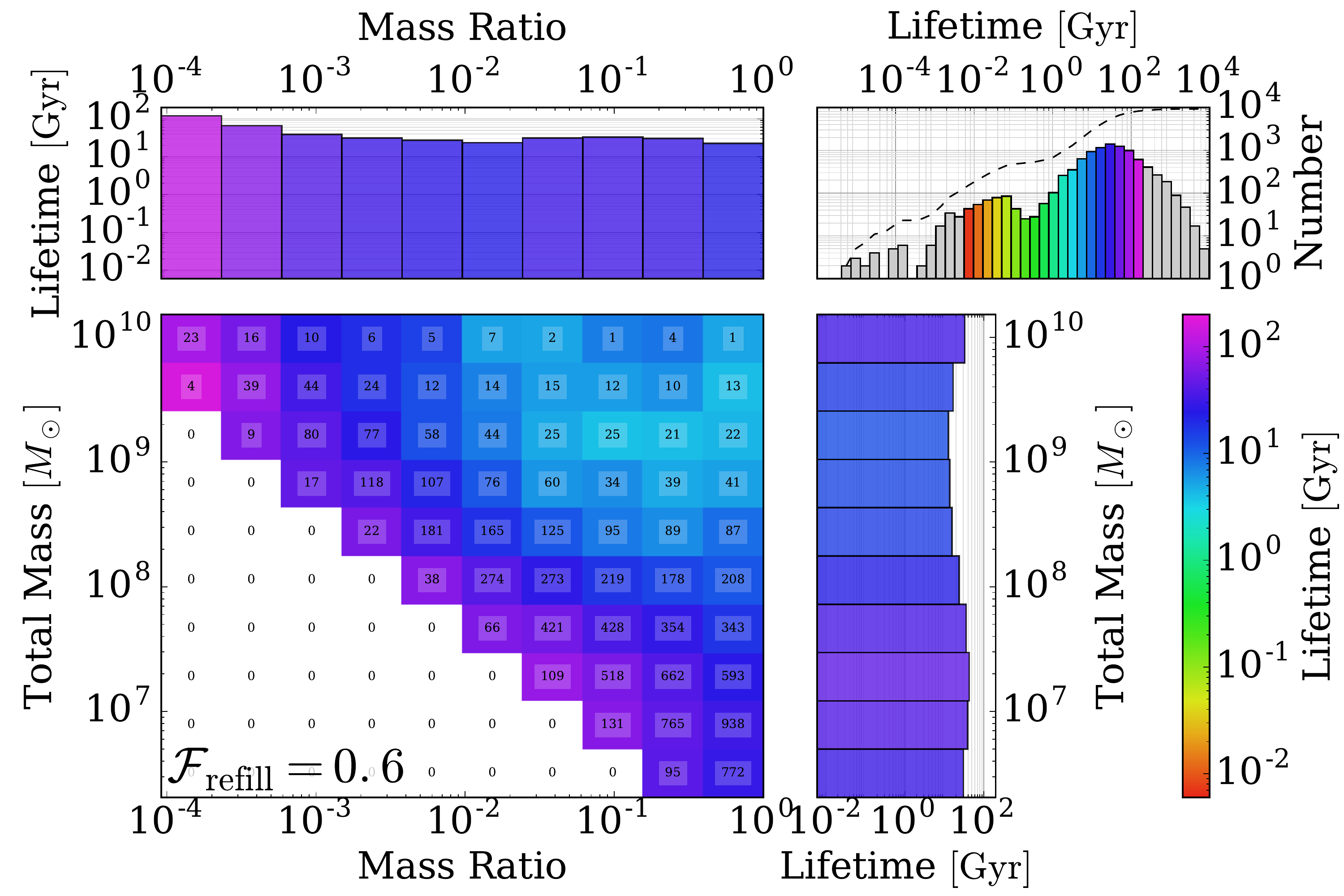}}
        \subfloat{\includegraphics[width=\columnwidth]{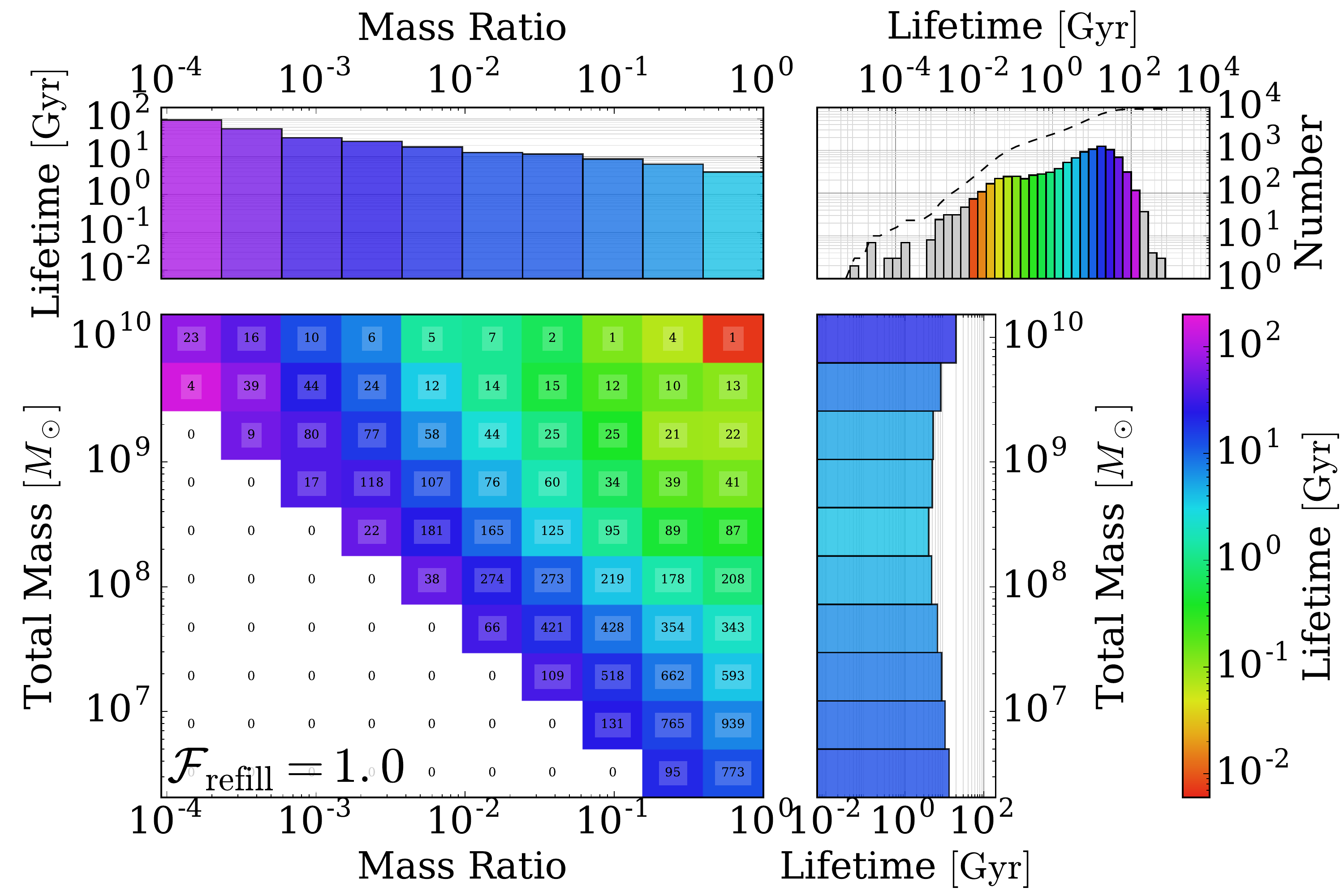}} \\

        \subfloat{\includegraphics[width=\columnwidth]{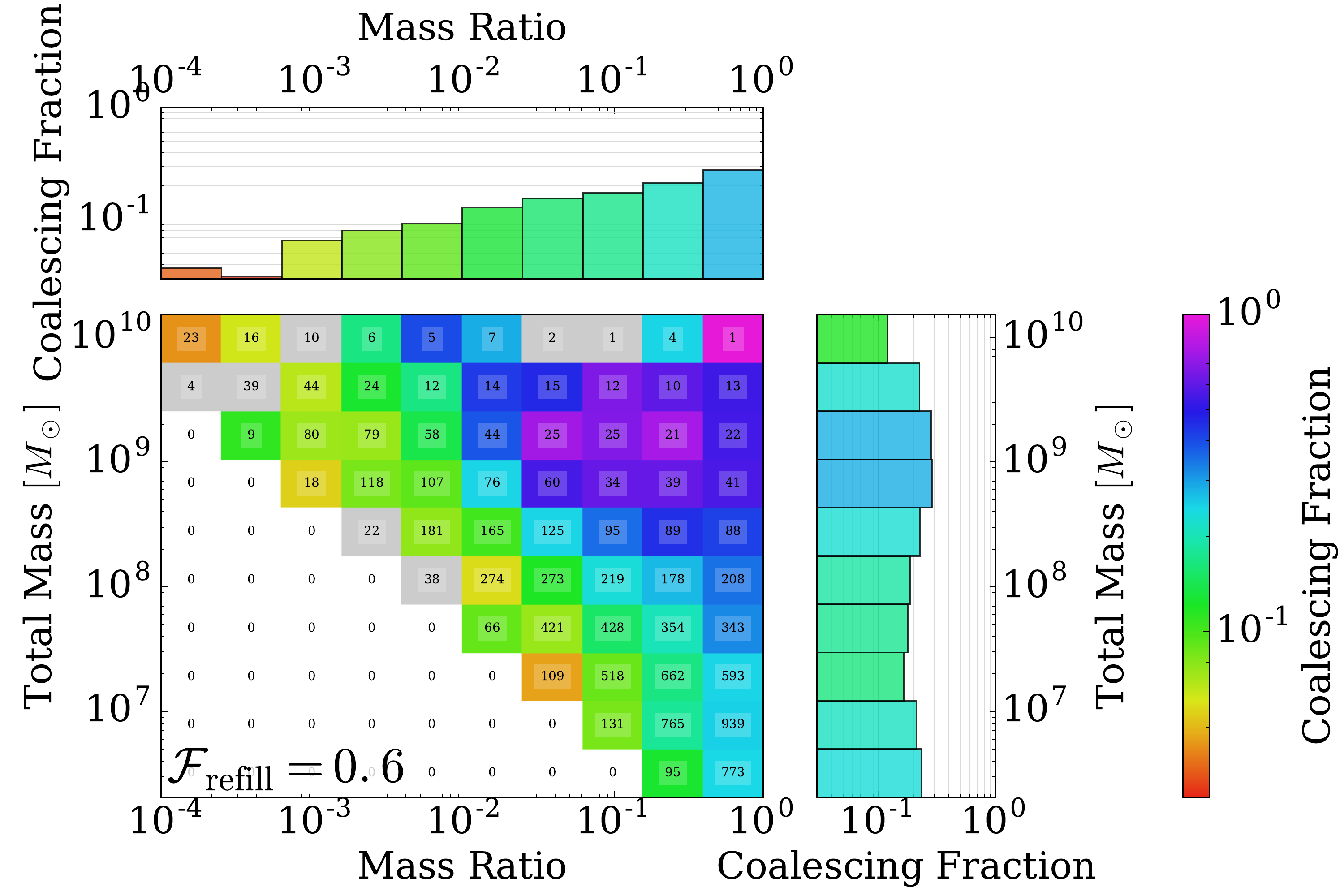}}
        \subfloat{\includegraphics[width=\columnwidth]{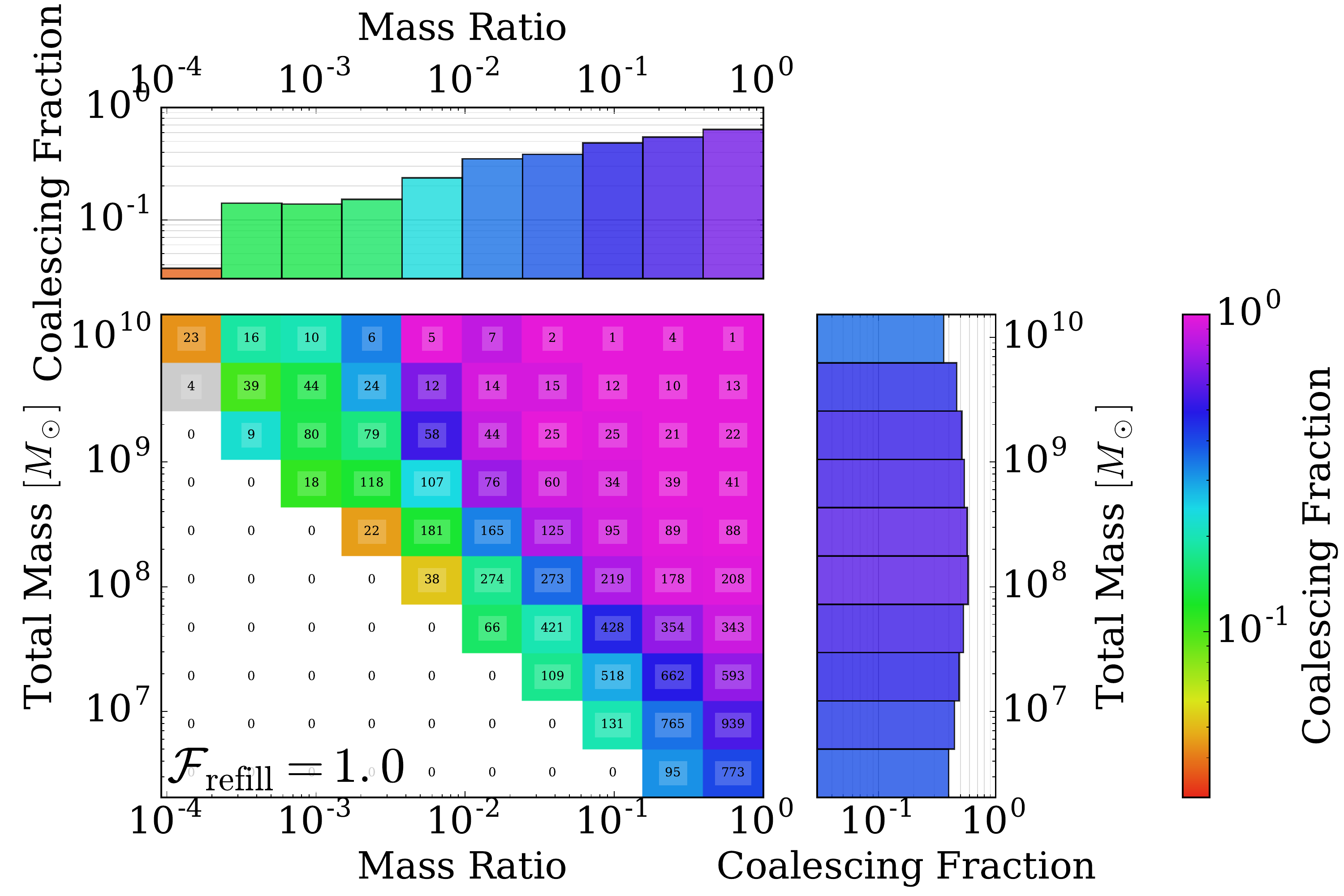}}
    \caption{Binary lifetimes (upper row) and coalescing fractions (lower row) for our fiducial,
    moderately refilled loss cone ($\frefill = 0.6$; left) and an always full one
    ($\frefill = 1.0$; right).   Both simulations use the `Enh-Stellar' DF model.
    The overall distribution of MBHB lifetimes are shown in the upper-right-most panel for each
    simulation,  with the cumulative distribution plotted as the dashed line.
    For $\frefill=0.6$, most binaries need to be both high total mass ($M \gs 10^8 \msol$), and
    moderately high mass ratio ($\mu \gs 10^{-2}$) to have lifetimes short enough to coalesce
    by redshift zero.  In the $\frefill = 1.0$ case, on the other hand, either criteria is
    sufficient---and the coalescing fraction in that parameter space approach unity.}
    \label{fig:life_lc10}
    \end{figure*}

	% Fig A5 ------ GWB Amplitude vs. Coalescing fraction at 10^{-2} 1/yr.
	\begin{figure}
	\centering
	\includegraphics[width=1.0\columnwidth]{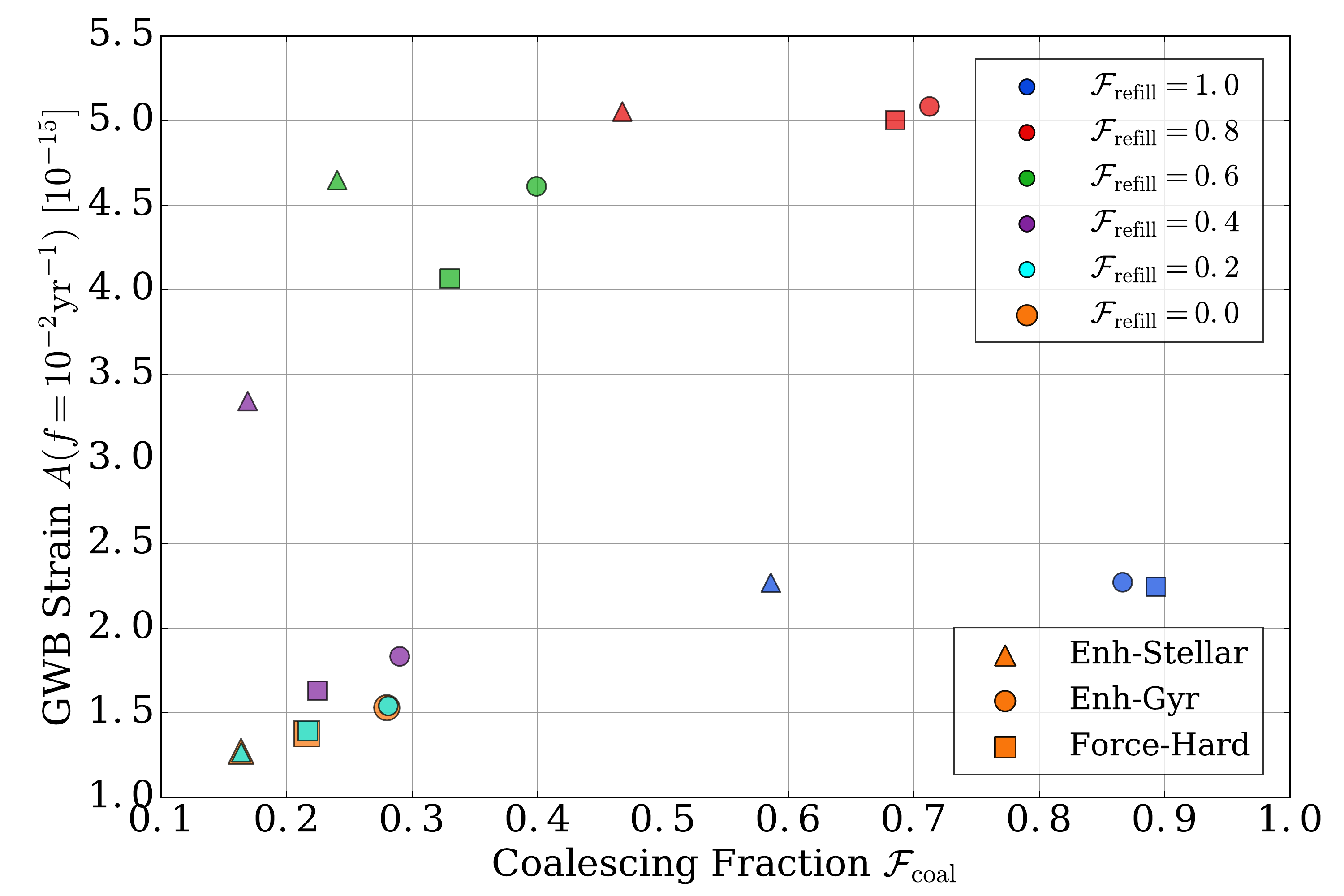}
	\caption{Dependence of GWB strain amplitude and binary coalescing fraction on LC refilling
	parameter and DF model.  The GW strain is measures at very low frequencies,
	$f = 10^{-2} \, \pyr$, and the coalescing fraction is defined using the population of
	high mass-ratio $\mu > 0.1$ systems.  Each symbol represents a different DF model,
	and each color a different LC refilling parameter.  $\ayr$ tends to increase monotonically
	with $\fcoal$, until the highest $\frefill$.  At $\frefill \approx 1.0$, LC stellar scattering
	is effective enough at these low frequencies to significantly attenuate the GWB amplitude.}
	\label{fig:gwb-coal-frac_f-2}
	\end{figure}

% \clearpage
\newpage
\section{Summary Of Quantitative Results for a Variety of Parameter Configurations}

    \begin{table*}\centering
    \renewcommand{\arraystretch}{1.2}
    \setlength{\tabcolsep}{4pt}
    \begin{tabularx}{\textwidth}{@{}ccc ccc lc cccc@{}}
    \toprule
								&					& \mc{4}{GWB}													&							&					&				&		&	&		\\
    Dynamical Friction			& \mr{2}{Loss Cone} & \mc{2}{Amplitude $[10^{-16}]$}& \mc{2}{Spectral Index}		& \mr{2}{Subset}& Lifetimes $[\textrm{Gyr}]$	& \mc{2}{\mr{2}{Coalescing Fraction}} & \mc{2}{Persisting Fraction} \\ 
	(Viscous Disk)				&					& $1\,\pyr$		& $0.1\,\pyr$	& $1\,\pyr$		& $0.1\,\pyr$	&							& Full (Coal)		& \mc{2}{}					& $> 1 \textrm{ pc}$ & $> 10^2 \textrm{ pc}$\\\midrule
	\mr{8}{\scell{Enh-Stellar\\(VD: None)}}
								& \mr{4}{\flcsix}	& \mr{4}{4.3}	& \mr{4}{18}	& \mr{4}{-0.64}	& \mr{4}{-0.55}	& All ($M>10^6 \, \msol$)	& 37  (1.3)			& $ 1641/ 9270$ & $(0.18)$	& $0.67$ & $0.46$ \\
								& 					& 				& 				& 				& 				& $\mu>0.2$					& 32  (1.1)			& $ 1089/ 4759$ & $(0.23)$	& $0.57$ & $0.33$ \\
								& 					& 				& 				& 				& 				& $M>10^8\,\msol$			& 26  (1.7)			& $  504/ 2610$ & $(0.19)$	& $0.68$ & $0.45$ \\
								& 					& 				& 				& 				& 				& $M>10^8\,\msol$,$\mu>0.2$	& 11  (0.89)		& $  184/  478$ & $(0.38)$	& $0.37$ & $0.01$ \\
								\cline{7-12}

								& \mr{4}{\flcten}	& \mr{4}{6.6}	& \mr{4}{25}	& \mr{4}{-0.65}	& \mr{4}{-0.39}	& All						& 8.0 (1.3)			& $ 4556/ 9270$ & $(0.49)$	& $0.47$ & $0.46$ \\
								& 					& 				& 				& 				& 				& $\mu>0.2$					& 5.1 (1.1)			& $ 2840/ 4759$ & $(0.60)$	& $0.35$ & $0.33$ \\
								& 					& 				& 				& 				& 				& $M>10^8\,\msol$			& 5.2 (0.80)		& $ 1414/ 2610$ & $(0.54)$	& $0.45$ & $0.45$ \\
								& 					& 				& 				& 				& 				& $M>10^8\,\msol$,$\mu>0.2$	& 0.37 (0.35)		& $  469/  478$ & $(0.98)$	& $0.01$ & $0.01$ \\
								\cline{2-12}

	\mr{8}{\scell{\textbf{Enh-Stellar}\\\textbf{(VD: Fiducial)}}}	
								& \mr{4}{\flcsix}	& \mr{4}{\textbf{3.7}}	& \mr{4}{\textbf{15}}	& \mr{4}{\textbf{-0.63}}	& \mr{4}{\textbf{-0.59}}	& All				& \textbf{29  (2.7)}			& $ \mathbf{1875/ 9270}$ & $\mathbf{(0.20)}$	& $\mathbf{0.66}$ & $\mathbf{0.46}$ \\
								& 					& 				& 				& 				& 				& $\mu>0.2$					& \textbf{26  (2.2)}			& $ \mathbf{1225/ 4759}$ & $\mathbf{(0.26)}$	& $\mathbf{0.55}$ & $\mathbf{0.33}$ \\
								& 					& 				& 				& 				& 				& $M>10^8\,\msol$			& \textbf{17  (4.0)}			& $ \mathbf{608 / 2610}$ & $\mathbf{(0.23)}$	& $\mathbf{0.64}$ & $\mathbf{0.45}$ \\
								& 					& 				& 				& 				& 				& $M>10^8\,\msol$,$\mu>0.2$	& \textbf{6.9 (3.3)}			& $ \mathbf{214 /  478}$ & $\mathbf{(0.45)}$	& $\mathbf{0.28}$ & $\mathbf{0.01}$ \\
								\cline{7-12}

								& \mr{4}{\flcten}	& \mr{4}{4.7}	& \mr{4}{17}	& \mr{4}{-0.61}	& \mr{4}{-0.38}	& All 						& 7.7 (1.2)			& $ 4634/ 9270$ & $(0.50)$	& $0.47$ & $0.46$ \\
								& 					& 				& 				& 				& 				& $\mu>0.2$					& 4.8 (1.0)			& $ 2900/ 4759$ & $(0.61)$	& $0.35$ & $0.33$ \\
								& 					& 				& 				& 				& 				& $M>10^8\,\msol$			& 4.9 (0.79)		& $ 1422/ 2610$ & $(0.54)$	& $0.45$ & $0.45$ \\
								& 					& 				& 				& 				& 				& $M>10^8\,\msol$,$\mu>0.2$	& 0.35 (0.35)		& $  472/  478$ & $(0.99)$	& $0.01$ & $0.01$ \\
								\cline{2-12}

	\mr{8}{\scell{Enh-Gyr\\(VD: Fiducial)}}
								& \mr{4}{\flcsix}	& \mr{4}{3.8}	& \mr{4}{16}	& \mr{4}{-0.63}	& \mr{4}{-0.62}	& All						& 13  (1.1)			& $ 3536/ 9270$ & $(0.38)$	& $0.38$ & $0.11$ \\
								& 					& 				& 				& 				& 				& $\mu>0.2$					& 13  (0.33)		& $ 1970/ 4759$ & $(0.41)$	& $0.30$ & $0.06$ \\
								& 					& 				& 				& 				& 				& $M>10^8\,\msol$			& 8.3 (3.5)			& $ 1091/ 2610$ & $(0.42)$	& $0.39$ & $0.15$ \\
								& 					& 				& 				& 				& 				& $M>10^8\,\msol$,$\mu>0.2$	& 6.4 (3.0)			& $  229/  478$ & $(0.48)$	& $0.26$ & $0.01$ \\
								\cline{7-12}

								& \mr{4}{\flcten}	& \mr{4}{4.7}	& \mr{4}{17}	& \mr{4}{-0.61}	& \mr{4}{-0.38}	& All						& 0.42 (0.30)		& $ 7783/ 9270$ & $(0.84)$	& $0.12$ & $0.11$ \\
								& 					& 				& 				& 				& 				& $\mu>0.2$					& 0.37 (0.28)		& $ 4122/ 4759$ & $(0.87)$	& $0.07$ & $0.06$ \\
								& 					& 				& 				& 				& 				& $M>10^8\,\msol$			& 0.32 (0.25)		& $ 2200/ 2610$ & $(0.84)$	& $0.15$ & $0.15$ \\
								& 					& 				& 				& 				& 				& $M>10^8\,\msol$,$\mu>0.2$	& 0.20 (0.20)		& $  474/  478$ & $(0.99)$	& $0.01$ & $0.01$ \\
								\cline{2-12}

	\mr{8}{\scell{Force-Hard\\(VD: Fiducial)}}
								& \mr{4}{\flcsix}	& \mr{4}{3.6}	& \mr{4}{15}	& \mr{4}{-0.64}	& \mr{4}{-0.61}	& All						& 14  (3.0)			& $ 3089/ 9270$ & $(0.33)$	& $0.32$ & $0.02$ \\
								& 					& 				& 				& 				& 				& $\mu>0.2$					& 17  (1.9)			& $ 1627/ 4759$ & $(0.34)$	& $0.24$ & $0.01$ \\
								& 					& 				& 				& 				& 				& $M>10^8\,\msol$			& 7.8 (4.4)			& $ 1093/ 2610$ & $(0.42)$	& $0.30$ & $0.04$ \\
								& 					& 				& 				& 				& 				& $M>10^8\,\msol$,$\mu>0.2$	& 7.0 (3.1)			& $  212/  478$ & $(0.44)$	& $0.23$ & $0.00$ \\
								\cline{7-12}

								& \mr{4}{\flcten}	& \mr{4}{4.7}	& \mr{4}{17}	& \mr{4}{-0.61}	& \mr{4}{-0.38}	& All						& 0.42 (0.30)		& $ 7783/ 9270$ & $(0.84)$	& $0.03$ & $0.02$ \\
								& 					& 				& 				& 				& 				& $\mu>0.2$					& 0.37 (0.28)		& $ 4122/ 4759$ & $(0.87)$	& $0.02$ & $0.01$ \\
								& 					& 				& 				& 				& 				& $M>10^8\,\msol$			& 0.32 (0.25)		& $ 2200/ 2610$ & $(0.84)$	& $0.05$ & $0.04$ \\
								& 					& 				& 				& 				& 				& $M>10^8\,\msol$,$\mu>0.2$	& 0.20 (0.20)		& $  474/  478$ & $(0.99)$	& $0.01$ & $0.00$ \\ 
    \bottomrule
    \end{tabularx}
    \caption{Summary of quantitative results for the gravitational wave background (GWB), and MBH binary lifetimes \& coalescing/persisting fractions.
    Results are shown for the three dynamical friction (DF) models described in the text: `Enh-Stellar', `Enh-Gyr' \& `Force-Hard'.  For our fiducial
    DF model (`Enh-Stellar'; the most conservative case), results are shown for both a case with no circumbinary, viscous disks (`VD: None'),
    and with a disk using our standard parameters (`VD: Fiducial').  Two different loss-cone (LC) scattering states are also compared, in which the
    LC is always full ($\frefill = 1.0$) and our fiducial case of moderately refilled ($\frefill = 0.6$).  Amplitudes and spectral indices are
    presented at both $f = 1\, \pyr$ and $f = 0.1 \, \pyr$.  For lifetime and fractional statistics, binaries are split into four subsets: `All'
    binaries included in our analysis, and those with mass ratios $\mu > 0.2$, total masses $M > 10^8 \, \msol$, and both criteria ($\mu > 0.2$,
    $M > 10^8 \, \msol$).  The lifetimes shown are median values of systems in each the `Full' subset, and only those which coalesce by redshift zero
    (`Coal')---the number (and fraction) of such systems are given in the `Coalescing Fraction' column.  Finally, the fraction of systems which
    remain uncoalesced at separations $r > 1 \textrm{ pc}$ and $r > 10^2 \textrm{ pc}$ are shown in the `Persisting Fraction' column.  Results for
    a simulation using models with all fiducial parameters are shown in bold.}
    \label{tab:results}
    \end{table*}

% -----------------------------
% ------- The Loss Cone -------
% -----------------------------

%\clearpage
\section{Stellar, Loss-Cone (LC) Scattering Calculations}
\label{sec:app-lc-calc}

The rate at which stars can refill the loss cone is governed by the `relaxation time'
($\trel$).  Following \citet{bt87}, consider a system of $N$ masses $m$, with number
density $n$, and characteristic velocities $v$.  The relaxation time can be written as,
	\begin{equation}
	\label{eq:relax}
	\trel 	\approx \frac{N}{8 \ln \Lambda} \tcross 
			\approx \frac{v^3}{8 \pi G^2 m^2 n \ln \Lambda},
	\end{equation}
where $\ln \Lambda$ is again the Coulomb Logarithm, and $\tcross \equiv r/v$ is the
crossing-time.  $\trel$ represents the characteristic time required to randomize a
particle's velocity via scatterings, i.e.,~\refeq{eq:relax} can be used to define the
diffusion coefficient  $\diffco$ as, $\trel \approx v^2 / \mathcal{D}_{v^2}$.  If
$t/\trel \ll 1$, then two-body encounters (and relaxation) haven't been important.

Consider a distribution function (or phase-space density)
$f = \nobreak f(\vec{x},\vec{v})$, such that the number of stars in a small
spatial-volume $d^3\vec{x}$ and velocity-space volume $d^3\vec{v}$ is given as
$f(\vec{x},\vec{v}) \, d^3\vec{x} \, d^3\vec{v}$.  In a spherical system in which
there are the conserved energy $E$ and angular momentum $\vec{L}$, the six
independent position and velocity variables can be reduced to these four independent,
conserved quantities via the Jeans theorem.  Furthermore, if the system is perfectly
spherically symmetric---which we assume in our analysis, then the three independent
angular momentum components can be replaced with the angular momentum magnitude,
i.e.~$f = f(E,L)$.

If we define a relative potential and relative energy, $\Psi \equiv - \Phi + \Phi_0$,
and, $\mce \equiv - E + \Phi_0 = \Psi - \frac{1}{2} v^2$, then we can calculate the
number density as\footnote{$\Phi_0$ is arbitrary, but
$\Phi_0 \equiv E(r \rightarrow \infty)$ may be convenient.},
	\begin{equation}
	\label{eq:num-dens-int}
	\begin{split}
	n(\vec{x}) = n(x) 	& = 4\pi \int_0^{\sqrt{2\Psi}} f(x,v) v^2 \, dv \\
						& = 4\pi \int_0^\Psi f(\mce) \left[ 2 (\Psi - \mce) \right]^{1/2} \, d \mce.
	\end{split}
	\end{equation}
Inverting this relationship, the distribution function can be calculated from an
isotropic density profile using,
	\begin{equation}
	\label{eq:dist-func}
	\begin{split}
	f(\mce) 	= 	&	\frac{1}{\pi^2 \sqrt{8}} \frac{d}{d \mce} \int_0^\mce
						\frac{ d n}{d \Psi} \frac{ d \Psi}{\left( \mce - \Psi \right)^{1/2} } \\
				=	&	\frac{1}{\pi^2 \sqrt{8}} \left[ \mce^{-1/2}
						\left(\frac{dn}{d\Psi}\right)_{\Psi=0} + \int_0^\mce \frac{ d^2n}{d\Psi^2}
						\frac{ d \Psi}{\left( \mce - \Psi \right)^{1/2} } \right].
	\end{split}
	\end{equation}
We have found the latter form of \eqref{eq:dist-func} to be much simpler and more reliable
to implement.

We follow the discussion and prescription for loss-cone scattering given by
\citet{magorrian1999}, corresponding to a single central object in a spherical
(isotropic) background of stars.  We adapt this prescription simply by
modifying the radius of interaction to be appropriate for scattering with a binary
instead of being tidally disrupted by a single MBH.  A more extensive discussion
of loss-cone dynamics---explicitly considering MBH binary systems and
asphericity---can be found in \citet{merritt2013}.

Stars with a pericenter distance smaller than some critical radius $\rcrit$ will interact
with the binary.  For a fixed energy ($\mce$) orbit, there is then a critical angular
momentum,
$J_{lc}(\mce) = \rcrit \left( 2 \left[ \mce - \Psi(\rcrit) \right] \right)^{1/2}
\approx \rcrit \left( 2 \left[ - \Psi(\rcrit) \right] \right)^{1/2}$, 
which defines the loss-cone \citep{fr76,ls77}.  In general, the number of stars with
energy and angular momentum in the range $d\mce$ and $dJ^2$ around $\mce$ and $J^2$ can
be calculated as,
	\begin{equation}
	N(\mce,J^2) \, d\mce \, d J^2 = 4 \pi^2 \, f(\mce,J^2) \cdot P(\mce, J^2) \, d\mce \, d J^2,
	\end{equation}
where $P(\mce,J^2)$ is the stellar orbital period.  For an isotropic stellar distribution
$f(\mce,J^2) = f(\mce)$, and $P(\mce,J^2) \approx P(\mce)$.  The total number of stars
can be calculated as,
	\begin{equation}
	\label{eq:numstars}
	N_i(\mce) \, d\mce = 4\pi^2 \, f(\mce) \, P(\mce) \, J_i^2(\mce) \, d\mce.
	\end{equation}
For the number of stars in the loss-cone $N_{lc}(\mce)$, this uses the LC angular
momentum, $J_i^2(\mce) = J_{lc}^2(\mce)$; and for all stars $N(\mce)$, the circular
(and thus maximum) angular momentum, $J_i^2(\mce) = J_{c}^2(\mce)$.  When we initially
calculate the distribution function, we use the stellar density profile from Illustris
galaxies which have recently hosted a MBH `merger' event (see: \secref{sec:host-gals}).
We assume that the resulting distribution function $f(\mce)$ is (so-far) unperturbed by
the MBH binary, i.e.~it does not take into account stars already lost (scattered). The
resulting $N_{lc}(\mce)$ from \refeq{eq:numstars} then corresponds to the number of
stars in the `full' loss-cone specifically.

% Consumption / Refilling timescales
Stars in the LC are consumed on their orbital timescale $\torb = \nobreak P(\mce)$.
The rate of flux of stars to within $\rcrit$ is then,
	\begin{equation}
	\label{eq:lc-flux-full}
	\fluxflc(\mce) \, d\mce = 4\pi^2 \, f(\mce) \, J_{lc}^2(\mce) \, d\mce,
	\end{equation}
coming almost entirely from within the central objects sphere of influence $\rinfl$,
defined as $M(r < \rinfl) \approx \mbh$.  Refilling of the loss-cone occurs on the
characteristic relaxation timescale $\trel$.  From \refeq{eq:relax}, it is clear that
$\torb/\trel \approx \tcross/\trel \ll 1$, i.e.~the loss-cone is drained significantly
faster than it is refilled---and the loss-cone will, in general, be far from `full'.

% Stead-State LC Flux
To calculate the steady-state flux of the loss-cone, the Fokker-Planck equation must
be solved with a fixed (unperturbed) background stellar distribution at the outer edge
of the LC and no stars surviving within the scattering region at the inner-edge. A full
derivation can be found in \citet{magorrian1999}, which yields a equilibrium flux of stars,
	\begin{equation}
	\label{eq:lc-flux-eq}
	\fluxsslc (\mce) \, d\mce = 4\pi^2 P(\mce) J^2_c(\mce) f(\mce) \frac{\mu(\mce)}{\ln R_0^{-1} (\mce)},
	\end{equation}
where the angular momentum diffusion parameter $\mu \equiv 2 r^2 \diffco / J_c^2$, and,
	\begin{equation}
	\ln R_0^{-1} = -\ln \rcrit + \begin{cases} q 						& q \geq 1 \\
											   0.186q + 0.824\sqrt{q}	& q < 1 \end{cases},
	\end{equation}
describes the effective refilling radius depending on which refilling regime
\citep[`pin-hole' or `diffusive', see Fig.~1 of][]{ls77} is relevant, for a refilling parameter
$q(\mce) \equiv P(\mce) \, \mu(\mce) / \rcrit(\mce)$.

Equations~\ref{eq:lc-flux-full}~\&~\ref{eq:lc-flux-eq} give the full and steady-state LC fluxes,
which are interpolated between using a logarithmic, `refilling fraction' (\refeq{eq:frefill})
which then determines the hardening rate of each binary in our simulations.

\label{lastpage}

\end{document}